\documentclass[12pt]{article}
\RequirePackage[OT1]{fontenc}
\RequirePackage{amsthm,amsmath}
\RequirePackage[authoryear]{natbib}

\usepackage{url}

\usepackage{graphicx}
\numberwithin{equation}{section}
\usepackage{subcaption}
\usepackage{multirow}
\usepackage{multicol}
\usepackage{threeparttable}
\usepackage{times}
\usepackage{bbm} 
\usepackage{amsfonts} 
\usepackage{mathrsfs}  
\usepackage{bm}
\usepackage{physics} 
\usepackage[plain,noend]{algorithm2e}
\usepackage{enumitem}
\usepackage{float} 

\newtheorem{corollary}{Corollary}[section]

\newtheorem{lemma}{Lemma}[section]

\newtheorem{remark}{Remark}[section]
\newtheorem{theorem}{Theorem}[section]

\begin{document}

\title{\large\bf Score-Matching Representative Approach for Big Data Analysis with Generalized Linear Models}
\author{Keren Li and Jie Yang\\
	Northwestern University and University of Illinois at Chicago}

\maketitle

\begin{abstract}
We propose a fast and efficient strategy, called the representative approach, for big data analysis with generalized linear models, 
especially for distributed data with localization requirements or limited network bandwidth. 
With a given partition of massive dataset, this approach constructs a representative data point for each data block and fits the target model using the representative dataset. In terms of time complexity, it is as fast as the subsampling approaches in the literature. As for efficiency, its accuracy in estimating parameters given a homogeneous partition is comparable with the divide-and-conquer method. 
Supported by comprehensive simulation studies and theoretical justifications, we conclude that mean representatives (MR) work fine for linear models or generalized linear models with a flat inverse link function and moderate coefficients of continuous predictors. For general cases, we  recommend the proposed score-matching representatives (SMR), which may improve the accuracy of estimators significantly by matching the score function values. As an illustrative application to the Airline on-time performance data, we show that the MR and SMR estimates are as good as the full data estimate when available. 
\end{abstract}


\section{Introduction}

In the past decade, big data or massive data has drawn dramatically increasing attention all over the world. It was in the 2009 ASA Data Expo competition when people found out that no statistical software was available to analyze the massive {\it Airline on-time performance data}. At that time, the airline data file, about 12GB in size, consists of 123,534,969 records of domestic flights in the United States from October 1987 to April 2008 \citep{kane2013}. Up to December 2019, the Airline on-time performance data collected from the Bureau of Transportation Statistics consists of 387 files and $N=188,690,624$ valid records in total.

The response in the Airline on-time performance data was treated as a binary variable {\tt Late Arrival} with $1$ standing for late by $15$ minutes or more \citep{wang2015}. Generalized linear models (GLMs) have been widely used for modeling binary responses, as well as Poisson, Gamma, and Inverse Gaussian responses  \citep{pmcc1989, dobson2018}.  In order to fit a GLM with $p$ predictors, a typical algorithm searching for the maximum likelihood estimate (MLE) based on the full data of size $N$ requires $O(\zeta_N Np^2)$ time to run, where $\zeta_N$ is the number of iterations required for the convergence of the full data MLE algorithm \citep{wang2017}. 

Starting in 2009, substantial efforts have been made on developing both methodologies and algorithms towards big data analysis (see, for example, \cite{wang2015}, for a good survey on relevant statistical methods and computing). Divide-and-conquer, also known as divide-and-recombine, split-and-conquer, or split-and-merge, first partitions a big dataset into $K$ blocks, fits the target model block by block, and then aggregates the $K$ fitted models to form a final one \citep{wang2015}. A divide-and-conquer algorithm proposed by  \cite{lin2011} reaches the time complexity of $O(\zeta_{N/K} N p^2)$, where $\zeta_{N/K}$ is the number of iterations required by a GLM MLE algorithm with $N/K$ data points. The accuracy of the estimated parameters based on the divide-and-conquer algorithm relies on the block size $N/K$, which typically depends on the computer memory. Therefore, as $N$ increases, $K$ has to increase accordingly. Typically, its accuracy is not as good as the full data estimate. 
The divide-and-conquer idea  has been applied to numerous statistical problems \citep{chen2014,schifano2016,zhao2016,lee2017,battey2018,shi2018,chen2019}. For example, \cite{chen2014} made an innovative attempt of majority voting on variable selection based on a divide-and-conquer framework that is similar in spirit to combining confidence distributions in meta-analysis \citep{singh2005,xie2011}.  \cite{schifano2016} extended the divide-and-conquer idea for online updating problems motivated from a Bayesian inference perspective.

Another popular strategy for big data analysis is subsampling. For example, leveraging technique has been used to sample a more informative subset of the full data for linear regression problems  \citep{ma2014}. Inspired by D-optimality in optimal design theory,  \cite{wang2017b} proposed an information-based subsampling technique, called IBOSS, for big data linear regression problems. Its time complexity is $O(Np)$ while the ordinary least square (OLS) estimate for linear models takes $O(Np^2)$ time complexity. Motivated by A-optimality,  \cite{wang2017} developed an efficient two-step subsampling algorithm for large sample logistic regression, which is a special case of generalized linear models. The time complexity of the A-optimal subsampler is also $O(Np)$. Compared with the divide-and-conquer strategy, the subsampling approach requires much less computational cost. Nevertheless, its accuracy relies on the subsample size and is typically not as good as the divide-and-conquer estimate.

Other developments include stochastic gradient descent  and the polynomial approximate sufficient statistics approaches. Stochastic gradient descent algorithms (\citealt{tran2015, lin2017}) update in a sequential manner based on a noisy gradient.  Polynomial approximate sufficient statistics methods (\citealt{huggins2017, keeley2019, zoltowski2018}) construct polynomial approximate sufficient statistics for GLMs for any combination of batch, parallel, or streaming.
In terms of accuracy, their performances are comparable with subsampling techniques.  

The success of divide-and-conquer methods relies on the similarity between blocks, 
which can be achieved by randomly partitioning the original dataset. In practice, however, a massive dataset is often stored in multiple files, within which data points have sort of similarity in some fields. We call it {\it homogeneous partition}. For example, the Airline on-time performance data (see Section~\ref{sec:case})  consists of 387 files labeled by month. In each file, all data points share the same values in fields {\tt YEAR}, {\tt MONTH}, and {\tt QUARTER}, while between files these values are different. A divide-and-conquer strategy would not work directly with the original homogeneous partition due to distinct blocks. Even worse, a massive dataset is sometimes distributed in multiple hard disks, or even a network of interconnected computers, known as {\it nodes} (see, for example, a {\it distributed database}  \citep{ozsu2011}). If one needs to re-partition the data randomly, it may involve intensive communication between nodes and require extensive network bandwidth. 

Another critical issue 
on massive data analysis is data localization requirements. Due to security or privacy concerns, {\it data localization} or {\it data residency} law prohibits health records, personal data, payment system data, etc, from being transferred freely (\citeauthor{bowman2017}, 2017; \citeauthor{fefer2020}, 2020). Technology companies, such as Microsoft, also use data storage locale controls in their cloud services (\citeauthor{vogel2014}, 2014). 
The Irish Data Protection Commission sent Facebook a preliminary order to suspend moving data from its European users to the United States (\citeauthor{schechner2020}, 2020). Under such kind of circumstances, the partition of data is predetermined and raw data exchange between nodes is prohibited.
We call such kind of data partition a {\it natural partition}. 

In the computer science literature, algorithms of the CoCoA family have been developed for optimization problems with distributed dataset and a central server (see \cite{smith2018cocoa} for a good review), which combine local solvers for subproblems in an efficient manner. \cite{he2018} further extended CoCoA to the COLA algorithm for decentralized environment, that is, distributed computing without a central coordinator. 

To avoid intensive data communications between nodes and even avoid any raw data transfer, we propose a different  data analysis strategy for distributed massive dataset with data localization requirements, named the {\it representative approach}.  When a massive data is provided in data blocks, either naturally or partitioned by a data binning procedure, we construct a representative data point for each data block using only the data within the block and initial parameter values, and then run regression analysis on the representative data points. The constructed representatives are typically artificial ones which may not belong to the original dataset. The collection of representative data points may be used for further analysis and inference. Compared with the original data, its data volume is significantly reduced.

The representative approach provides an ideal solution for the analysis of naturally partitioned massive data.  By exchanging only the estimated parameters and the representative data points among parallel computing computers, the representative approach can work well even with slow-speed or restricted network connection. It fulfills user privacy or security requirements since analysts perform regression analysis on the representatives without direct access to the raw data.

The representative approach is inspired by the data binning technique in the computer science literature. By discretizing continuous variables into categorical variables (see, for example,  \cite{kot2006}), a data binning procedure partitions a continuous feature space into blocks labeled by so-called smoothing values. It focuses on how to partition data into blocks or bins, while the smoothing values are typically chosen from class labels, boundary points, centers, means, or medians of the data blocks. A data binning procedure is often used for data pre-processing, whose performance is not guaranteed, especially for nonlinear models. 

Different from the data binning technique, the representative approach proposed in this paper assumes a given data partition and concentrates on constructing the best smoothing values, which we call {\it representatives}, more efficiently for a pre-specified regression model. 
With a given partition, the goal of the representative approach is to run as fast as subsampling approaches, while estimating model parameters as accurately as the divide-and-conquer method.

Actually, using the proposed representative approach, a GLM is fitted on $K$ representatives constructed from the original $N$ data points ($K \ll N$). Its time complexity is also $O(Np)$, 
same as the subsampling approaches. On the other hand, the $K$ representatives are not a subset of the $N$ data points, but summarize the information from each single one of the original $N$ data points. 
Based on our comprehensive simulation studies, by matching the score function values of GLMs, the proposed score-matching representative (SMR) approach is often comparable with the full data estimate.

The data binning technique that we consider here is to partition the data according to the predictors. It is different from the sufficient dimension reduction techniques which start with a data partition based on the responses or the conditional class probabilities of binary responses \citep{shin2014,shin2017}. 

The representatives described in this paper are different from the representative points (rep-points) developed in the statistical literature, which aims to represent a distribution the best. Research work on rep-points includes the principal points in \cite{flury1990}, the mse-rep-points in \cite{fang1994}, and the recent support points in \cite{mak2018a, mak2018b}. Although the rep-points can capture the covariate distribution better, they may not be more informative in estimating model parameters, and thus are not as efficient as subsampling methods or the proposed representative approaches (see Section~\ref{sec:support} in the Supplementary Material for comparisons).

The remainder of this paper proceeds as follows. In Section~\ref{sec:rm} we describe the general framework of the representative approach and  data partitions.  After comparing mean, median and mid-point representatives, we recommend mean representative (MR) for big data linear regression analysis. In Section~\ref{sec:smr} we develop SMR along with its theoretical justifications. We recommend SMR for big data general linear regression analysis.   In Section~\ref{sec:case}, we use the Airline on-time performance data as an illustrative example for real big data analysis. We show that the MR and SMR estimates are as accurate as the full data estimate when available. We conclude in Section~\ref{sec:conc}. The proofs of theorems and corollaries, more corollaries,  more  simulations, as well as more details about the real case study, are relegated to the Supplementary Material.

\section{GLM, Massive Data and Mean Representative}\label{sec:rm}

\subsection{Generalized Linear Model and Score Function}\label{sec:glm}

Given the original data set $\{ (\mathbf{x}_i,y_i), i = 1,2,\dots,N \}$ with covariates $\mathbf{x}_i =(x_{i1}, \ldots,$ $x_{id})^T \in \mathbb{R}^d$ and response $y_i \in \mathbb{R}$, we consider a generalized linear model assuming independent response random variable $Y_i$'s and the corresponding predictors $\mathbf{X}_i = (h_1(\mathbf{x}_i), \ldots, h_p(\mathbf{x}_i))^T \in \mathbb{R}^p$. For model-based data analysis with fairly general known functions $h_1(\cdot), \ldots, h_p(\cdot)$, we would rather regard the data set as $\mathscr{D} = \{(\mathbf{X}_i, y_i), i= 1, \ldots, N\}$. 
For many applications, $h_1({\mathbf x}_i)\equiv 1$ corresponds to the intercept. For examples, $\mathbf{X}_i = (1, x_{i1}, \ldots, x_{i7})^T$ for a main-effects model or $\mathbf{X}_i=(1, x_{i1}, x_{i2}, x_{i3}, x_{i1}x_{i2}$, $x_{i1}x_{i3}, x_{i2}x_{i3}, x_{i1}x_{i2}x_{i3})^T$ for a model with interactions.

Following \cite{pmcc1989}, there exists a link function $g$ and regression parameters $\boldsymbol\beta = (\beta_1,...,\beta_p)^T$, such that $\mathbb{E}(Y_i)=\mu_i$ and $\eta_i=g(\mu_i)=\boldsymbol{\mathrm{X}}_i^T \boldsymbol\beta$,
where $Y_i$ has a distribution 
in the exponential family. For typical applications, the link function $g$ is one-to-one and differentiable.

According to \citeauthor{pmcc1989} (1989, Section~2.5), the maximum likelihood estimate of $\boldsymbol\beta$ solves the score equation $s(\boldsymbol\beta; {\mathbf y}, {\bm X}) = 0$, 
where ${\mathbf y} = (y_1, \ldots, y_N)^T$, ${\bm X} = ({\mathbf X}_1, \ldots, {\mathbf X}_N)^T$, and the score function
\begin{equation}\label{eq:score}
	s(\boldsymbol\beta; {\mathbf y}, {\bm X}) = \sum_{i=1}^N (y_i - G(\eta_i)) \nu(\eta_i) {\mathbf X}_i
\end{equation}
with $G(\eta) = g^{-1}(\eta)$ and $\nu(\eta) = G'(\eta)/h(\eta)$, where $h(\eta_i)=\mathrm{Var}(Y_i)$.
For commonly used GLMs, their $\nu$ and $G$ are listed  in Table~\ref{tab:egeta}.

\begin{table}[bt]
	\caption{Examples $\nu(\eta)$ and $G(\eta)$ of commonly used GLMs}\label{tab:egeta}
	\begin{threeparttable}
		\resizebox{\columnwidth}{!}{
			\begin{tabular}{lccc}
				\hline
				\bf{Distribution of $Y$} & \bf{ link function $g$ } & \bf{$\nu(\eta)$ (up to a constant)} & \bf{$G(\eta)$}\\
				\hline
				Normal$(\mu, \phi)$ & {\it identity} & $1 $ & $\eta$\\ 
				Bernoulli$(\mu)$ & {\it logit} & 1 & $\exp(\eta)\{1+\exp(\eta)\}^{-1}$\\ 
				Bernoulli$(\mu)$& {\it progit} & $\phi(\eta)\{\Phi(\eta)\Phi(-\eta)\}^{-1}$ & $\Phi(\eta)$  \\ 
				Bernoulli$(\mu)$ & {\it cloglog} & $\exp(\eta)\{1-\exp[-\exp(\eta)]\}^{-1}$ & $1-\exp\{-\exp(\eta)\} $\\ 
				Bernoulli$(\mu)$ & {\it loglog} &  $\exp(-\eta)\{1-\exp[-\exp(-\eta)]\}^{-1}$ & $\exp\{-\exp(-\eta)\}$ \\ 
				Bernoulli$(\mu)$& {\it cauchit} &  $\pi\{(1+\eta^2)(\pi^2/4-\arctan^2(\eta))\}^{-1}$&$\arctan(\eta)/\pi+1/2$ \\ 
				Poisson$(\mu)$ & {\it log} & 1 & $\exp(\eta)$\\ 
				Gamma$(k, \mu/k)$ & {\it reciprocal} & 1 & $1/\eta$ \\ 
				Inverse Gaussian$(\mu, \phi)$ & {\it inverse squared} & 1 & $1/\sqrt{\eta}$\\ 
				\hline  
			\end{tabular}
		}
	\end{threeparttable}
\end{table}

\subsection{Representative Approaches and Mean Representatives}\label{sec:mrlinear}

In this paper, we assume that the data is given along with a data partition. Let $I=\{1,2,\dots,N \}$ be the data index set. The data partition can be denoted by a partition $\{I_1, I_2,$ $\ldots,$ $I_K \}$ of $I$. The $k$th data block $\mathscr{D}_k = \{({\mathbf X}_i, y_i), i \in I_k\}$ has block size $n_k=\abs{I_k}$. 

A {\it representative approach} for model-based regression analysis  constructs a representative data point $(\tilde{\mathbf X}_k, \tilde{y}_k)$ for data block $\mathscr{D}_k$, $k=1,\dots,K$, and then fit the regression model based on the weighted representative dataset $\tilde{\mathscr{D}} = \{(n_k, \tilde{\mathbf X}_k, \tilde{y}_k), k=1, \ldots, K\}$.  This procedure could be repeated to achieve a desired accuracy level of estimated model parameters. 

Unlike subsampling approaches, a representative data point may not belong to  the original data set. With a given data partition, the goal of the representative approach is to make the model parameter estimate $\tilde{\boldsymbol\beta}$ based on the weighted representative dataset close enough to the full data estimate $\hat{\boldsymbol\beta}$. Given an initial value of parameter estimate, the construction of representative data points in one block shall not be affected by another to facilitate parallel computing. 

In practice, a massive dataset is often provided in multiple data files, which forms a {\it natural partition} with each block representing a data file. In order to improve the efficiency of the proposed representative approach, one may construct a sub-partition within each natural data block or data file such that the predictors have similar values in each finer block.  Many partitioning methods have been proposed in the literature (see, for example, \cite{fahad2014} for a good survey).
From our point of view, there are two types of partitioning methods, grid or clustered. Grid partition is based on the feature space in $\mathbb{R}^p$, with cut points obtained from the summary information of data, such as quantiles (called {\it equal-depth}) or equal-width points, which is usually feasible with a moderate number of predictors \citep{kot2006}. Its time complexity is $O(Np)$.  Clustered partition is based on clustering algorithms.  For example, \cite{pakhira2014} proposed a linear $k$-means algorithm with time complexity $O(Np)$, which is especially useful with large $p$. When the massive data consists of multiple natural blocks or data files, these partitioning methods could be applied on the natural blocks one-by-one to obtain a finer partition of the whole dataset. In practice the sizes of natural blocks or data files are typically fixed. Only the number of data files increases when the massive dataset is updated in a time order. Therefore, the overall time complexity for obtaining all sub-partitions is still $O(Np)$. Improving the efficiency of sub-partitioning is important but out of the scope of this paper.

There are lots of representative choices that could possibly work for the representative approach. A naive choice of the representative is the block center, which is popular in the data binning literature. More specifically, given the $k$th data block $\mathscr{D}_k = \{({\mathbf X}_i, y_i), i \in I_k\}$ with block size $n_k$, 
the options for its representative $\tilde{\mathbf X}_k$ include 
(1) mid-point of the rectangular block, when a grid partition is given; (2) (component-wise) median; (3) (vector) mean, 
that is, $\tilde{\mathbf X}_k = n_k^{-1} \sum_{i\in I_k} {\mathbf X}_i$. Then the weighted representative data for the $k$th block is defined as $(n_k,\tilde{\mathbf X}_k, \tilde{y}_k)$ with $\tilde{y}_k =n_k^{-1} \sum_{i\in I_k} y_i$, which will be justified in Theorem~\ref{thm:mrglm}. 

A comprehensive simulation study with linear models  shows that using block means,  the {\it mean representative approach} (MR), is more efficient than the mid-point  and median options, as well as the IBOSS subsampling approach \citep{wang2017b}. For details please see Section~\ref{sec:linear} in the Supplementary Material.

As for time complexity, the full data ordinary least squares (OLS) estimate takes $O(Np^2)$. The IBOSS \citep{wang2017b} costs $O(Np)$.  For typical applications, $K \ll N$ and $p\ll N$, then the overall time complexity including constructing representatives and fitting models is $O(N)$ for mid-point or $O(Np)$ for median and mean representative approaches for both linear regression models and GLMs. 

The  simulation  studies  in    Section~\ref{sec:linear} of the Supplementary Material also imply  that the maximum Euclidean distance within data blocks  
$\Delta = {\max}_k {\max}_{i,j \in I_k} \|{\mathbf X}_i - {\mathbf X}_j\|$ plays an important role on the efficiency of mean representatives.  For general representatives, the key quantity is the maximum deviation from the corresponding representatives
$\tilde{\Delta} = \max_k \max_{i\in I_k}$ $\|{\mathbf X}_i - \tilde{\mathbf X}_k\|$.

\section{Score-Matching Representative Approach for GLMs}\label{sec:smr}

The MR approach works very well for linear models and can be validated for GLMs when $\Delta$ is sufficiently small. Nevertheless, for moderate $\Delta$ with general GLMs, MR approach is not so satisfactory (see Section~\ref{sec:simulatelogistic}).
In this section, we propose a much more efficient representative approach, called {\it score-matching representative} (SMR) approach for GLMs. 

Recall that in Section~\ref{sec:glm} the maximum likelihood estimate $\hat{\boldsymbol\beta}$ solves the score equation $s(\boldsymbol\beta)=0$. It is typically obtained numerically by the Fisher scoring method \citep{pmcc1989}, which iteratively updates the score function with the current estimate of $\boldsymbol\beta$. 
Inspired by the Fisher scoring method, given some initial values of the estimated parameters, our score-matching representative approach constructs data representatives by matching the values of the score function block by block, and then applies the Fisher scoring method on the representative dataset and gets estimated parameter values for the next iteration. We may repeat this procedure for a few times till a certain accuracy level is achieved. According to our comprehensive simulation studies (see Section~\ref{sec:t} in the Supplementary Material), three iterations are satisfactory for typical applications.

\subsection{Score-Matching Representative Approach}\label{sec:smralgo}

Let $s_k(\boldsymbol\beta) = \sum_{i \in I_k} (y_i - G(\eta_i)) \nu(\eta_i) {\mathbf X}_i$ denote the value of the score function contributed by the $k$th data block $\mathscr{D}_k=\{({\mathbf X}_i, y_i), i\in I_k\}$, and $\tilde{s}_k(\boldsymbol\beta) = n_k (\tilde{y}_k - G(\tilde{\eta}_k)) \nu(\tilde{\eta}_k) \tilde{\mathbf X}_k$ denote the value of the score function based on the weighted representative data $(n_k, \tilde{\mathbf X}_k, \tilde{y}_k)$, where $\eta_i = {\mathbf X}_i^T \boldsymbol\beta$ and $\tilde{\eta}_k = \tilde{\mathbf X}_k^T \boldsymbol\beta$ are functions of $\boldsymbol\beta$.

Suppose the estimated parameter value is $\tilde{\boldsymbol\beta}^{(t)}$ at the $t$th iteration. For the $(t+1)$th iteration, our strategy is  to find the representative $(\tilde{\mathbf X}_k,\tilde{y}_k)$ carrying the same score as the $k$th data block at $\tilde{\boldsymbol\beta}^{(t)}$, that is, $s_k(\tilde{\boldsymbol\beta}^{(t)}) = \tilde{s}_k(\tilde{\boldsymbol\beta}^{(t)})$, or
\begin{equation}\label{eq:ss}
	\sum_{i \in I_k} \nu({\mathbf X}_i^T \tilde{\boldsymbol\beta}^{(t)}) (y_i  -G({\mathbf X}_i^T \tilde{\boldsymbol\beta}^{(t)}))   {\mathbf X}_i = n_k\ \nu(\tilde{\mathbf X}_k^T \tilde{\boldsymbol\beta}^{(t)})(\tilde{y}_k  - G(\tilde{\mathbf X}_k^T \tilde{\boldsymbol\beta}^{(t)}))   \tilde{\mathbf X}_k 	
\end{equation}
Multiplying by $\tilde{\boldsymbol\beta}^{(t)}$ on both sides of \eqref{eq:ss}, we get 
\begin{equation}\label{eq:ssb}
	\sum_{i \in I_k} \nu(\eta_i) (y_i-G(\eta_i))  \eta_i= n_k\ \nu(\tilde{\eta}_k) (\tilde{y}_k -G(\tilde{\eta}_k))  \tilde{\eta}_k
\end{equation}
where $\eta_i = {\mathbf X}_i^T\tilde{\boldsymbol\beta}^{(t)}$ and $\tilde{\eta}_k = \tilde{\mathbf X}_k^T \tilde{\boldsymbol\beta}^{(t)}$. The weight $\nu(\eta_i)\eta_i$ of $y_i$ in \eqref{eq:ssb} suggests that we take $\tilde{y}_k$ as a weighted average of $y_i$'s for the SMR approach, that is,
\begin{equation}\label{eq:ytilde}
	\tilde{y}_k=\frac{\sum_{i\in I_k} \nu(\eta_i) \eta_i y_i}{\sum_{i\in I_k} \nu(\eta_i) \eta_i}
\end{equation}

\begin{remark}\label{rem:ytildek}
{\rm	The $\tilde{y}_k$ defined by \eqref{eq:ytilde} is a natural generalization of the mean representative. Actually, let $\bar{y}_k = n_k^{-1} \sum_{i\in I_k} y_i$, $\bar{\mathbf X}_k = n_k^{-1} \sum_{i \in I_k} {\mathbf X}_i$, and $\bar{\eta}_k = n_k^{-1} \sum_{i \in I_k} \eta_i$ $= \bar{\mathbf X}_k^T \tilde{\boldsymbol\beta}^{(t)}$ denote the mean representative. Since $|\eta_i - \bar{\eta}_k| \leq \|{\mathbf X}_i - \bar{\mathbf X}_k\|\cdot \|\tilde{\boldsymbol\beta}^{(t)}\| = O(\Delta)$, then it can be verified that $\tilde{y}_k = \bar{y}_k + O(\Delta)$ as $\Delta$ goes to $0$, given that $\bar{\eta}_k$ is bounded away from $0$.  
In order to avoid $0 \in (\min_{i\in I_k} \eta_i, \max_{i \in I_k} \eta_i)$, which may lead to unbounded $\tilde{y}_k$ when $\sum_{i\in I_k} \nu(\eta_i)\eta_i$ is close to $0$, 
we split such a block into two pieces by the signs of  $\eta_i$'s and generate two representatives, one for positive $\eta_i$'s and the other for negative $\eta_i$'s. 
}\end{remark}

Since $\tilde{y}_k$ in \eqref{eq:ytilde} does not rely on $\tilde{\eta}_k$, we can further obtain $\tilde{\eta}_k$ by solving \eqref{eq:ssb}. 

\begin{theorem}\label{thm:eta}
	There exists an $\tilde{\eta}_k \in [\min_{i\in I_k} \eta_i,\ \max_{i\in I_k} \eta_i]$ that solves equation~\eqref{eq:ssb}.
\end{theorem}

The existence of the score-matching representative is guaranteed by Theorem~\ref{thm:eta}, whose proof is relegated to the Appendix. Since the solution solving \eqref{eq:ssb} may not be unique, we choose the $\tilde{\eta}_k$ closest to the mean representative to keep a smaller $\tilde\Delta$.
Different from mean representatives, the coordinates of the score-matching representatives corresponding to the intercept term may not be exactly one.

Plugging the $\tilde{\eta}_k$ solving \eqref{eq:ssb} into \eqref{eq:ss}, we get the predictor representative ${\tilde{\mathbf X}_k}$ for SMR:
\begin{equation}\label{eq:xtilde}
	\tilde{\mathbf X}_k = \left[n_k \nu(\tilde{\eta}_k) (\tilde{y}_k  - G(\tilde{\eta}_k))\right]^{-1} \sum_{i \in I_k} \nu({\mathbf X}_i^T\tilde{\boldsymbol\beta}^{(t)})(y_i  - G({\mathbf X}_i^T\tilde{\boldsymbol\beta}^{(t)})) \mathbf{X}_i  
\end{equation}

The $k$th weighted representative data point  $\tilde{\mathscr{D}}_k=(n_k,\tilde{\mathbf X}_k,\tilde{y}_k)$, which carries the same score value at $\tilde{\boldsymbol\beta}^{(t)}$ as the $k$th original data block, will be used for fitting $\tilde{\boldsymbol\beta}^{(t+1)}$ via the Fisher scoring method. 
For typical applications, we may repeat this procedure for $T=3$ times to achieve the desired accuracy level (see Figure~\ref{fig:iter} in the Supplementary Material for a trend of SMR iterations). The complete procedure of the $T$-iteration SMR approach is described by  Algorithm~\ref{alg:smr}.

\begin{algorithm}
	\caption{Score-Matching Representative Approach}	\label{alg:smr}
	\SetAlgoLined
	\KwData{$\mathscr{D} = \{(\mathbf{X}_i, y_i), i= 1, \ldots, N\}$ with a partition $\{I_1, \ldots, I_K\}$ of $I=\{1, \ldots, N\}$}
	\KwResult{SMR estimate $\tilde{\boldsymbol\beta}$ for a generalized linear model and a pre-specified number $T$ of iterations}
	Use the mean representatives as the initial weighted dataset $\tilde{\mathscr{D}}^{(0)}=\{(n_k,\tilde{\mathbf X}_k^{(0)},\tilde{y}_k^{(0)})\}_{k=1}^K$;\\
	Apply the Fisher scoring method on $\tilde{\mathscr{D}}^{(0)}$ and obtain the initial estimate $\tilde{\boldsymbol\beta}^{(0)}$;\\
	\For{$t=1, \ldots, T$}
	{
		\For{$k=1,\dots,K$}{
			Calculate $\eta_i :=\boldsymbol{\mathrm{X}}_i^T \tilde{\boldsymbol\beta}^{(t-1)}$ for $i \in I_k$;\\
			Calculate $\tilde{y}_k^{(t)}$ by \eqref{eq:ytilde};\\
			Solve the one-dimensional equation \eqref{eq:ssb} for $\tilde{\eta}_k^{(t)}$;\\
			Calculate $\tilde{\mathbf X}_k^{(t)}$ by \eqref{eq:xtilde};\\
		}
		Apply the Fisher scoring method on the weighted data set $\tilde{\mathscr{D}}^{(t)}=\{(n_k,\tilde{\mathbf X}_k^{(t)},\tilde{y}_k^{(t)})\}_{k=1}^K$ and obtain $\tilde{\boldsymbol\beta}^{(t)}$;\\
	}
	$\tilde{\boldsymbol\beta}:=\tilde{\boldsymbol\beta}^{(T)}$
\end{algorithm}

For a GLM, the time cost is $O(Np)$ for calculating all $\eta_i$'s, $O(N)$ for calculating all $\tilde{y}_k$'s using \eqref{eq:ytilde}, $O(N)$ plus $\zeta_r$ iterations for solving \eqref{eq:ssb}, and $O(N p)$ for calculating all $\tilde{\mathbf X}_k$'s by \eqref{eq:xtilde}. Along with the time cost $O(\zeta_K Kp^2)$ for finding the MLE baesd on $K$ representative points, a 3-iteration SMR requires $O(Np+N\zeta_r+\zeta_K Kp^2)$. Since $\zeta_r,\zeta_K,K,p \ll N$, the time complexity of SMR for a GLM is essentially $O(Np)$.

\subsection{Simulation Studies with Logistic Regression Model}\label{sec:simulatelogistic}

Logistic regression model is one of the most widely used generalized linear models. \cite{wang2017} proposed an A-optimal subsampling approach for big data logistic regression.  
Following \cite{wang2017}, we run a comprehensive simulation study with the model
\begin{equation}\label{eq:logistic}
	{\rm logit} \left(P(Y_i=1\mid {\mathbf x}_i)\right) = \beta_0 + \beta_1 x_{i1} + \cdots + \beta_7 x_{i7}
\end{equation}

Analogue to the simulation setup in \cite{wang2017}, we choose $\beta_0=0$, $\beta_1 = \cdots = \beta_7= 0.5$. For simulating $\mathbf{x}_i = (x_{i1}, \ldots, x_{i7})^T$, we consider six unbounded distributions plus a bounded one: (1) {\bf mzNormal},  $N_7({\mathbf 0}, \boldsymbol\Sigma)$ with $\boldsymbol\Sigma$ having diagonal $1$ and off-diagonal $0.5$; (2) {\bf nzNormal}, $N_7(1.5\cdot {\mathbf 1}, \boldsymbol\Sigma)$, a case with imbalanced responses, where ${\mathbf 1}$ is a vector of all ones; (3) {\bf ueNormal}, $N_7({\mathbf 0}, \boldsymbol\Sigma_u)$ with $\boldsymbol\Sigma_u$ having diagonals $\{1^2, \ldots, 7^2\}$ and off-diagonal $0.5$; (4) {\bf mixNormal}, $0.5 N_7({\mathbf 1}, \boldsymbol\Sigma) + 0.5 N_7(-{\mathbf 1}, \boldsymbol\Sigma)$, a case with bimodal ${\mathbf x}_i$; (5) $\boldsymbol{T}_3$, Multivariate $t$ with $3$ degrees of freedom ${\mathbf t}_3({\mathbf 0}, \boldsymbol\Sigma)/10$, a case with heavy tails; (6) {\bf EXP}, $\exp(\lambda=2)$, an iid case with a heavier tail on the right; (7) {\bf BETA}, ${\rm Beta}(\alpha=0.5, \beta=0.5)$, a bounded iid case with ``U'' shaped distribution.

For illustration purpose, we choose a moderate population size $N=10^6$ in this simulation study. In the absence of a natural partition, we use two data-driven partitions: (1) An equal-depth partition with $m=4$ splits for each predictor, that is, using the three sample quartiles (25\%, 50\%, and 75\%) as the cut points for each predictor and partitioning the whole data into up to $4^7 = 16,384$ blocks (after removing empty blocks, the number of blocks $K$ is actually between $11,488$ and $16,384$); (2) a $k$-means partition with the number of blocks $K=1000$. The proposed SMR approach starts with MR estimates as its initial values and repeats the iterations for 3 times. 

For comparison purpose, stochastic gradient descent (SGD) proposed by \cite{tran2015},  A-optimal subsampling (A-opt) by \cite{wang2017} with subsample size  $20,000$, and divide-and-conquer (DC) by \cite{lin2011}  with $1000$ blocks from a random partition are applied to the simulated data. 

Table~\ref{tab:logitdd} shows the average and standard deviation of the root mean squared errors (RMSEs, $(\sum_{i=1}^7 (\tilde\beta_i-\beta_i)^2/7)^{1/2}$) between the estimated parameter value $\tilde{\beta}_i$'s and the true value $\beta_i$'s across different simulation settings and each with 100 independent simulations. 

\begin{table}[bt]
	\centering
	\caption{Average (std) of RMSEs ($10^{-3}$) of 100 simulations for logistic regression model}\label{tab:logitdd}
	\begin{threeparttable}
		\resizebox{\columnwidth}{!}{
			\begin{tabular}{cr|rr|rr|rrr}
				\hline
				{\bf Simulation}  &\multicolumn{1}{c}{\bf Full}  &\multicolumn{2}{c}{{\bf Equal-depth}} & \multicolumn{2}{c}{{\bf $k$-means }} & \\
				\cline{3-6}
				{\bf setup}	   &  \multicolumn{1}{c}{\bf data} & \multicolumn{1}{c}{\bf MR} & \multicolumn{1}{c}{\bf SMR}  & \multicolumn{1}{c}{\bf MR} & \multicolumn{1}{c}{\bf SMR}& \multicolumn{1}{c}{\bf A-opt} & \multicolumn{1}{c}{\bf DC}  &\multicolumn{1}{c}{\bf SGD}\\
				\hline
				{\bf mzNormal}  & 3.7 (1.1) & 20.4 (1.0) & 3.9 (1.2) & 17.9 (1.0) & 4.1 (1.2) & 20.4 (5.7) & 7.9 (1.1) & 55.9 (14.8) \\ 
				{\bf nzNormal} & 7.2 (2.0) & 20.4 (1.7) & 9.6 (2.6) & 17.4 (1.7) & 8.3 (2.0) & 24.4 (6.1) & 21.5 (1.6) & 99.2 (27.8) \\ 
				{\bf ueNormal} & 2.1 (0.8) & 170.4 (0.7) & 4.4 (1.1) & 195.7 (3.8) & 14.8 (7.4) & 9.0 (3.3) & 13.2 (1.3) & 62.9 (77.0) \\ 
				{\bf mixNormal} & 5.0 (1.3) & 19.9 (1.1) & 5.4 (1.6) & 13.8 (1.2) & 5.7 (1.6) & 22.9 (5.5) & 12.1 (1.2) & 79.3 (21.0) \\ 
				$\boldsymbol{T}_3$ & 16.0 (4.4) & 30.5 (4.2) & 20.9 (5.8) & 25.1 (7.5) & 23.4 (7.3) & 97.9 (30.1) & 19.6 (3.9) & 177.8 (17.8) \\ 
				{\bf EXP}  & 6.2 (1.7) & 23.0 (2.6) & 13.4 (2.4) & 20.3 (3.1) & 8.8 (2.5) & 31.3 (7.9) & 18.2 (2.3) & 147.9 (20.9) \\ 
				{\bf BETA}  & 7.5 (2.4) & 7.6 (2.5) & 7.6 (2.5) & 11.4 (3.0) & 11.4 (3.7) & 39.9 (11.6) & 9.3 (2.5) & 179.4 (20.0) \\ 
				\hline  
			\end{tabular}
		}
		\begin{tablenotes}
			\item {\footnotesize Sample size $N=10^6$; MR, SMR: equal-depth partition, $m=4$ or $k$-means partition, $K=1000$;}
			\item {\footnotesize A-opt: A-optimal subsampler, $20,000$ subsamples;  DC: Divide-and-Conquer, $1000$ random blocks;}
			\item {\footnotesize SGD: stochastic gradient descent}
		\end{tablenotes}
	\end{threeparttable}
\end{table}

According to Table~\ref{tab:logitdd}, MR, A-opt and SGD do not match either DC or SMR in terms of accuracy.
SMR performs either the best or comparable with DC. For more than half of the scenarios, SMR is  even comparable with the full data estimates.   
With the $k$-means partition, SMR achieves roughly the same accuracy level with only $1000$ representatives. A justification under linear models, which is relegated to the Supplementary Material (Section~\ref{sec:linear}), shows that a best partition keeps the cluster size $\delta_k = \max_{i,j\in I_k} \|{\mathbf X}_i - {\mathbf X}_j\|$ the same for $k=1, \ldots, K$, which partially explains the better performance of $k$-means partitions.

As a conclusion, when the predictors are bounded,
MR is a fast and low-cost (computationally cheaper) solution for big data analysis with generalized linear models. 
For more general cases, especially when a higher accuracy level is desired, MR can be used as a pre-analysis for SMR, while the latter has a significant improvement across different scenarios and different partitions.

\subsection{Other GLM Examples}\label{sec:simuom}

Commonly used GLMs include binary responses with logit, probit, cloglog, loglog, and cauchit links, Poisson responses with log link, Gamma responses with reciprocal link, Inverse Gaussian responses with inverse squared link (see detailed formulae 
in Table~\ref{tab:egeta}). 

\begin{table}[bt]
	\centering
	\caption{Average (std) of RMSEs  ($10^{-3}$) of 100 simulations for three GLMs }\label{tab:3models}
	\begin{threeparttable}
		\resizebox{\columnwidth}{!}{
			\begin{tabular}{lrrr|rrr|rrr}
				\hline
				&\multicolumn{3}{c}{\bf Binary with cloglog} & \multicolumn{3}{c}{\bf Poisson with log} & \multicolumn{3}{c}{\bf Logistic with interactions} \\
				\hline
				&  \multicolumn{1}{c}{\bf Full} &  \multicolumn{1}{c}{\bf MR} & \multicolumn{1}{c}{\bf SMR}   &  \multicolumn{1}{c}{\bf Full} &  \multicolumn{1}{c}{\bf MR} & \multicolumn{1}{c}{\bf SMR} &  \multicolumn{1}{c}{\bf Full} &  \multicolumn{1}{c}{\bf MR} & \multicolumn{1}{c}{\bf SMR} \\
				\hline
				From true  & 2.81 & 43.73 & 3.88 & 0.22 & 29.87 & 12.17 & 3.51 & 7.40 & 5.12 \\ 
				& (0.75)&(0.99)&(1.04)&(0.06)&(9.86)&(10.41)&(1.20)&(2.01)&(1.52) \\ 
				From full & 0 & 43.57 & 2.57 & 0 & 29.88 & 12.17 & 0 & 6.45 & 3.79 \\ 
				& -&(0.55)&(0.67)&-&(9.86)&(10.41)&-&(0.42)&(0.99) \\ 
				\hline
			\end{tabular}
		}
		\begin{tablenotes}
			\item {\footnotesize Sample size $N=10^6$; Covariate distribution: {\bf mzNormal}; MR, SMR: $k$-means partition, $K=1000$}
		\end{tablenotes}
	\end{threeparttable}
\end{table}

In Table~\ref{tab:3models}, we show the RMSEs ($(\sum_{i=1}^p (\tilde\beta_i-\beta_i)^2/p)^{1/2}$) from the true parameter $\boldsymbol\beta$  and the RMSEs ($(\sum_{i=1}^p (\tilde\beta_i-\hat\beta_i)^2/p)^{1/2}$) from the full data estimate $\hat{\boldsymbol\beta}$. The representative approaches, MR and SMR, are based on $k$-means partition with $K=1000$ for the following three models: (1) Binary response with complementary log-log (or cloglog) link $g(\mu)=\log (-\log(1-\mu))$. Since $G(\eta) = 1-\exp\{-\exp(\eta)\}$ is relatively flat, SMR estimate is comparable with the full data estimate even with a not-so-good MR estimate.
(2) Poisson response with the canonical link  $g(\mu)=\log \mu$. Since $G(\eta) = \exp(\eta)$ increases exponentially, the improvement of SMR is slowed down with a not-so-good MR estimate. The variances of MR and SMR estimates are both high. Thus a good initial value for Poisson regression is crucially important (see Section~\ref{sec:asym} for detailed discussion on how the slope of $G(\eta)$ affects the performance of SMR). 
(3) Logistic model with interactions. 
In this case, we simulate $\mathbf{x} = (x_1, x_2, x_3)^T$ from {\bf mzNormal} and assume the non-intercept predictors to be $(h_1(\mathbf{x}), \ldots, h_7(\mathbf{x})) =(x_1, x_2, x_3, x_1 x_2, x_1 x_3, x_2 x_3, x_1 x_2 x_3)$. Both MR and SMR estimates work well.

\subsection{Theoretical Justification of SMR}

First of all, for the proposed SMR approach in Section~\ref{sec:smralgo}, the full data estimate $\hat{\boldsymbol\beta}$ is a stationary point of the SMR iteration.
That is, if the current estimate $\tilde{\boldsymbol\beta}^{(t)} = \hat{\boldsymbol\beta}$, then the representative dataset achieves score $0$ at $\tilde{\boldsymbol\beta}^{(t)}$ and thus $\tilde{\boldsymbol\beta}^{(t+1)} = \hat{\boldsymbol\beta}$ as well.

Recall that $\hat{\boldsymbol\beta}$  is the maximum likelihood estimate (MLE) based on the full data $\mathscr{D} = \{({\mathbf X}_i, y_i), i=1, \ldots, N\}$. Let $\tilde{\boldsymbol\beta}$ be the MLE based on a weighted representative data $\tilde{\mathscr{D}} = \{(n_k, \tilde{\mathbf X}_k, \tilde{y}_k), k=1, \ldots, K\}$, which could be obtained by MR, SMR, or other representative approaches. 
Since $\|{\mathbf X}_i - {\mathbf X}_j\| \leq \|{\mathbf X}_i - \tilde{\mathbf X}_k\| + \|{\mathbf X}_j - \tilde{\mathbf X}_k\|$ for any $i, j \in I_k$, then $\Delta = \max_k \max_{i,j\in I_k} \|{\mathbf X}_i - {\mathbf X}_j\| \leq 2\tilde{\Delta}$, where $\tilde{\Delta} = \max_k \max_{i\in I_k} \|{\mathbf X}_i - \tilde{\mathbf X}_k\|$. 
Theorem~\ref{thm:mrglm} below provides asymptotic results for fairly general representative approaches, whose proof is relegated to the Supplementary Material (Section~\ref{sec:proofs}).

\begin{theorem}\label{thm:mrglm}
	For a generalized linear model with a given dataset, suppose its log-likelihood function $l(\boldsymbol\beta)$ is strictly concave and twice differentiable on a compact set $B \subset \mathbb{R}^p$ and its maximum can be attained in the interior of $B$. Suppose the representatives satisfies $\tilde{y}_k = n_k^{-1} \sum_{i\in I_k} y_i$. Then $\tilde{\boldsymbol\beta} \rightarrow \hat{\boldsymbol\beta}$ as $\tilde{\Delta} \to 0$. Furthermore, $\|\tilde{\boldsymbol\beta}-\hat{\boldsymbol\beta}\|=O(\tilde\Delta^{1/2})$.
\end{theorem}

Theorem~\ref{thm:mrglm} covers mean, median, and mid-point representatives. For mean and mid-point representatives, we actually have $\tilde{\Delta} \leq \Delta$. For median representatives, we also have $\tilde{\Delta} \leq p^{1/2}\Delta$. Thus as direct corollaries, $\|\tilde{\boldsymbol\beta} - \hat{\boldsymbol\beta}\| = O(\Delta^{1/2})$ for all the three center representatives (see Corollaries~\ref{cor:center} \& \ref{cor:cate} in Section~\ref{sec:proofs} of the Supplementary Material).

\begin{theorem}\label{thm:smrglm}
	Suppose $\tilde{y}_k = n_k^{-1} \sum_{i\in I_k} y_i + O(\tilde{\Delta})$. Under the same conditions as in Theorem~\ref{thm:mrglm}, for any given $t$, the SMR  estimate $\tilde{\boldsymbol\beta}^{(t)}$ converges to  $\hat{\boldsymbol\beta}$ if $\tilde{\Delta}$ goes to zero, and $\|\tilde{\boldsymbol\beta}^{(t)} - \hat{\boldsymbol\beta}\| = O(\tilde{\Delta}^{1/2})$.
\end{theorem}	

Technically speaking, any representative approach satisfying \eqref{eq:ss} could be called a score-matching approach. The proposed SMR approach which satisfies \eqref{eq:ytilde} and \eqref{eq:xtilde} is one of the possible solutions for \eqref{eq:ss}, which is a natural extension of the MR approach. Both Theorem~\ref{thm:smrglm} and the following theorem provide consistency results for general score-matching approaches, whose proofs are relegated to Section~\ref{sec:proofs} of the Supplementary Material.

\begin{theorem}\label{thm:smr1}
	Consider a more general iterative representative approach with estimated parameter $\tilde{\boldsymbol\beta}^{(t)}$ at its $t$th iteration. Suppose for the $(t+1)$th iteration, for each $k=1, \ldots, K$, the obtained weighted representative data $(n_k, \tilde{\mathbf X}_k^{(t+1)}, \tilde{y}_k^{(t+1)})$ satisfies the following two conditions:
	\begin{enumerate}[label=\textnormal{(\alph*)}]
		\item The representative matches the score function at $\tilde{\boldsymbol\beta}^{(t)}$, that is, ~\eqref{eq:ss} is true;  
		\label{itm:cond1}
		\item The representative response 
		$\tilde{y}_k^{(t+1)} = \bar{y}_k + O(\tilde{\Delta})$, where $\bar{y}_k = n_k^{-1} \sum_{i \in I_k} y_i$.
		\label{itm:cond2}
	\end{enumerate}
	Then the estimated parameter $\tilde{\boldsymbol\beta}^{(t+1)}$ based on the weighted representative data satisfies
	\begin{equation}\label{eq:convergence}
		\|\tilde{\boldsymbol\beta}^{(t+1)} - \hat{\boldsymbol\beta}\| \leq \rho(\tilde{\Delta}) \
		\|\tilde{\boldsymbol\beta}^{(t)} - \hat{\boldsymbol\beta}\| + o(\tilde{\Delta}^{1/2})
	\end{equation}
	where $\rho(\tilde{\Delta}) = O(\tilde{\Delta}) < 1$ for small enough $\tilde{\Delta}$.
	Therefore, $\tilde{\boldsymbol\beta}^{(t)} \rightarrow \hat{\boldsymbol\beta}$ as $t\rightarrow \infty$ and $\tilde{\Delta} \rightarrow 0$.
\end{theorem}

\begin{remark}\label{rem:rho}
{\rm	We call $\rho(\tilde{\Delta})$ in ~\eqref{eq:convergence}  the {\it global rate of convergence}, which depends on the size of $\tilde{\Delta}$. Its specific form can be found in the proof of Theorem~\ref{thm:smr1}. Based on our experience, even for moderate size of $\tilde{\Delta}$, $\rho(\tilde{\Delta})$ can be significantly smaller than $1$ and the first few iterations can improve the accuracy level significantly. 
}\end{remark}

For score-matching representatives, condition~\ref{itm:cond1} of Theorem~\ref{thm:smr1}  holds instantly.
As for condition~\ref{itm:cond2} of Theorem~\ref{thm:smr1}, if $|\bar{\eta}_k| > \delta$ for some $\delta > 0$ as $\Delta$ goes to $0$, then $\tilde{y}_k = \bar{y}_k + O(\Delta)$ (see Remark~\ref{rem:ytildek}), and thus $\tilde{y}_k = \bar{y}_k + O(\tilde{\Delta})$ since $\Delta \leq 2 \tilde\Delta$. After splitting blocks according to the signs of $\eta_i$'s, the cases of data blocks with $\bar{\eta}_k$ close to $0$ are rare. For those blocks, we may simply define $\tilde{y}_k = \bar{y}_k$. Thus condition~\ref{itm:cond2} can be guaranteed in practice.

The conditions and conclusions of Theorem~\ref{thm:smr1} are expressed in terms of $\tilde{\Delta}$. When applying the SMR approach, a slight modification may guarantee $\tilde{\Delta} \leq \Delta \leq 2\tilde{\Delta}$. Actually in our simulation studies, it is almost always the case for the proposed SMR approach. Occasionally, $\tilde{\mathbf X}_k$ could be out of the convex hull of $\{{\mathbf X}_i, i\in I_k\}$ due to $\nu(\tilde{\eta}_k)(\tilde{y}_k  - G(\tilde{\eta}_k)) \approx 0$. For such kind of cases, we replace $\tilde{\mathbf X}_k$ with the MR representative $\bar{\mathbf X}_k$. The difference caused for the value of score function is negligible. By this way, the conclusions of Theorem~\ref{thm:smr1} hold for $\Delta \rightarrow 0$ as well.

Overall, Theorem~\ref{thm:smr1} justifies why the proposed SMR approach works well.

\begin{corollary}\label{cor:cate2}
	When $\Delta=0$ or $\tilde{\Delta}=0$, MR and SMR generate the same set of representatives. Both SMR and MR estimates are equal to the full data estimate for GLMs. 
\end{corollary}

A special case of Corollary~\ref{cor:cate2}, whose proof is relegated to Section~\ref{sec:proofs} of the Supplementary Material, is when all covariates are categorical and the dataset is naturally partitioned by distinct covariate values. When most covariates are categorical except for  a few continuous variables, for example, the Airline on-time performance data analysis in Section~\ref{sec:case}, the partition could be chosen such that $\Delta$ is fairly small and thus both MR and SMR estimates work very well.

\subsection{Asymptotic Properties of MR and SMR for Big Data}\label{sec:asym}

In order to study the asymptotic properties of MR and SMR estimates as $N$ goes to $\infty$, we assume that the predictors ${\mathbf X}_1, \ldots, {\mathbf X}_N \in \mathbb{R}^p$ are iid $\sim F$ with a finite expectation, and the partition $\{B_1, \ldots, B_K\}$ of the predictor space $\mathbb{R}^p$ is fixed. To avoid trivial cases, we assume $p_k = F(B_k) > 0$ for each $k=1, \ldots, K$.  Then the index block $I_k = \{i\in \{1, \ldots, N\}\mid {\mathbf X}_i \in B_k\}$ with size $n_k$. By the strong law of large numbers (see, for example, \citeauthor{resnick1999} (1999, Corollary~7.5.1)),  as $N\rightarrow \infty$, $n_k/N \rightarrow p_k >0$ almost surely. In order to investigate asymptotic properties, we consider the discrepancy from the true parameter value $\boldsymbol\beta$ instead of the estimate $\hat{\boldsymbol\beta}$ based on the full data.  

For the MR approach, as $N\rightarrow \infty$,  
\begin{align}\label{eq:asymr}
	\tilde{\mathbf X}_k \rightarrow p_k^{-1} \int_{B_k} {\mathbf x}\ F(d{\mathbf x}),\>\>\> 
	\tilde{y}_k \rightarrow p_k^{-1} \int_{B_k} G({\boldsymbol\beta}^T {\mathbf x})\ F(d{\mathbf x})
\end{align}
almost surely. If the link function $g$ or $G=g^{-1}$ is linear, then $g(\tilde{y}_k) - \tilde{\mathbf X}_k^T{\boldsymbol\beta} \rightarrow 0$ and thus the MR estimate $\tilde{\boldsymbol\beta} \rightarrow \boldsymbol\beta$. Nevertheless, in general $g$ is nonlinear, and the accuracy of the MR estimate mainly depends on the size $\Delta$ of blocks, not the sample size $N$. In other words, for a general GLM and a fixed partition of the predictor space, the accuracy of the MR estimate is restricted by \eqref{eq:asymr} and thus will not benefit from an increased sample size.

Different from MR, by matching the score function of the full data, the proposed SMR approach can still improve its accuracy as the sample size increases, even with a fixed partition of the predictor space.
Actually, for a general GLM, $\mathbb{E} (Y_i) = G(\eta_i)$ and $Y_i - G(\eta_i) \overset{ind}{\sim} (0, \sigma_i^2)$, where $\sigma_i^2 = \mathrm{Var}(Y_i) = h(\eta_i) > 0$. For a bounded block $B_k$, $\max_{i\in I_k} \sigma_i^2$ is also bounded. By the strong law of large numbers for independent sequence of random variables (see, for example, \citeauthor{resnick1999} (1999, Corollary~7.4.1)) and the first-order Taylor expansion, as $N \rightarrow \infty$ and thus $n_k \rightarrow \infty$, the left hand side (LHS) of \eqref{eq:ssb} after divided by $n_k$ is
\begin{eqnarray}
	& & n_k^{-1} \sum_{i \in I_k} \nu(\eta_i) \eta_i (y_i-G(\eta_i)) \nonumber \\
	&=& n_k^{-1} \sum_{i\in I_k} \nu(\eta_i)\eta_i \left[y_i-G({\mathbf X}_i^T {\boldsymbol\beta}) - G'({\mathbf X}_i^T {\boldsymbol\beta}){\mathbf X}_i^T (\tilde{\boldsymbol\beta}^{(t)}-\boldsymbol\beta) + O(\|\tilde{\boldsymbol\beta}^{(t)}-\boldsymbol\beta\|^2)    \right] \nonumber\\
	&\overset{a.s.}{\rightarrow}&  n_k^{-1} \sum_{i\in I_k} \nu(\eta_i)\eta_i \left[- G'({\mathbf X}_i^T {\boldsymbol\beta}){\mathbf X}_i^T (\tilde{\boldsymbol\beta}^{(t)}-\boldsymbol\beta) + O(\|\tilde{\boldsymbol\beta}^{(t)}-\boldsymbol\beta\|^2)    \right] \label{eq:lhs1}\\
	&=& -\nu(\tilde{\eta}_k)\tilde{\eta}_k G'(\tilde{\mathbf X}_k^T\boldsymbol\beta) \tilde{\mathbf X}_k^T (\tilde{\boldsymbol\beta}^{(t)}-\boldsymbol\beta)+O(\Delta \|\tilde{\boldsymbol\beta}^{(t)}-\boldsymbol\beta\|)+ O(\|\tilde{\boldsymbol\beta}^{(t)}-\boldsymbol\beta\|^2) \label{eq:lhs2}
\end{eqnarray}
From \eqref{eq:lhs1} we see that as $N$ increases, the leading discrepancy of LHS caused by response $y_i$'s vanishes. Even if the maximum block size $\Delta$ is fixed, when $\tilde{\boldsymbol\beta}^{(t)}$ is close to $\boldsymbol\beta$, the LHS of \eqref{eq:ssb} is small, and so is its right hand side. For blocks with $\tilde{\eta}_k$ away from $0$, it indicates that $\tilde{y}_k - G(\tilde{\eta}_k)$ and thus $\tilde{y}_k - G(\tilde{\mathbf X}_k^T\boldsymbol\beta)$ are small. That is, when $N\rightarrow \infty$, the SMR representatives $\{(\tilde{\mathbf X}_k, \tilde{y}_k), k=1, \ldots, K\}$ stay close to the true curve, $\mu = E(Y) = G({\mathbf X}^T \boldsymbol\beta)$, which leads to a faster convergence rate of the SMR estimate towards $\boldsymbol\beta$ than MR's.

From \eqref{eq:lhs2} we conclude that a relatively large $G'(\tilde{\mathbf X}_k^T\boldsymbol\beta)$ may slow down the convergence of the SMR estimate. For example, under a Poisson regression model with log link (see Model~(2) in Section~\ref{sec:simuom}), $G(\eta)=e^{\eta}$. If the initial estimate of the regression parameter is not so accurate, SMR may converge slowly. For such kind of cases, we suggest a finer partition or smaller $\Delta$ to obtain a good initial estimate. For models with fairly flat $G$ functions, such as models with logit link, $G'$ is small for most blocks. For this kind of cases, even if the initial estimate for SMR is not so accurate, we can still improve the accuracy of the estimate significantly after a few iterations.

To reveal the performance of SMR visually over sample size $N$, we use the first simulation setup {\bf mzNormal} for illustration purpose, varying $N=10^5, 10^6, 10^7, 10^8$.  For MR and SMR, the number of blocks is fixed at $K=1000$ with a $k$-means partition. For the divide-and-conquer method, the block size typically restricted by the computer memory and thus will not change as $N$ increases. For illustration purpose,  we fix the same block size $1000$ as in Section~\ref{sec:simulatelogistic}. As $N$ increases, the number of blocks for the divide-and-conquer method increases proportionally.

\begin{figure}[bt]
	\centering
	\begin{subfigure}[b]{0.45\textwidth}
		\includegraphics[width=\textwidth]{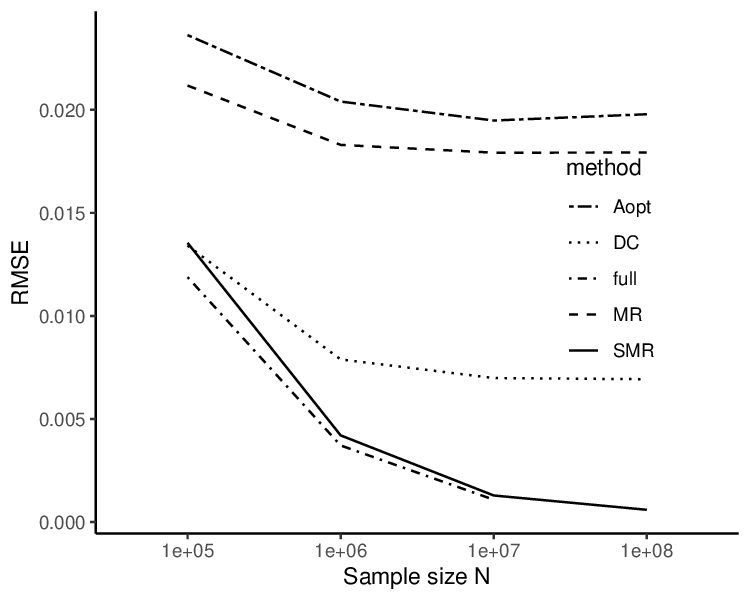}
		\caption{RMSE from true $\boldsymbol\beta$}
	\end{subfigure}
	~ 
	\begin{subfigure}[b]{0.45\textwidth}
		\includegraphics[width=\textwidth]{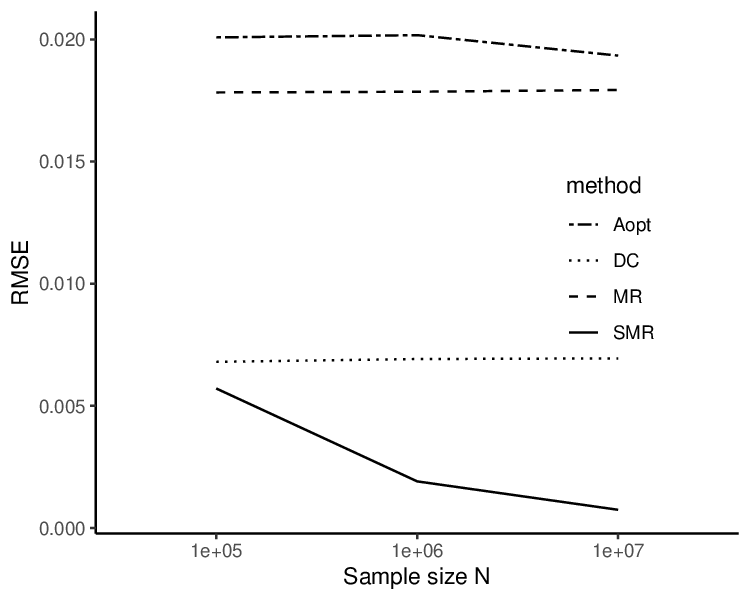}
		\caption{RMSE from full data $\hat{\boldsymbol\beta}$}
	\end{subfigure}
	~ 	
	\caption{RMSE vs $N$ of MR, SMR, A-opt, and divide-conquer for logistic model}		\label{fig:logitdd}
\end{figure}

Figure~\ref{fig:logitdd} (see also Table~\ref{tab:popu}) shows how much those methods benefit from increased sample size $N$ over 100 simulations for each $N$. Figure~\ref{fig:logitdd}(a) shows that in terms of RMSE from the true parameter value, SMR's estimate is comparable with the full data estimate and converges to $\boldsymbol\beta$ much faster than other methods. Figure~\ref{fig:logitdd}(b) shows that SMR's estimate quickly gets closer to the full data estimate $\hat{\boldsymbol\beta}$ as $N$ increases, while other methods' estimates do not show a clear pattern getting closer. 
This simulation study confirms our conclusion in Section~\ref{sec:asym}.

\begin{table}[bt]
	\centering
	\caption{Average (std) of RMSEs ($10^{-3}$) of 100 simulations  with various $N$}\label{tab:popu}
	\begin{threeparttable}
		\resizebox{\columnwidth}{!}{
			\begin{tabular}{rrrrrr|rrrr}
				\hline
				\multicolumn{1}{c}{ $N$  }&\multicolumn{5}{c}{\bf RMSE from true $\boldsymbol\beta$} & \multicolumn{4}{c}{\bf RMSE from full $\hat{\boldsymbol\beta}$} \\
				\cline{2-10}
				& \multicolumn{1}{c}{\bf Full data} &\multicolumn{1}{c}{\bf MR}  & \multicolumn{1}{c}{\bf SMR}  & \multicolumn{1}{c}{\bf Aopt} & \multicolumn{1}{c}{\bf DC} & \multicolumn{1}{c}{\bf MR}  & \multicolumn{1}{c}{\bf SMR} & \multicolumn{1}{c}{\bf Aopt} & \multicolumn{1}{c}{\bf DC}\\
				\hline
				$10^5$ & 11.9 (3.5) & 21.2 (3.3) & 13.5 (3.6) & 23.6 (7.1) & 13.4 (3.5) & 17.8 (1.1) & 5.7 (1.7) & 20.1 (6.0) & 6.8 (0.4) \\ 
				$10^6$ & 3.7 (1.1) & 18.3 (1.0) & 4.2 (1.2) & 20.4 (5.7) & 7.9 (1.1) & 17.9 (0.3) & 1.9 (0.6) & 20.2 (5.7) & 6.9 (0.1) \\ 
				$10^7$ & 1.1 (0.3) & 17.9 (0.3) & 1.3 (0.4) & 19.5 (5.2) & 7.0 (0.3) & 17.9 (0.1) & 0.7 (0.2) & 19.3 (5.1) & 6.9 (0.0) \\ 
				$10^8$ & - & 17.9 (0.1) & 0.6 (0.7) & 19.8 (5.0) & 6.9 (0.1) & - & - & - & - \\ 
				\hline
			\end{tabular}
		}
		\begin{tablenotes}[noitemsep]
			\item {\footnotesize Notes: Logistic model; predictor distribution: {\bf mzNormal;}}
			\item {\footnotesize MR, SMR: $k$-means with $K=1000$; DC: $1000$ observations per block; Aopt: $20,000$ subsample size;}
			\item {\footnotesize Full data estimate is not available at $N=10^8$ due to memory limit}
		\end{tablenotes}
	\end{threeparttable}
\end{table}

\subsection{CPU Time}

We use R (version 3.6.1) for all simulation studies listed in this paper. For the IBOSS \citep{wang2017b} and A-optimal \citep{wang2017} subsampling methods, we use the R functions provided by the authors. We also use data.table format in R package {\tt data.table}  for calculating  representatives or fitting by group, and  {\tt KMeans\_arma} in R package {\tt ClusterR} for faster $k$-means clustering. All computations are carried out on a single thread of a MAC Pro running macOS 10.15.6 with 3.5 GHz 6-Core Intel Xeon E5 and 32GB 1866 MHz DDR3 memory. 

The CPU time costs for MR and 1-iteration SMR with a given $k$-means ($K=1000$) partition, A-optimal subsampling with subsample size $20,000$, and  divide-and-conquer with $1000$ random blocks are shown in Table~\ref{tab:glmtime1}  and Figure~\ref{fig:glmtime} in the Supplementary Material.  
With the given partition, MR is comparable to A-optimal subsampling method in terms of computational time; SMR is a little slower than MR but faster than the divide-and-conquer method.  

One drawback of $k$-means clustering algorithm is that its time cost grows fast as $N$ increases. In practice, we recommend a {\it subset clustering} strategy. That is, the $K$ clustering centers are determined by a subset of $M$ data points ($M \ll N$), and the partition of the full data is determined by measuring the distance between the data points and the $K$ centers. Since the predetermined $M$ mainly depends on the number $p$ of predictors, not the sample size $N$, the time cost of subset clustering is $O(Np)$. According to our simulation study (see Section~\ref{sec:subset} of the Supplementary Material), the subset clustering strategy can significantly reduce the clustering time cost, while keeping the efficiency of representative approaches.

\section{A Case Study: Airline On-time Performance Data}\label{sec:case} 

The Airline on-time performance data for the US domestic flights of arrival time from October 1987 to December 2019 were collected from the Bureau of Transportation Statistics (\url{https://www.transtats.bts.gov/}) as a real example for big data analysis.  The original dataset consists of 387 csv files with the total number of records $192,555,789$ (see Table~\ref{tab:flightorg} in the Supplementary Material for more details). After cleaning, the total number of valid records is $N=188,690,624$.

For illustration purpose, we consider a main-effects logistic regression model for binary response {\tt ArrDel15} (arrival delay for $15$ minutes or more, 1=YES) with three categorical covariates and one continuous covariate:
{\tt QUARTER} (season, $1 \sim 4$) instead of {\tt MONTH} for simplification purpose;  {\tt DayOfWeek} (day of week, $1 \sim 7$); {\tt DepTimeBlk} (departure time block, $1 \sim 4$) following the convention of the O'Hare International Airport; and {\tt DISTANCE} (distance of flight, $8 \sim 4983$ miles). More details about the data and the fitted models could be found in Section~\ref{sec:cases} of the Supplementary Material.

In order to evaluate the performance of MR and SMR given that the full data estimate is not available, we choose the MR estimate of the main-effects logistic model on the last 5 years' data (from January 2015 to  December 2019) as the ``oracle" regression coefficients (denoted by $\boldsymbol\beta$, see Table~\ref{tab:oracle} in the Supplementary Material), and simulate 10 independent replicates of responses using the logistic model with the oracle parameter values $\boldsymbol\beta$. 

To show how MR and SMR work on the natural partition, we treat each data file (labeled by {\tt MONTH}) as a node and prohibit raw data exchanging between nodes. Note that the natural partition in this example is pre-determined. The data from different nodes follow different distributions since the predictor {\tt QUARTER} is a constant in each node but varies across different nodes.
We further split each data file into $7 \times 4\times 8 = 224$ sub-partitions by cutting the only continuous covariate {\tt DISTANCE} at $8$ equal-depth points and combining distinct values of {\tt DayOfWeek} and {\tt DepTimeBlk}.

In order to show the change of estimate accuracy along with increased data size,  we run a sequence of four experiments using the first $60$ months, $120$ months, $240$ months and all $387$ months of data, respectively. 
In each experiment, we obtain the full data estimate (not available for $240$ months and $353$ months due to too big data size), as well as the MR and SMR estimates, which are listed in Table~\ref{tab:case}. The average and standard deviation (std) of RMSEs are obtained from $10$ independent simulations.

In terms of RMSE from the oracle $\boldsymbol\beta$ (see Table~\ref{tab:case}),  MR and SMR perform as good as the full data estimate for 60 months and 120 months. The main reason is that the oracle coefficient of the only continuous predictor {\tt DISTANCE} is as small as $7.955\times 10^{-5}$. Even multiplied by the largest value of {\tt DISTANCE}, $4983$, the contribution of {\tt DISTANCE} is still less than $0.4$, which is too small compared with the oracle intercept $-2.322$. In other words, this scenario is fairly close to a case where all predictors are categorical. According to Corollary~\ref{cor:cate2}, both MR and SMR estimates match the full data estimate very well. Table~\ref{tab:case} also shows that as the data size gets bigger, including 240 months and 387 months, both MR and SMR estimates are better than the last available full data estimate obtained at 120 months.

It is interesting that if we amplify the effect of the continuous predictor {\tt DISTANCE}, say enlarge its oracle coefficient from $7.955\times 10^{-5}$ to $7.955\times 10^{-4}$, SMR estimate will show clearly higher accuracy than MR estimate (see Section~\ref{sec:cases} in the Supplementary Material for more details).

\begin{table}[bt]
	\centering
	\caption{Average (std) of RMSEs ($10^{-3}$) from oracle $\boldsymbol\beta$ for airline on-time performance data}\label{tab:case}
	\begin{threeparttable}
		\begin{tabular}{rrrr}
			\hline
			{\bf Number of months} &  {\bf Full} & {\bf MR} &  {\bf SMR} \\ 
			\hline
			60 months& 13.768 (5.155) & 13.763 (5.164) & 13.760 (5.175) \\ 
			120 months& 11.391 (4.417) & 11.384 (4.423) & 11.370 (4.415) \\ 
			240 months& - & 10.234 (4.068) & 10.225 (4.066) \\ 
			387 months& - & 9.058 (3.967) & 9.051 (3.966) \\ 
			\hline
		\end{tabular}
		\begin{tablenotes}[noitemsep]
			\item {\footnotesize Note: Full data estimates for 240 months and 387 months are not available due to memory limit.}
		\end{tablenotes}
	\end{threeparttable}
\end{table}

\section{Discussion and Conclusion} \label{sec:conc}

\begin{figure}[H]
	\centering
	\includegraphics[width=.4\textwidth]{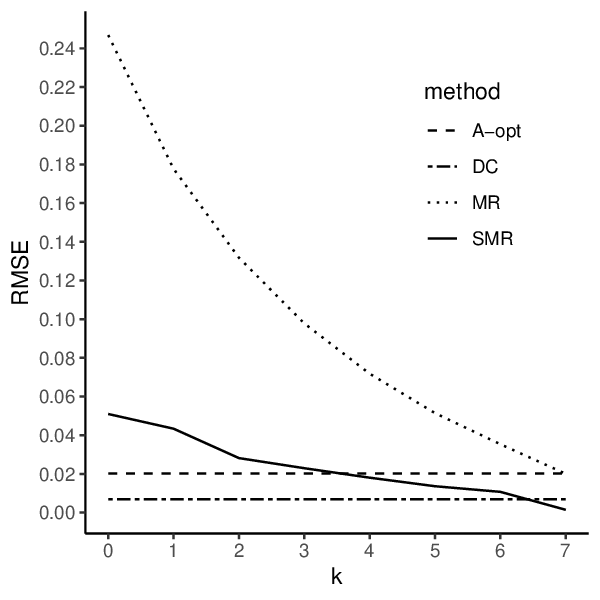}
	\caption{Average RMSE from full data estimate $\hat{\boldsymbol\beta}$ vs $k$, the number of covariates under equal-depth partitions (100 {\bf mzNormal} simulations for $7$ predictors udner logistic  model with $N=10^6$)} \label{fig:randomsplit}
\end{figure}

In practice, a natural partition may be provided with partially homogeneous blocks. 
In order to investigate the dependence of the proposed SMR approach on homogeneous partitions, we run another simulation study letting the given partition gradually  change from a random partition to a homogeneous partition. More specifically, under the logistic regression model \eqref{eq:logistic} with 7 covariates, for $k=0,1,2,\dots,7$, we first partition the data according to the first $k$ covariates into $4^k$ blocks using the equal-depth criterion. We then, for each of the $4^k$ blocks, randomly divide the data in this block into $4^{7-k}$ sub-blocks. By this way, $k=0$ actually leads to a random partition of $4^7$ blocks, and $k=7$ corresponds to an equal-depth homogeneous partition. As $k$ increases from $0$ to $7$, the partition becomes more and more homogeneous. The performance of MR and SMR on these partitions are displayed in Figure~\ref{fig:randomsplit}. In this simulation study, both MR and SMR are affected by the homogeneous level of the natural partition. The higher $k$ is, the better performance they have. Overall, MR seems to be the worst, which does not meet A-optimal sampler until $k=7$. SMR is not as good as A-optimal one and DC when $k$ is small, while it surpasses A-optimal one at $k\geq 4$ and does better than DC at $k=7$. An implication is that sub-partitions are necessary for representative approaches when the natural partition is not homogeneous.

When all predictors of the GLMs are categorical or discrete, the best solution would be partitioning the data according to distinct predictor values if applicable. In this case, $\Delta=0$, both the MR and SMR estimates exactly match the full data estimate. 

For GLMs with flat $G(\eta)$ (that is, $G'(\eta)$ is bounded by some moderate number), such as logit, probit, cloglog, loglog, and cauchit links for binomial models, one may check the coefficients of the continuous variables fitted by MR. 
If all linear predictors contributed by continuous variables are relative small comparing to the intercept or linear predictors contributed by categorical ones, then the MR estimate might be good enough. Otherwise, we recommend the SMR solution. For GLMs with unbounded or large $G'(\eta)$, such as Poisson model and Gamma model, we recommend SMR over MR with a finer partition (see Section~\ref{sec:asym} for more detailed discussion).

Data partition, or more specifically, the maximum block size $\Delta$, is critical for both MR and SMR. Asymptotically, the accuracy of MR estimate depends on $\Delta$ and will not benefit from an increased sample size unless for linear models, while the accuracy of SMR estimate can still be improved with increased sample size and fixed partition of the predictor space (see  Section~\ref{sec:asym} and Section~\ref{sec:part}  in the Supplementary Material for more detailed discussion).

\begin{figure}[bt]
	\centering
	\begin{subfigure}[b]{0.45\textwidth}
		\includegraphics[width=\textwidth]{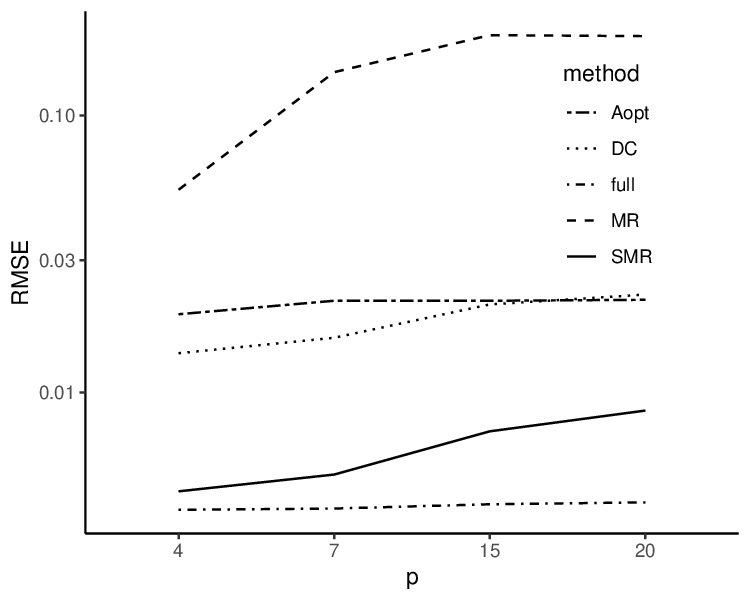}
		\caption{RMSE from true $\boldsymbol\beta$}
	\end{subfigure}
	~ 
	\begin{subfigure}[b]{0.45\textwidth}
		\includegraphics[width=\textwidth]{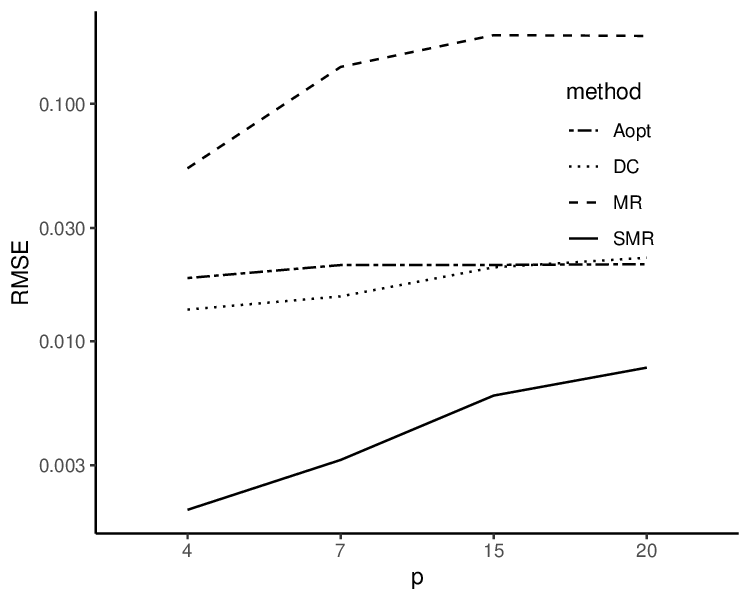}
		\caption{RMSE from full data $\hat{\boldsymbol\beta}$}
	\end{subfigure}
	~ 	
	\caption{RMSE vs $d$, the number of main effects, under logistic model with $\|\boldsymbol\beta\|=3$ and $N=10^6$}		\label{fig:para}
\end{figure}

To illustrate how SMR scales with dimension $d$, we also run simulations towards various covariate dimension $d$ using MR, SMR, A-opt, and DC. To avoid the increment of linear predictor along with dimension $d$, we randomly generated a $(d+1)$-dimensional $\boldsymbol\beta$ for each simulation such that $\|\beta\|=3$ as the true regression coefficients. Figure~\ref{fig:para} shows that, as the covariate dimension $d$ in the main-effects logistic model increases, the performance of the SMR estimate, using MR estimate as initial value, gets away from the full data estimate. As we expected, $\Delta$ increases with $d$, which leads to a challenge for both MR and SMR. 
It seems that A-optimal sampler is fairly robust across different dimensions. One strategy for solving large-$d$ problems is to use A-optimal estimate as the initial parameter value for SMR iterations.

When $d$ is large, it is also difficult to reduce the maximum block size $\Delta$ efficiently due to the curse of dimensionality. How we can obtain an efficient partition or do variable selection for large dimension $d$ is critical for representative approaches, but is out of the scope of this paper.

The framework of representative approaches allows the data analysts to work with the representative data instead of the raw data. In the scenario where different sources of data are owned by different  individuals, organizations, companies, or countries (also regarded as nodes) with competing interests, the exchange of raw data between nodes might be prohibited. In this scenario, neither divide-and-conquer nor subsampling approaches are feasible, while representative approaches may provide an ideal solution since the representative data points typically could not be used to track the raw data.

Compared with the CoCoA algorithms~\citep{smith2018cocoa}, the representative approaches introduced in this paper also require a central server to collect representative data points and process regression analysis. Utilizing a similar idea as in the COLA algorithm~\citep{he2018}, the representative approaches may be applied to decentralized environment as well.  

\appendix

\section*{Appendix}\label{app}

\noindent
{\bf Proof of Theorem~\ref{thm:eta}:}
Plugging \eqref{eq:ytilde} into the left hand side of \eqref{eq:ssb}, we get
\[
\sum_{i \in I_k} \nu(\eta_i) (\tilde{y}_k-G(\eta_i))  \eta_i= n_k\ \nu(\tilde{\eta}_k) (\tilde{y}_k -G(\tilde{\eta}_k))  \tilde{\eta}_k
\]
Let $S(\eta) =\nu(\eta) (\tilde{y}_k  -  G(\eta)) \eta$, a one-dimensional function of $\eta$. Then \eqref{eq:ssb} is equivalent to
\begin{align}
	n_k^{-1} \sum_{i \in I_k} S(\eta _i) = S(\tilde{\eta}_k) \label{eq:SS} 
\end{align}
Let $a = \min_{i\in I_k} \{\eta_i\}$ and $b = \max_{i \in I_k} \{\eta_i\}$. Since $S(\eta)$ is continuous on $[a, b]$, there exist $\eta_{\min}, \eta_{\max}\in [a, b]$, such that, $S(\eta_{\min}) = \min_{\eta \in [a,b]} S(\eta)$ and $S(\eta_{\max}) = \max_{\eta \in [a,b]} S(\eta)$. Then $n_k^{-1} \sum_{i \in I_k} S(\eta _i) \in [S(\eta_{\min}), S(\eta_{\max})]$ and there exists a $\tilde{\eta}_k$ between $\eta_{\min}$ and $\eta_{\max}$ solving \eqref{eq:SS} and thus \eqref{eq:ssb}. 
\hfill{$\Box$}

\section*{Acknowledgments}
This work was supported in part by the U.S. National Science Foundation. The authors thank the Editor, the Associate Editor and the referee for their constructive comments and suggestions. The authors also thank Mr.~Lie He from the \'{E}cole Polytechnique F\'{e}d\'{e}rale de Lausanne (EPFL) in Switzerland for his help with the COLA package.

\clearpage
\setcounter{page}{1}
\def\thepage{A\arabic{page}}

\vspace{0.8pc}
\centerline{\large\bf Score-Matching Representative Approach for}
\centerline{\large\bf Big Data Analysis with Generalized Linear Models}
\vspace{.25cm}

\centerline{Keren Li and Jie Yang}
\vspace{.4cm}

\vspace{.55cm}
\centerline{\bf Supplementary Material}
\vspace{.55cm}

\renewcommand{\thesection}{A}
\setcounter{equation}{0}
\setcounter{table}{0}
\setcounter{figure}{0}
\setcounter{subsection}{0}
\renewcommand{\theequation}{A.\arabic{equation}}
\renewcommand{\thetable}{A.\arabic{table}}
\renewcommand{\thefigure}{A.\arabic{figure}}

\subsection{SMR and MR for Linear Model}\label{sec:linear}

The simulation study in this section is based on the linear regression model
\begin{equation}\label{eq:linearmodel}
	y_i = \beta_0 + \beta_1 x_{i1} + \cdots + \beta_d x_{id} + \epsilon_{i}
\end{equation}
where  $i=1, \ldots, N$ and $\epsilon_i$'s are iid $\sim N(0, \sigma^2)$. 
Note that linear regression models are actually special cases of the generalized linear models with normally distributed responses and identity link (see Table~\ref{tab:egeta}).
Analogue to the simulation setup in Section~\ref{sec:simulatelogistic}, we take $N=10^6$, $d=7$, $\beta_0=0$, $\beta_1 = \cdots = \beta_d= 0.5$ and $\sigma^2=1$, as well as the same distributions for simulating $\mathbf{x}_i = (x_{i1}, \ldots, x_{i7})^T$. The main-effects predictors in \eqref{eq:linearmodel} are for illustration purpose. The representative approach can actually work with general predictors including, for example, interactions of covariates. 

In the absence of a natural partition, we again use an equal-depth partition with $m=4$ splits for each predictor, and a $k$-means partition with the number of blocks $K=1000$.

\begin{table}[h!]
	\centering
	\caption{Average (std) of RMSEs ($10^{-3}$) of 100 simulations for linear model ($N=10^6$)}\label{tab:linear}
	\begin{threeparttable}
		\resizebox{\columnwidth}{!}{
			\begin{tabular}{cr|rrrr|rr|r}
				\hline
				{\bf Simulation}  &\multicolumn{1}{c}{\bf Full}  &\multicolumn{4}{c}{{\bf Equal-depth ($m=4$)}} & \multicolumn{2}{c}{{\bf $k$-means ($K=1000$)}} & \\
				\cline{3-6} \cline{7-8}
				{\bf setup}	   &  \multicolumn{1}{c}{\bf data} & \multicolumn{1}{c}{\bf Mid} & \multicolumn{1}{c}{\bf Med}& \multicolumn{1}{c}{\bf MR} &  \multicolumn{1}{c}{\bf SMR} & \multicolumn{1}{c}{\bf MR}  & \multicolumn{1}{c}{\bf SMR}& \multicolumn{1}{c}{\bf IBOSS} \\
				\hline
				{\bf mzNormal}  & 1.2 (0.3) & 239.4 (2.8) & 28.7 (0.2) & 1.4 (0.4) & 1.4 (0.4) & 1.5 (0.4) & 1.5 (0.4) & 6.8 (1.9) \\ 
				{\bf nzNormal} & 1.2 (0.3) & 239.4 (2.8) & 28.7 (0.2) & 1.4 (0.4) & 1.4 (0.4) & 1.6 (0.4) & 1.5 (0.4) & 6.8 (1.9) \\ 
				{\bf ueNormal} & 0.5 (0.2) & 251 (3.4) & 43.6 (0.3) & 0.5 (0.2) & 0.5 (0.2) & 0.9 (0.6) & 0.9 (0.6) & 2.3 (1.0) \\ 
				{\bf mixNormal} & 1.3 (0.3) & 202.1 (2.9) & 16.4 (0.2) & 1.6 (0.5) & 1.6 (0.5) & 1.5 (0.4) & 1.5 (0.4) & 7.5 (2.0) \\ 
				$\boldsymbol{T}_3$ & 7.4 (2.1) & 483 (4.2) & 106.8 (1.7) & 10.3 (2.8) & 10.2 (2.8) & 11.1 (3.1) & 10.7 (3.1) & 12.0 (4.1) \\ 
				{\bf EXP} & 1.9 (0.5) & 368.9 (4.1) & 77 (1.1) & 2.2 (0.6) & 2.2 (0.6) & 2.3 (0.6) & 2.1 (0.6) & 6.0 (1.8) \\ 
				{\bf BETA} & 2.9 (0.7) & 27.7 (0.9) & 12.9 (0.9) & 2.9 (0.8) & 2.9 (0.8) & 3.7 (0.9) & 3.3 (0.9) & 18.2 (5.2) \\ 
				\hline 
			\end{tabular}
		}
		\begin{tablenotes}
			\item {\footnotesize Mid: mid-point representative; Med: median representative;} \item
			{\footnotesize IBOSS: information-based optimal subdata selection.}
		\end{tablenotes}
		
	\end{threeparttable}
\end{table}

Table~\ref{tab:linear} shows both the average and standard deviation of the root mean squared errors (RMSEs, $(\sum_{i=1}^7 (\tilde\beta_i-\beta_i)^2/7)^{1/2}$) between the estimated parameter value $\tilde{\beta}_i$'s and the true value $\beta_i$'s across different simulation settings and each with 100 independent simulations.

In terms of RMSE, Table~\ref{tab:linear} clearly shows that MR outperforms both mid-point (Mid) and median (Med) representative approaches, as well as information-based optimal subdata selection (IBOSS) proposed by \cite{wang2017b} with $20,000$ subsamples, which is larger than the largest possible number of non-empty blocks or representatives. Compared with the true parameter value, MR estimates are comparable even with the estimates based on the full data.

From Table~\ref{tab:linear}, we also see that the RMSE of MR based on the equal-depth partition obtained from $11,488 \sim 16,384$ non-empty blocks or representatives on average are comparable with the RMSE of MR from the $k$-means partition with $1000$ representatives. It implies that representative approaches based on clustered partition are more efficient. Actually, the maximum distance within data blocks  $\Delta = {\max}_k {\max}_{i,j \in I_k} \|{\mathbf X}_i - {\mathbf X}_j\|$, may play an important role in extracting data information more efficiently. 
The following theorem shows that for linear models, the MR estimate is unbiased.
It is also asymptotically efficient as $\Delta \to 0$. 

\begin{theorem}\label{thm:linear}
	Suppose $\sum_{i=1}^N {\mathbf X}_i {\mathbf X}_i^T$ is positive definite. For linear model $y_i={\mathbf X}_i^T\boldsymbol\beta +\epsilon_i$, $i=1, \ldots, N$, 
	with $\epsilon_i$ iid $\sim N(0,\sigma^2)$, the MR estimator 
	$$\tilde{\boldsymbol\beta}=(\sum_{k=1}^K n_k\tilde{\mathbf X}_k \tilde{\mathbf X}_k^T )^{-1}  \sum_{k=1}^K n_k \tilde{\mathbf X}_k \tilde{y}_k$$
	has mean $\boldsymbol\beta$ and covariance $\mathrm{Cov}(\tilde{\boldsymbol\beta})=  \sigma^2  (\sum_{k=1}^K n_k \tilde{\mathbf X}_k^T \tilde{\mathbf X}_k)^{-1}$ given that $\Delta^2 < \|N^{-1} \sum_{i=1}^N {\mathbf X}_i$ ${\mathbf X}_i^T\|_2$, where the induced matrix norm $\|\cdot\|_2$ is the largest eigenvalue for positive semi-definite matrices. Furthermore, 
	$\|\mathrm{Cov}(\tilde{\boldsymbol\beta})-\mathrm{Cov}(\hat{\boldsymbol\beta})\|_2=O(\Delta^2)$, as $\Delta$ goes to zero.
\end{theorem}

\begin{proof}
	of {\bf Theorem~\ref{thm:linear}}:
	Recall that $\mathbf{X}_i \in \mathbb{R}^p$ is the $i$th predictor vector, $y_i \in \mathbb{R}$ is the $i$th response variable, $i = 1,2,\dots,N$; and ${\mathbf y} = (y_1, \ldots, y_N)^T$ is the response vector of the full data, ${\bm X} = ({\mathbf X}_1, \ldots, {\mathbf X}_N)^T$ is the predictor matrix of the full data.

	Given a data partition $\{I_1, \ldots, I_K\}$ of $I=\{1, \ldots, N\}$, we denote by $\bm X_k$, $\boldsymbol{y}_k$, $\boldsymbol{\epsilon}_k$ the predictor matrix, the response vector and the error vector of the $k$th block, $k=1, \ldots, K$, respectively.	Then $\sum_{k=1}^K \bm X_k^T\bm X_k = \sum_{i=1}^N {\mathbf X}_i {\mathbf X}_i^T$, which is positive definite according to our assumption.	
	
	Denote by $\|\cdot\|_2$ the induced matrix norm defined by $\|{\mathbf A}\|_2 = \max_{\|\mathbf x\|=1} \|\mathbf{A}\mathbf{x}\|$, which is actually the square root of the largest eigenvalue of $\mathbf{A}^T\mathbf{A}$. If $\mathbf{A}$ is positive semi-definite, then $\|{\mathbf A}\|_2$ is simply its largest eigenvalue.  
	\begin{align*}
		\norm{\sum_{k=1}^K \bm X_k^T\bm X_k- \sum_{k=1}^K n_k \tilde{\mathbf X}_k \tilde{\mathbf X}_k^T}_2= \norm{\sum_{k=1}^K \sum_{i\in I_k} (\mathbf X_i- \tilde{\mathbf X}_k)  (\mathbf X_i- \tilde{\mathbf X}_k)^T}_2	\\
		\leq\sum_{k=1}^K 		 \norm{ \sum_{i\in I_k} (\mathbf X_i- \tilde{\mathbf X}_k)  (\mathbf X_i- \tilde{\mathbf X}_k)^T}_2
	\end{align*}
	Denote by $\delta_k= \max_{i,j \in I_k} \|{\mathbf X}_i - {\mathbf X}_j\|$. Then $\sum_{i\in I_k} (\mathbf X_i- \tilde{\mathbf X}_k)  (\mathbf X_i- \tilde{\mathbf X}_k)^T=\delta_k ^2 \sum_{i\in I_k} \mathbf a_i \mathbf a_i^T$ for some ${\mathbf a_i}$'s satisfying $\|\mathbf a_i\|\leq 1$. By the definition of the matrix norm,
	\begin{align*}
		\norm{\sum_{i\in I_k} \mathbf a_i \mathbf a_i^T}_2&=\underset{\norm{\mathbf x}=1}{\max} \norm{\sum_{i\in I_k} \mathbf a_i \mathbf a_i^T \mathbf x}\\
		& \leq \underset{\norm{\mathbf x}=1}{\max} \sum_{i\in I_k}\norm{\mathbf a_i}^2 \norm{\mathbf x}\\
		&\leq n_k 
	\end{align*}
	Therefore we have
	\begin{align}\label{eq:3311}
		\norm{\sum_{k=1}^K \bm X_k^T \bm X_k- \sum_{k=1}^K n_k \tilde{\mathbf X}_k^T \tilde{\mathbf X}_k} _2
		\leq \sum_{k=1}^K n_k \delta_k^2  \leq \Delta^2N
	\end{align}
	
	Denote by $\lambda_1$ and $\lambda_1^*$ the smallest eigenvalues of $\sum_{k=1}^K \bm X_k^T \bm X_k$ and $\sum_{k=1}^K n_k \tilde{\mathbf X}_k^T \tilde{\mathbf X}_k$, respectively.  According to our assumption, $\lambda_1>0$ regardless of the partition. By \eqref{eq:3311}, we have $\lambda_1^* > \lambda_1- \Delta^2 N > 0$ if $\Delta^2<\lambda_1/N$. That is, $\sum_{k=1}^K n_k \tilde{\mathbf X}_k^T \tilde{\mathbf X}_k$ is invertible when $\Delta^2$ is sufficiently small.
	
	Therefore, we have the weighted least squares (WLS) estimate from mean representative dataset
	\begin{eqnarray*}
		\tilde{\boldsymbol\beta}&=&\left(\sum_{k=1}^K n_k\tilde{\mathbf X}_k \tilde{\mathbf X}_k^T \right)^{-1}  \sum_{k=1}^K n_k \tilde{\mathbf X}_k \tilde{y}_k\\
		&=&\left(\sum_{k=1}^K n_k^{-1} \bm X_k^T \mathbbm{1}_{n_k} \mathbbm{1}_{n_k}^T\bm X_k \right)^{-1}  \sum_{k=1}^K n_k^{-1} \bm X_k^T \mathbbm{1}_{n_k} \mathbbm{1}_{n_k}^T \boldsymbol{y}_k\\
		&=& \boldsymbol\beta+ \left(\sum_{k=1}^K n_k^{-1} \bm X_k^T \mathbb{J}_{n_k} \bm X_k \right)^{-1}  \sum_{k=1}^K  n_k^{-1} \bm X_k^T \mathbb{J}_{n_k}  \boldsymbol{\epsilon}_k
	\end{eqnarray*}
	where $\mathbbm{1}_{n_k}$ is an $n_k\times 1$ vector of all $1$'s, and $\mathbb{J}_{n_k}$ is an $n_k\times n_k$ matrix of all $1$'s. Then the MR estimator is unbiased since
	\[\mathbb{E} (\tilde{\boldsymbol\beta}) =\boldsymbol\beta+\mathbb{E}\left(\left(\sum_{k=1}^K n_k^{-1} \bm X_k^T \mathbb{J}_{n_k} \bm X_k \right)^{-1}  \sum_{k=1}^K n_k^{-1} \bm X_k^T \mathbb{J}_{n_k}  \epsilon_k\right) =\boldsymbol\beta  \]
	and the covariance matrix of the MR estimator is given by
	\begin{align*}
		&\mathrm{Cov}(\tilde{\boldsymbol\beta})= \sum_{k=1}^K\mathrm{Cov}{\left(\left(\sum_{k=1}^K n_k^{-1} \bm X_k^T \mathbb{J}_{n_k} \bm X_k \right)^{-1}  n_k^{-1} \bm X_k^T \mathbb{J}_{n_k}  \epsilon_k\right)}\\
		=& \sigma^2 \sum_{k=1}^K \left(\sum_{k=1}^K n_k^{-1}\bm X_k^T \mathbb{J}_{n_k} \bm X_k \right)^{-1}   n_k^{-2} \bm X_k^T \mathbb{J}_{n_k}^2 \bm X_k \left(\sum_{k=1}^K n_k^{-1}\bm X_k^T \mathbb{J}_{n_k} \bm X_k \right)^{-1}\\
		=& \sigma^2  \left(\sum_{k=1}^K \frac{1}{n_k}\bm X_k^T \mathbb{J}_{n_k} \bm X_k \right)^{-1}\\
		=& \sigma^2  \left(\sum_{k=1}^K n_k \tilde{\mathbf X}_k^T \tilde{\mathbf X}_k \right)^{-1}
	\end{align*}
	and the matrix norm of difference between the Fisher information matrices of  OLS and MR is given by
	\[\sum_{k=1}^K \bm X_k^T \bm X_k- \sum_{k=1}^K n_k \tilde{\mathbf X}_k^T \tilde{\mathbf X}_k\]

	Consider the induced matrix norm of difference between covariance matrices of OLS and MR estimators
	\begin{align*}
		&\norm{\mathrm{Cov}(\tilde{\boldsymbol\beta}) - \mathrm{Cov}(\hat{\boldsymbol\beta})}_2\\
		=&\sigma^2  \norm{( \sum_{k=1}^K n_k \tilde{\mathbf X}_k^T \tilde{\mathbf X}_k )^{-1} - (\sum_{k=1}^K\bm X_k^T \bm X_k )^{-1}}_2\\
		=&\sigma^2 \norm{ ( \sum_{k=1}^K n_k \tilde{\mathbf X}_k^T \tilde{\mathbf X}_k )^{-1} (\sum_{k=1}^K \bm X_k^T \bm X_k - \sum_{k=1}^K n_k \tilde{\mathbf X}_k^T \tilde{\mathbf X}_k)   (\sum_{k=1}^K \bm X_k^T \bm X_k )^{-1}}_2\\
		\leq & \sigma^2 \norm{ ( \sum_{k=1}^K n_k \tilde{\mathbf X}_k^T \tilde{\mathbf X}_k)^{-1} }_2\cdot \norm{\sum_{k=1}^K \bm X_k^T \bm X_k - \sum_{k=1}^K n_k \tilde{\mathbf X}_k^T \tilde{\mathbf X}_k }_2\cdot \norm{  (\sum_{k=1}^K \bm X_k^T \bm X_k )^{-1}}_2\\
		\leq & \sigma^2 ({\lambda_1^*})^{-1}\cdot \Delta^2 N \cdot \lambda_1^{-1}\\
		\leq & \sigma^2 ({\lambda_1 - \Delta^2 N})^{-1}\cdot \Delta^2 N \cdot \lambda_1^{-1}\\
		\leq & 2  \sigma^2 \lambda_1^{-2} N\Delta^2
	\end{align*}
	The last ``$\leq$'' holds if $\Delta^2 < \lambda_1/(2N)$.
	Therefore, as $\Delta$ goes to zero, $\mathrm{Cov}(\tilde{\boldsymbol\beta})$ converges to $ \mathrm{Cov}(\hat{\boldsymbol\beta})$ in terms of the largest eigenvalue.
\end{proof}

\medskip
{\bf Best Partition:}
From the proof of Theorem~\ref{thm:linear}, we know that the difference between the Fisher information matrices of $\tilde{\boldsymbol\beta}$ and $\hat{\boldsymbol\beta}$ is actually bounded by $\sum_{k=1}^K n_k \delta_k^2$ up to a constant, where $\delta_k= \max_{i,j \in I_k} \|{\mathbf X}_i - {\mathbf X}_j\|$ is the $k$th block size. Fixing the number $K$ of blocks, a natural question is to find a partition that minimizes the upper bound $\sum_{k=1}^K n_k \delta_k^2$.

Assume that the average density of the $k$th block is $f_k$, such that, $n_k \approx c\cdot\delta_k^r f_k$ for some constant $c>0$ and $r>0$. Typically $r=d$ while in general $r$ depends on the mapping from the covariate vector ${\mathbf x}_i$ to the predictor vector ${\mathbf X}_i$. Then the goal is to minimize $\sum_{k=1}^K n_k^{1+2/r} f_k^{-2/r}$ subject to $\sum_{k=1}^K n_k = N$. Let $w_k = n_k/N$. The goal is equivalent to minimize
\begin{align}\label{eq:lmdelta}
	\sum_{k=1}^K  w_k^{1+2/r} f_k^{-2/r}
\end{align}
subject to $\sum_{k=1}^K w_k = 1$. Plugging $w_K = 1 - \sum_{k=1}^{K-1} w_k$ into \eqref{eq:lmdelta} and differentiating it with respect to $w_k$, we get
\begin{align*}
	\sum_{k=1}^{K-1}  w_k^{1+2/p} f_k^{-2/p} + \left(1-\sum_{k=1}^{K-1}  w_k\right)^{1+2/p} f_K^{-2/p}
\end{align*}
For $k=1,\dots,K-1$, 
\[
\pdv{}{w_k} \left[\sum_{k=1}^{K-1}  w_k^{1+2/r} f_k^{-2/r} + \left(1-\sum_{k=1}^{K-1}  w_k\right)^{1+2/r} f_K^{-2/r}\right] =0\]
$k=1, \ldots, w_{K-1}$, which implies $w_k/f_k =  w_K/f_K$. $k=1, \ldots, K-1$, which further implies  \[w_k = \frac{f_k}{f_1+\cdots+f_K}\] Therefore, the partition minimizing the upper bound satisfies \[\delta_k \approx \left(\frac{n_k}{cf_k}\right)^{1/r} = \left(\frac{N/c}{f_1+\cdots+f_K}\right)^{1/r} \] which is the same for $k=1, \ldots, K$. That is, the best partition keeps all the blocks about the same size. It explains why a $k$-means partition works usually better than a equal-depth partition for MR in linear regressions, because it minimizes $\sum_{k=1}^K \sum_{i \in I_k} \|{\mathbf X}_i - \tilde{\mathbf X}_i\|^2$ \citep{raykov2016}.

From Theorem~\ref{thm:linear}, we also see that if the grids of partition are coarse, the variance of MR estimate could be large and away from the variance of OLS.

\medskip
As for the comparison between MR and SMR, in terms of the RMSEs between $\tilde{\boldsymbol\beta}$ and $\boldsymbol\beta$ (such as in Table~\ref{tab:linear}), the performances of MR and SMR are similar and both comparable with the full data estimate $\hat{\boldsymbol\beta}$ (shown in Table~\ref{tab:linear}). Since for practical data ``true" model or ``true" parameter value may not exist, a more realistic goal is to match the full data estimate $\hat{\boldsymbol\beta}$.

In Table~\ref{tab:linear2}, we show the RMSEs between $\tilde{\boldsymbol\beta}$ and $\hat{\boldsymbol\beta}$. 
It confirms the conclusions in Theorem~\ref{thm:smr1} and Section~\ref{sec:asym}. That is, for linear models, both MR and SMR perform very well, while SMR is slightly better. Nevertheless, when the $\Delta$ is small (for example, in the partition obtained by $k$-means), the global rate $\rho(\Delta)$ of convergence (see Remark~\ref{rem:rho}) is close to 0 and the improvement from MR to SMR is significant.

It is also interesting to see different estimates on the intercept $\beta_0$. From Table~\ref{tab:linearint}, it seems that MR and SMR are among the best and comparable with the full data estimate.  

\begin{table}[bt]
	\centering
	\caption{Average (std) of RMSEs ($10^{-3}$) from $\hat{\boldsymbol\beta}$ of 100 simulations for linear model ($N=10^6$)}\label{tab:linear2}
	\begin{threeparttable}
		\begin{tabular}{ccc|cc}
			\hline
			\multicolumn{1}{c}{\bf Simulation}  &\multicolumn{2}{c}{{\bf Equal-depth ($m=4$)}} & \multicolumn{2}{c}{{\bf $k$-means ($K=1000$)}} \\
			\cline{2-5}
			\multicolumn{1}{c}{\bf setup}	   &  \multicolumn{1}{c}{\bf MR}  &  \multicolumn{1}{c}{\bf SMR}  &  \multicolumn{1}{c}{\bf MR}  &  \multicolumn{1}{c}{\bf SMR}  \\
			\hline
			{\bf mzNormal} & 0.664 (0.183) & 0.660 (0.182) & 0.761 (0.214) & 0.733 (0.207) \\ 
			{\bf nzNormal} & 0.664 (0.183) & 0.657 (0.181) & 1.067 (0.300) & 0.980 (0.299) \\ 
			{\bf ueNormal}  & 0.194 (0.106) & 0.192 (0.108) & 0.723 (0.482) & 0.699 (0.472) \\ 
			{\bf mixNormal} & 0.856 (0.240) & 0.845 (0.238) & 0.858 (0.243) & 0.823 (0.237) \\ 
			$\boldsymbol{T}_3$ & 6.793 (1.938) & 6.729 (1.942) & 7.709 (2.326) & 7.226 (2.198) \\ 
			{\bf EXP} & 1.175 (0.291) & 1.145 (0.288) & 1.250 (0.313) & 0.974 (0.276) \\ 
			{\bf BETA}  & 0.627 (0.149) & 0.610 (0.144) & 2.286 (0.650) & 1.461 (0.568) \\ 
			\hline  
		\end{tabular}
	\end{threeparttable}
\end{table}

\begin{table}[bt]
	\centering
	\caption{Average absolute value of intercept estimate ($10^{-3}$) of 100 simulations for linear model ($N=10^6$, using equal-depth partitions with $m=4$ for representative approaches)}\label{tab:linearint}
	\begin{threeparttable}
		\begin{tabular}{crrrrrr}
			\hline
			\bf{Simulation setup} & \bf{Full data} & \bf{Mid}& \bf{Med} & \bf{MR} &\bf{SMR} & \bf{IBOSS}\\ 
			\hline
			{\bf mzNormal}  & 0.7 & 23.8 & 0.9 & 0.7 & 0.7  & 6.1 \\ 
			{\bf nzNormal} & 1.7 & 2512.8 & 300.6 & 1.8 & 1.8  & 8.0 \\ 
			{\bf ueNormal}  & 0.7 & 104.3 & 1.9 & 0.7 & 0.7 & 4.0 \\ 
			{\bf mixNormal} & 0.8 & 29.5 & 0.8 & 0.8 & 0.8  & 5.2 \\ 
			$\boldsymbol{T}_3$ & 0.7 & 46.9 & 0.7 & 0.7 & 0.7 &  5.3 \\ 
			{\bf EXP} & 2.2 & 661.9 & 104.2 & 2.7 & 2.7 &  9.6 \\ 
			{\bf BETA} & 2.7 & 96.3 & 44.0 & 2.7 & 2.7 & 23.5 \\ 
			\hline
		\end{tabular}
	\end{threeparttable}
\end{table}

\subsection{Practical Number of Iterations for SMR}\label{sec:t}

\begin{figure}[H]
	\centering
	\begin{subfigure}[b]{0.25\textwidth}
		\includegraphics[width=\textwidth]{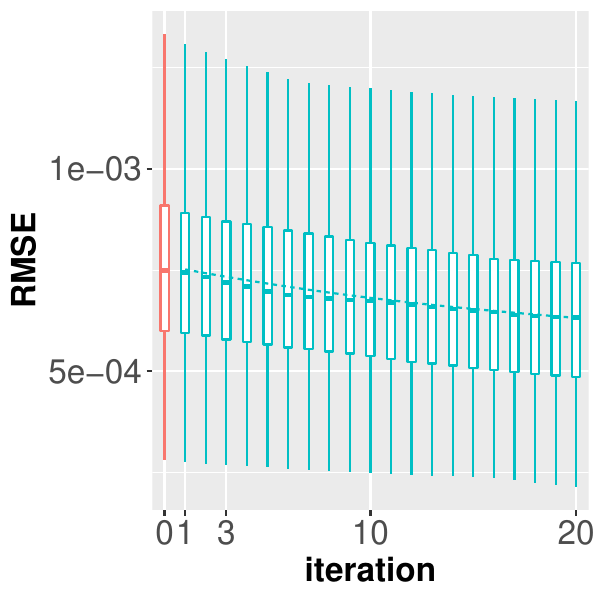}
		\caption{{\bf mzNormal}}
	\end{subfigure}
	~ 
	\begin{subfigure}[b]{0.25\textwidth}
		\includegraphics[width=\textwidth]{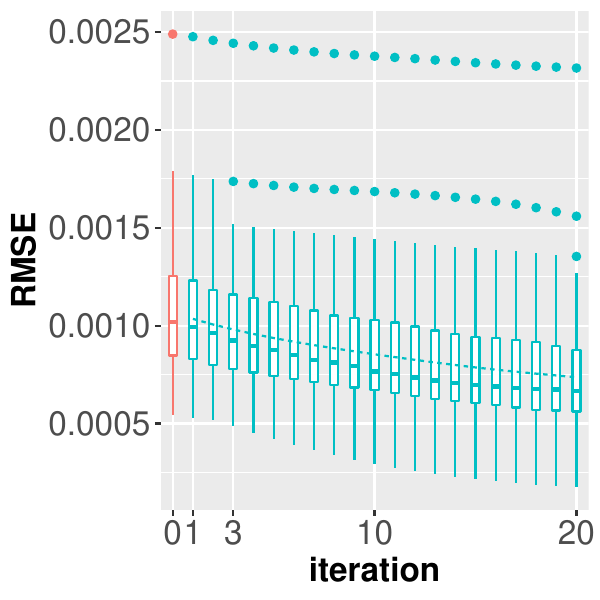}
		\caption{{\bf nzNormal}}
	\end{subfigure}
	~ 
	\begin{subfigure}[b]{0.25\textwidth}
		\includegraphics[width=\textwidth]{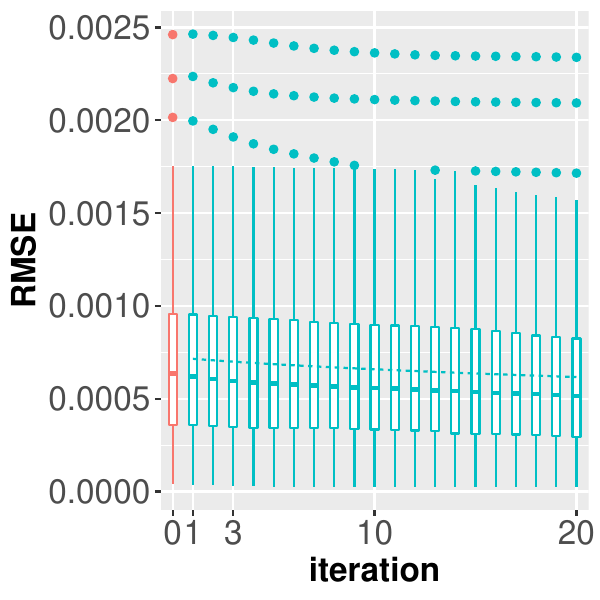}
		\caption{{\bf ueNormal}}
	\end{subfigure}
	~ 
	\begin{subfigure}[b]{0.25\textwidth}
		\includegraphics[width=\textwidth]{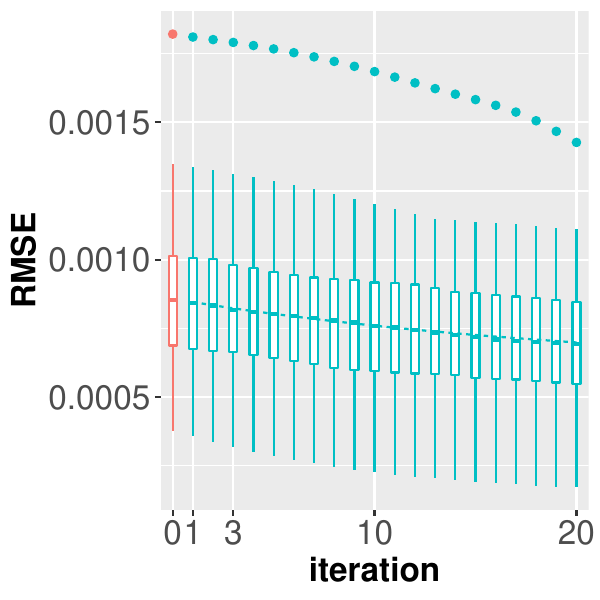}
		\caption{{\bf mixNormal}}
	\end{subfigure}
	~ 
	\begin{subfigure}[b]{0.25\textwidth}
		\includegraphics[width=\textwidth]{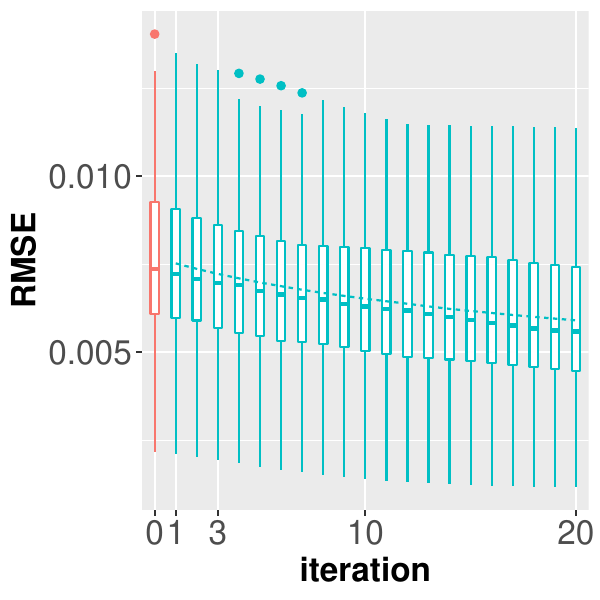}
		\caption{$\boldsymbol{T}_3$}
	\end{subfigure}
	~ 
	\begin{subfigure}[b]{0.25\textwidth}
		\includegraphics[width=\textwidth]{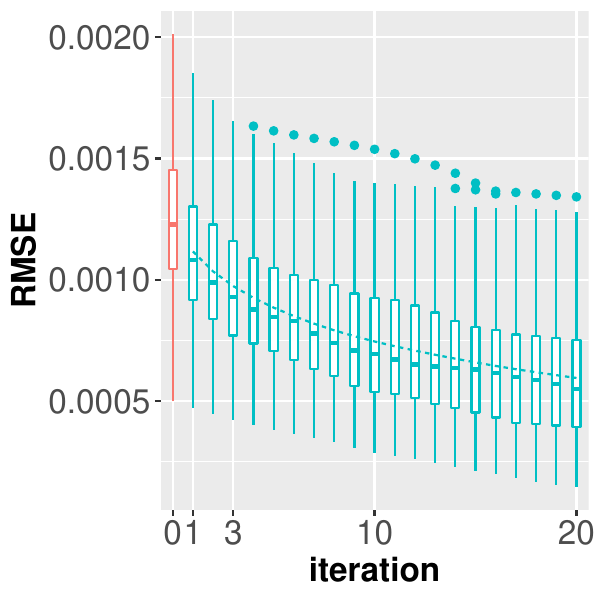}
		\caption{{\bf EXP}}
	\end{subfigure}
	~ 
	\begin{subfigure}[b]{0.32\textwidth}
		\includegraphics[width=\textwidth]{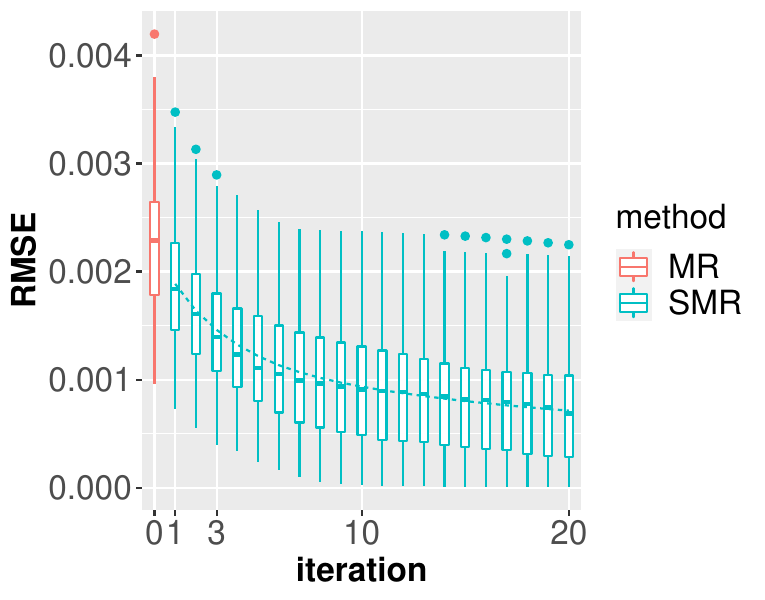}
		\caption{{\bf BETA}}
	\end{subfigure}   
	\caption{Box-plots of iterative SMR on linear model: $T=3$ achieves a reasonably good accuracy level; average RMSE from full data estimate $\hat{\boldsymbol\beta}$ based on 100 replicates; $N=10^6$ with $k$-means partition at $K=1000$; $x$-axis as the number of iterations of SMR from 0 to 20 with 0 representing MR.}\label{fig:linear}
\end{figure}

\begin{figure}[H]
	\centering
	\begin{subfigure}[b]{0.25\textwidth}
		\includegraphics[width=\textwidth]{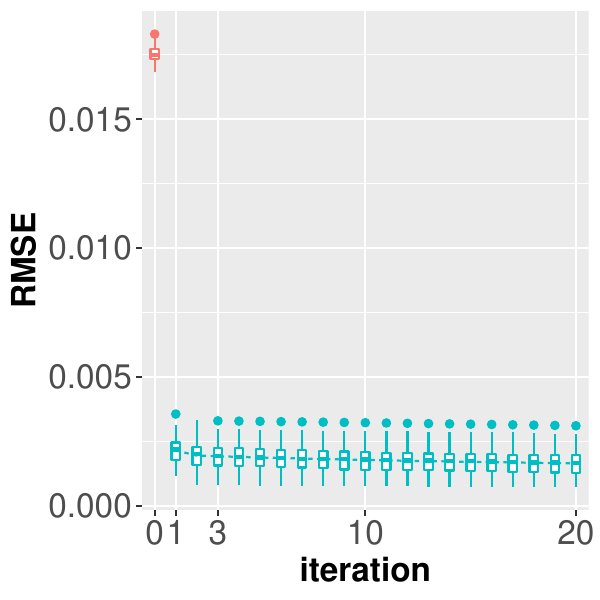}
		\caption{{\bf mzNormal}}
	\end{subfigure}
	~ 
	\begin{subfigure}[b]{0.25\textwidth}
		\includegraphics[width=\textwidth]{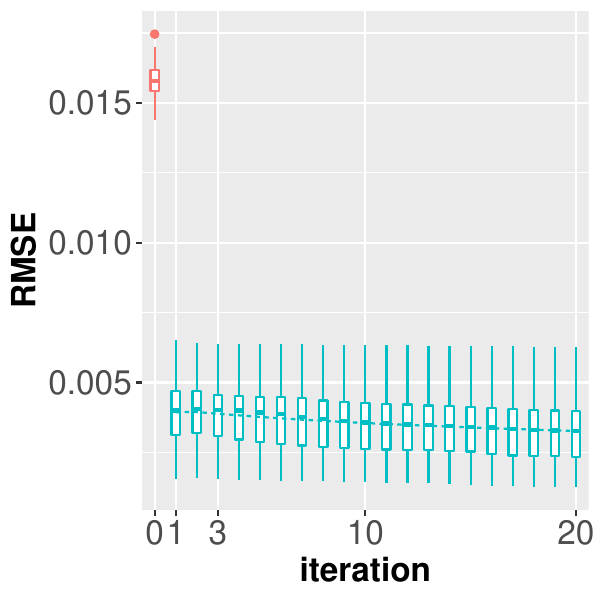}
		\caption{{\bf nzNormal}}
	\end{subfigure}
	~ 
	\begin{subfigure}[b]{0.25\textwidth}
		\includegraphics[width=\textwidth]{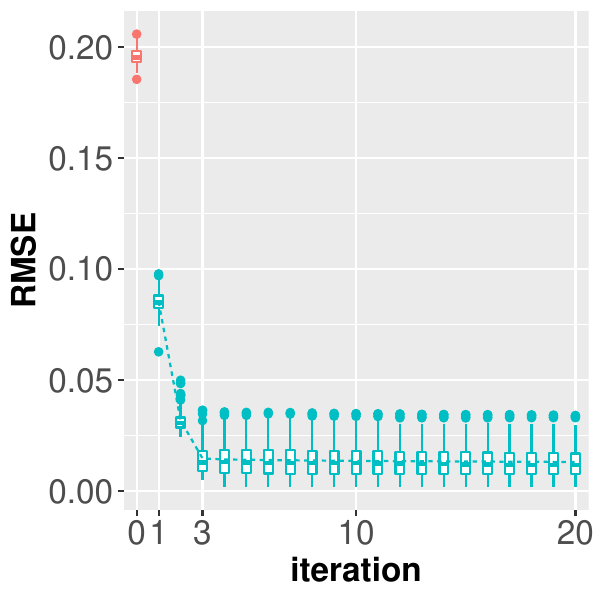}
		\caption{{\bf ueNormal}}
	\end{subfigure}
	~ 
	\begin{subfigure}[b]{0.25\textwidth}
		\includegraphics[width=\textwidth]{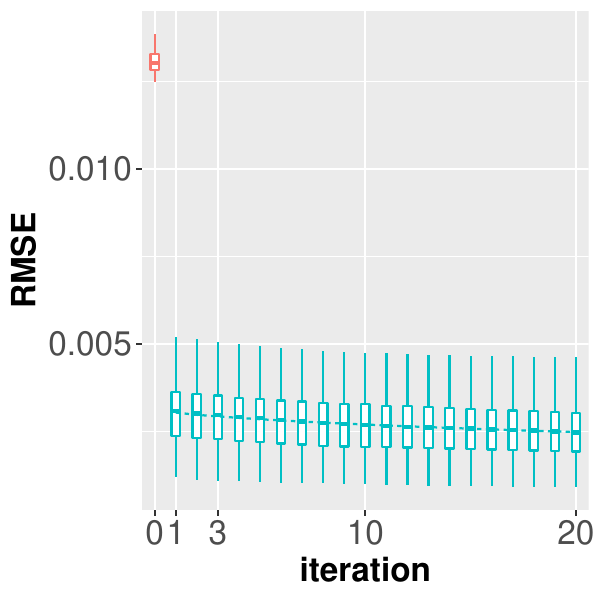}
		\caption{{\bf mixNormal}}
	\end{subfigure}
	~ 
	\begin{subfigure}[b]{0.25\textwidth}
		\includegraphics[width=\textwidth]{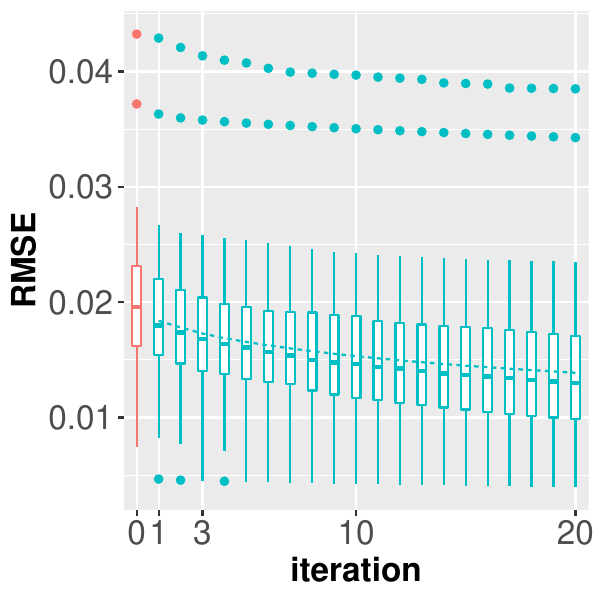}
		\caption{$\boldsymbol{T}_3$}
	\end{subfigure}
	~ 
	\begin{subfigure}[b]{0.25\textwidth}
		\includegraphics[width=\textwidth]{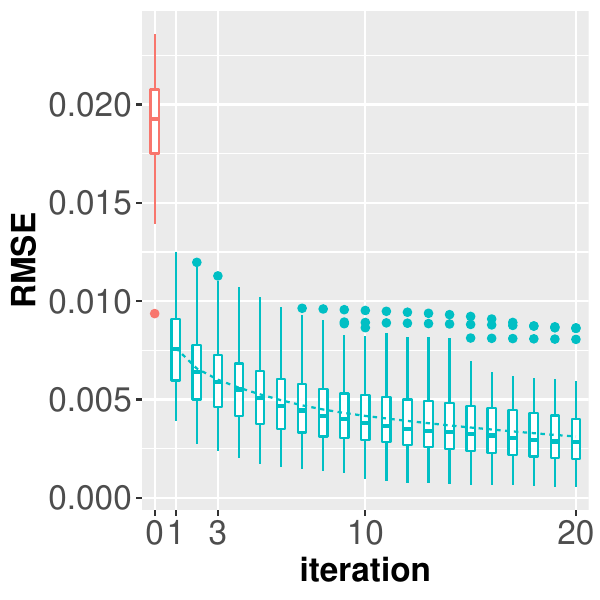}
		\caption{{\bf EXP}}
	\end{subfigure}
	~ 
	\begin{subfigure}[b]{0.32\textwidth}
		\includegraphics[width=\textwidth]{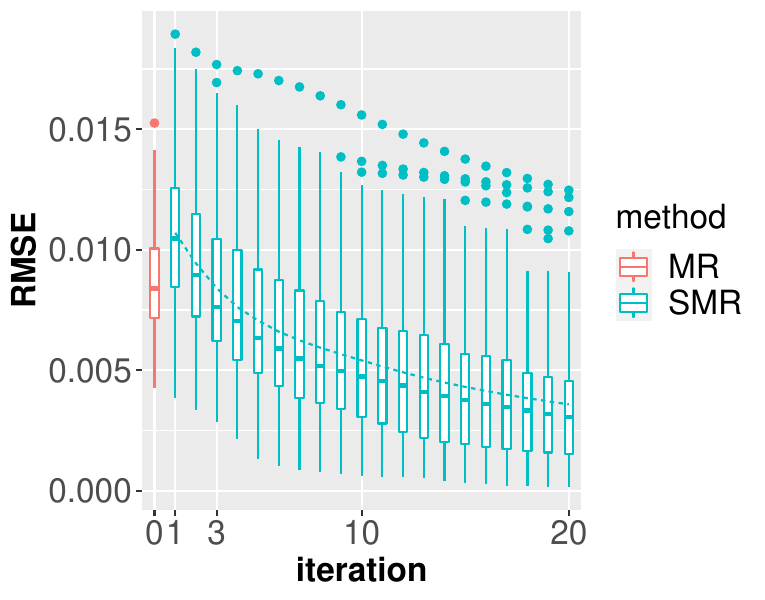}
		\caption{{\bf BETA}}
	\end{subfigure}   
	\caption{Box-plots of iterative SMR on logistic model: $T=3$ achieves a reasonably good accuracy level; average RMSE from full data estimate $\hat{\boldsymbol\beta}$ based on 100 replicates; $N=10^6$ with $k$-means partition at $K=1000$; $x$-axis as the number of iterations of SMR from 0 to 20 with 0 representing MR.
	}\label{fig:iter}
\end{figure}

In order to determine a practical number $T$ of iterations for SMR algorithm, we simulate 100 datasets of size $N=10^6$ for each of the seven distributions of covariates in  the linear model~\eqref{eq:linearmodel} and 
the logistic regression model~\eqref{eq:logistic}. We apply 20 iteration steps of SMR  for examining a practical number of iterations. Figure~\ref{fig:linear} and Figure~\ref{fig:iter} show that the iterative SMR based on a $k$-means partition with $K=1000$ improves the accuracy level significantly in the first 1 or 2 iterations. Starting from the 4th iteration, the relative improvements of RMSE are less than $5\%$ for {\bf ueNormal}, and less than $2\%$ for other distribution settings.

For GLMs with flat $G(\eta)$ such as linear model and logistic model, typically the first iteration based on the initial MR estimate improves the accuracy level significantly. For GLMs with steeper $G(\eta)$ such as Poisson model with log link, we recommend $T = 3$ (see also asymptotic properties discussed in Section~\ref{sec:asym}).

\subsection{SMR vs Divide-and-Conquer for Logistic Models}\label{sec:smrvsdc}

In this section, we use  simulation studies to show that in terms of accuracy level the 3-iteration SMR is comparable with the divide-and-conquer approach \citep{lin2011}, which is also known as divide and recombine, split and conquer, or split and merger in the literature \citep{wang2015}. When there is no ambiguity, we call the 3-iteration SMR simply SMR.	

\begin{figure}
	\centering
	\begin{subfigure}[b]{0.25\textwidth}
		\includegraphics[width=\textwidth]{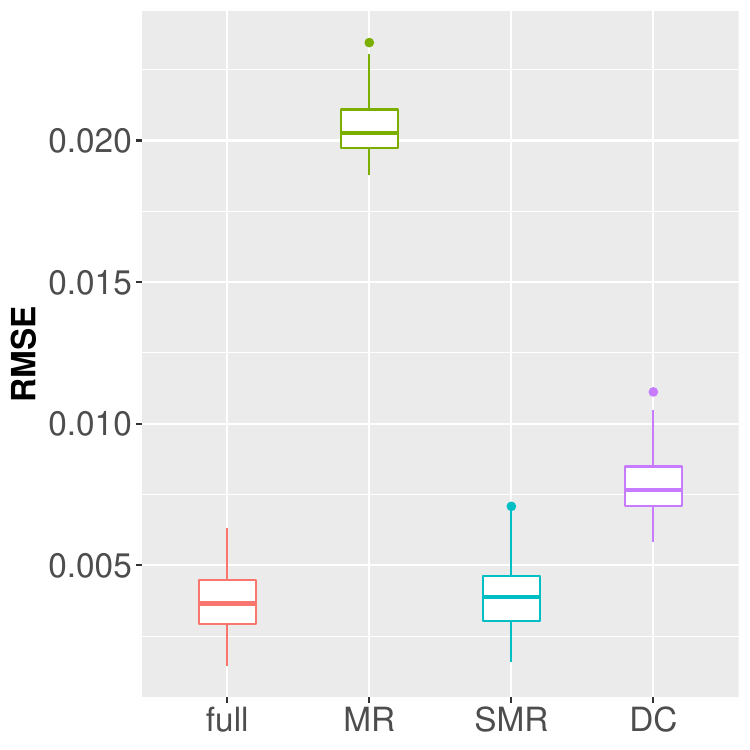}
		\caption{{\bf mzNormal}}
	\end{subfigure}
	~ 
	\begin{subfigure}[b]{0.25\textwidth}
		\includegraphics[width=\textwidth]{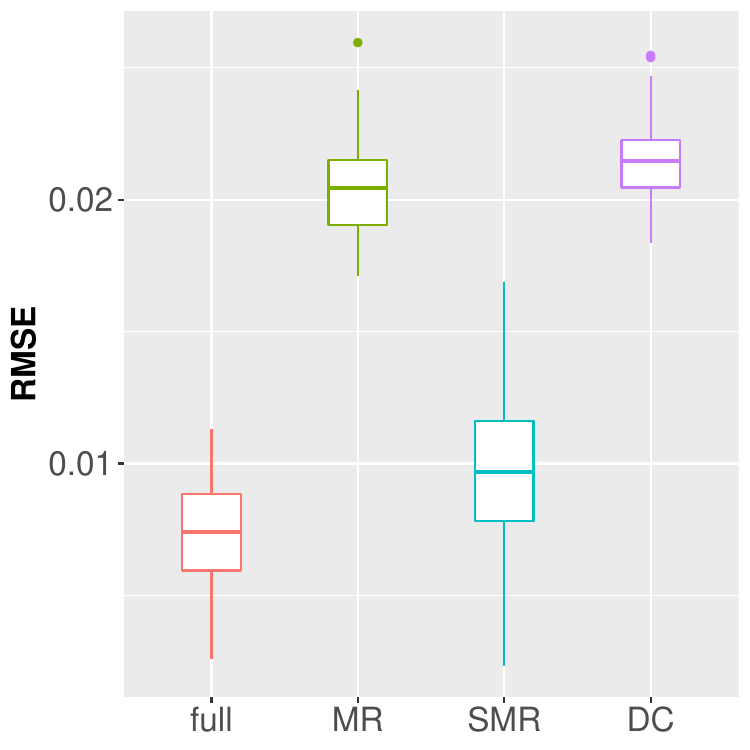}
		\caption{{\bf nzNormal}}
	\end{subfigure}
	~ 
	\begin{subfigure}[b]{0.25\textwidth}
		\includegraphics[width=\textwidth]{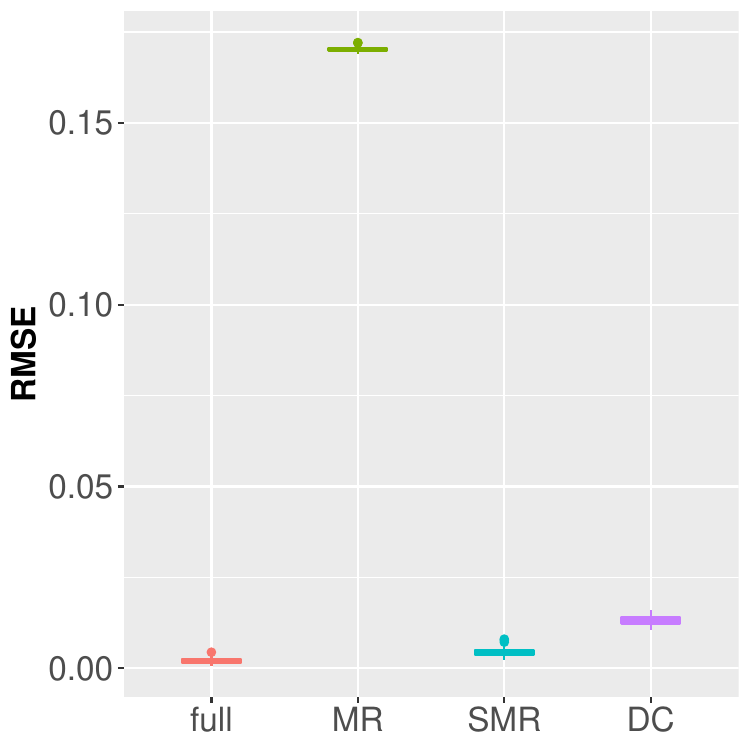}
		\caption{{\bf ueNormal}}
	\end{subfigure}
	~ 
	\begin{subfigure}[b]{0.25\textwidth}
		\includegraphics[width=\textwidth]{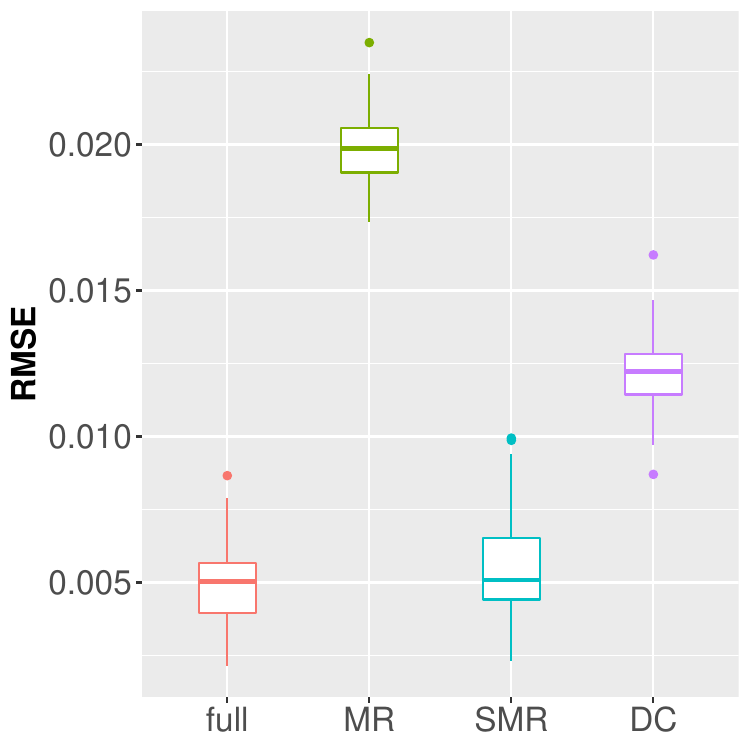}
		\caption{{\bf mixNormal}}
	\end{subfigure}
	~ 
	\begin{subfigure}[b]{0.25\textwidth}
		\includegraphics[width=\textwidth]{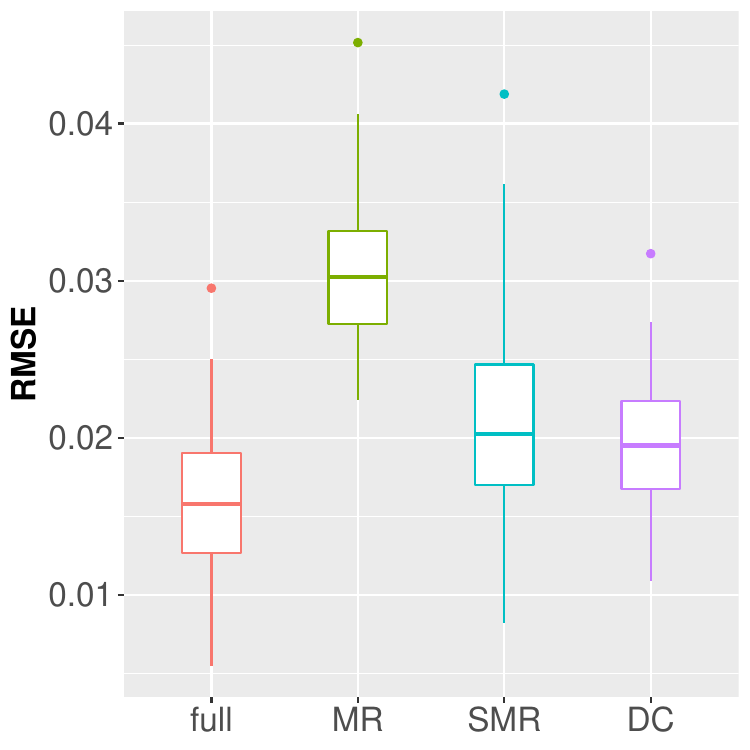}
		\caption{$\boldsymbol{T}_3$}
	\end{subfigure}
	~ 
	\begin{subfigure}[b]{0.25\textwidth}
		\includegraphics[width=\textwidth]{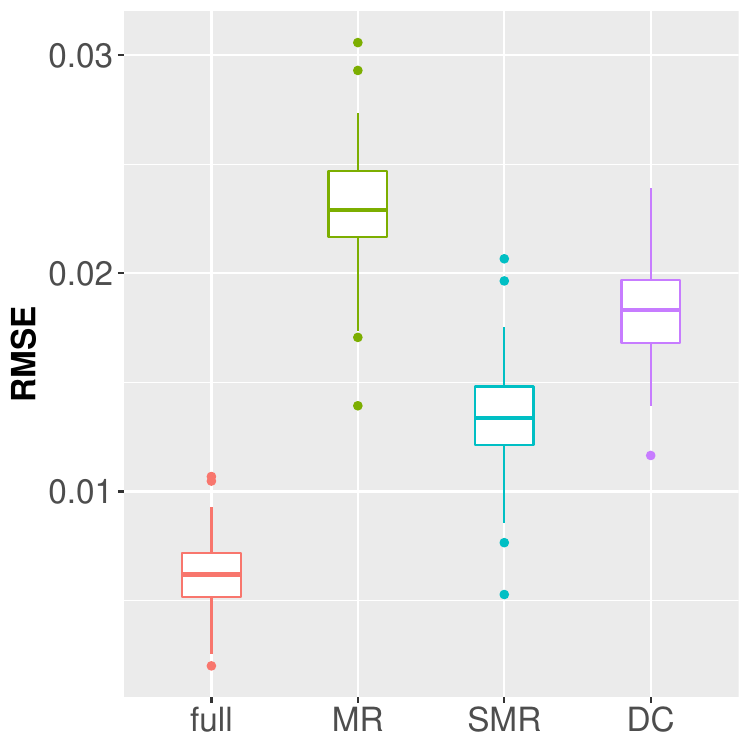}
		\caption{{\bf EXP}}
	\end{subfigure}
	~ 
	\begin{subfigure}[b]{0.25\textwidth}
		\includegraphics[width=\textwidth]{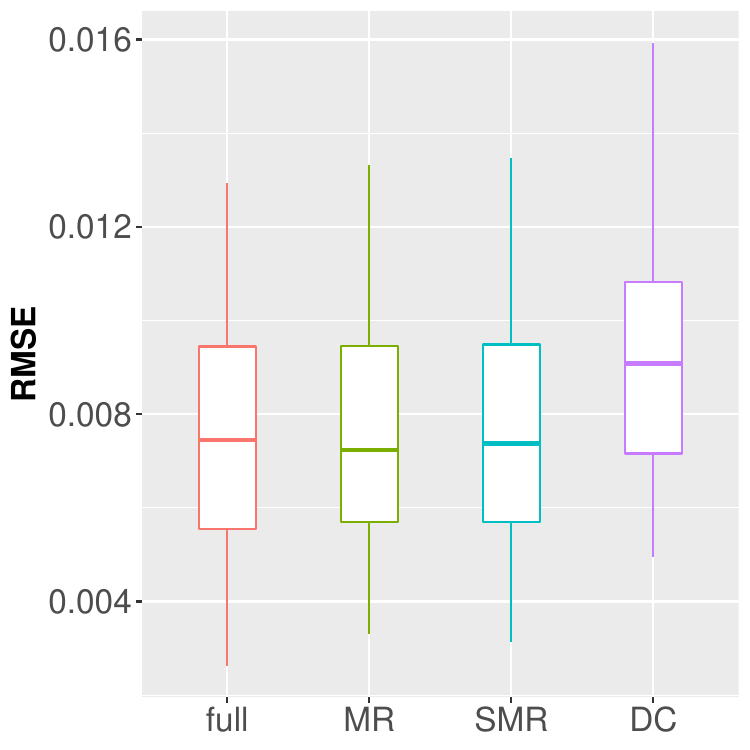}
		\caption{{\bf BETA}}
	\end{subfigure}   
	\caption{Boxplots of RMSEs from true parameter values based on 100 simulations under logistic model with $N=10^6$: full data estimate, MR and SMR using $k$-means partition with $K=1000$, and DC with 1000 points per block}\label{fig:logitkm}
\end{figure}

\begin{figure}[bt]
	\centering
	\begin{subfigure}[b]{0.25\textwidth}
		\includegraphics[width=\textwidth]{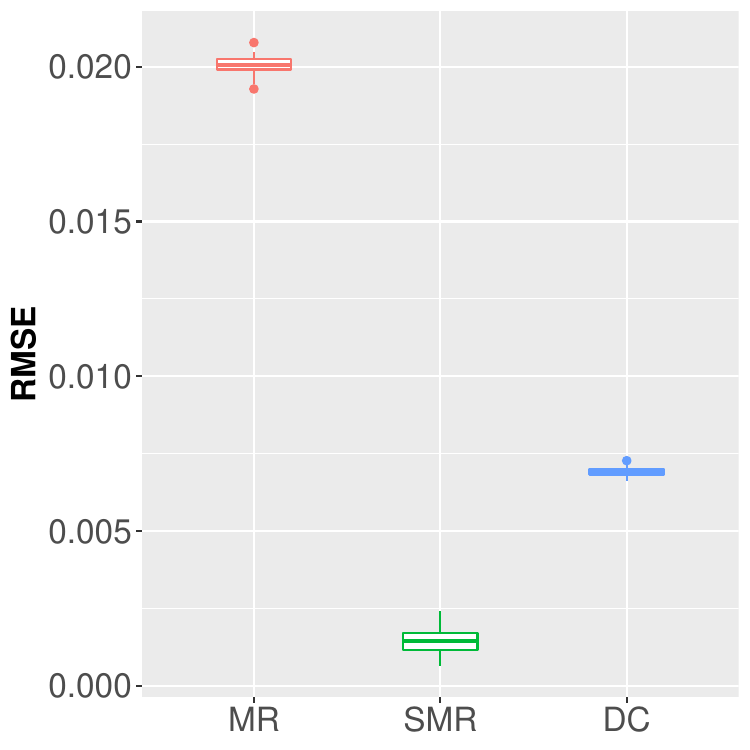}
		\caption{{\bf mzNormal}}
	\end{subfigure}
	~ 
	\begin{subfigure}[b]{0.25\textwidth}
		\includegraphics[width=\textwidth]{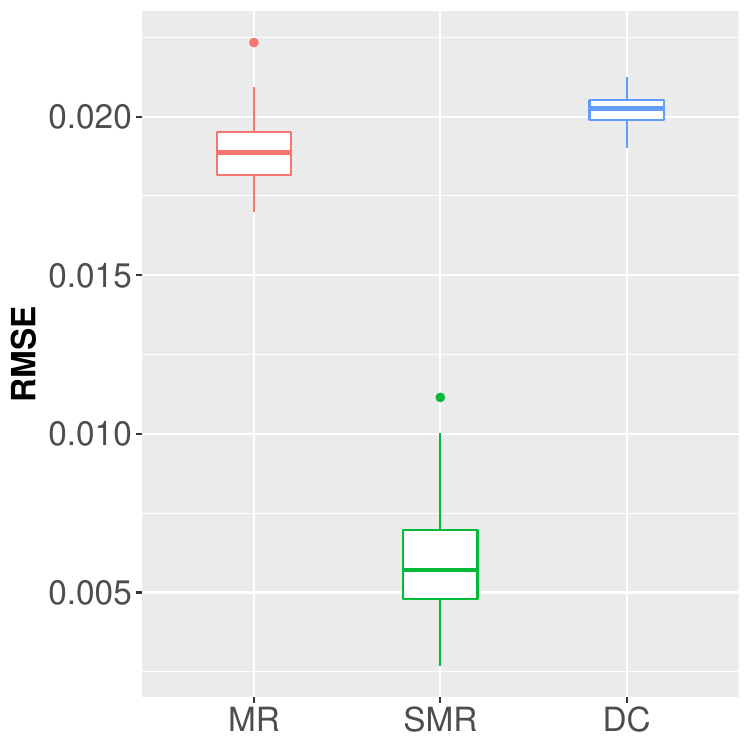}
		\caption{{\bf nzNormal}}
	\end{subfigure}
	~ 
	\begin{subfigure}[b]{0.25\textwidth}
		\includegraphics[width=\textwidth]{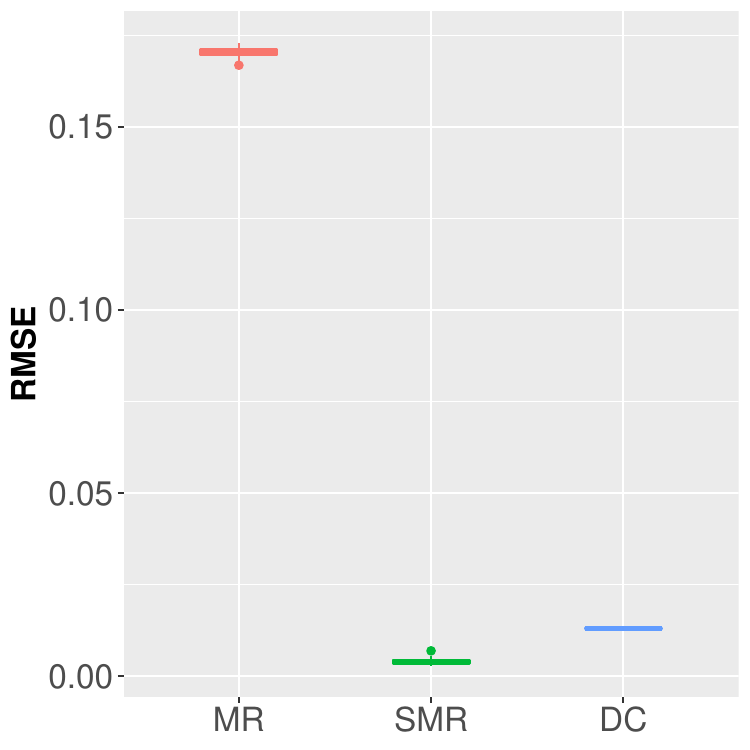}
		\caption{{\bf ueNormal}}
	\end{subfigure}
	~ 
	\begin{subfigure}[b]{0.25\textwidth}
		\includegraphics[width=\textwidth]{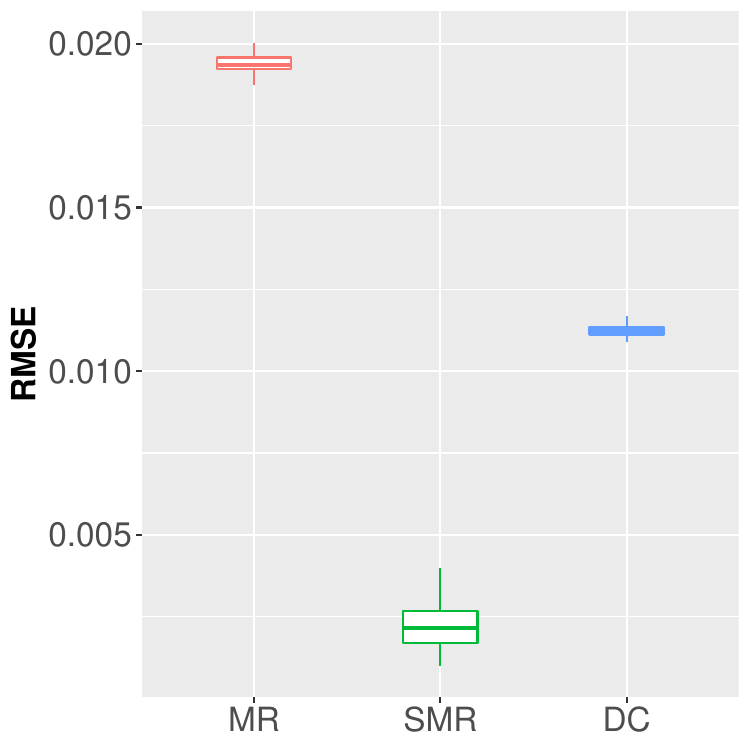}
		\caption{{\bf mixNormal}}
	\end{subfigure}
	~ 
	\begin{subfigure}[b]{0.25\textwidth}
		\includegraphics[width=\textwidth]{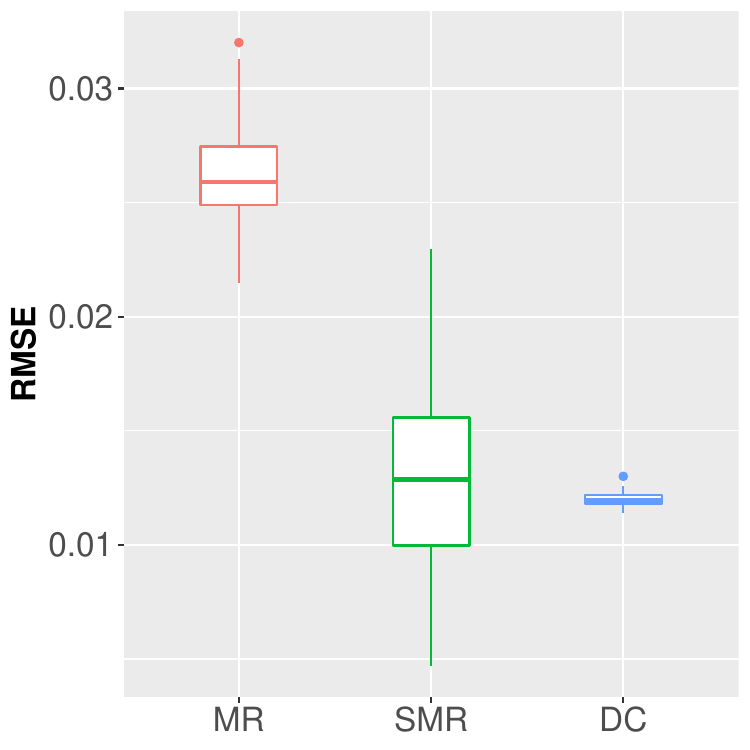}
		\caption{$\boldsymbol{T}_3$}
	\end{subfigure}
	~ 
	\begin{subfigure}[b]{0.25\textwidth}
		\includegraphics[width=\textwidth]{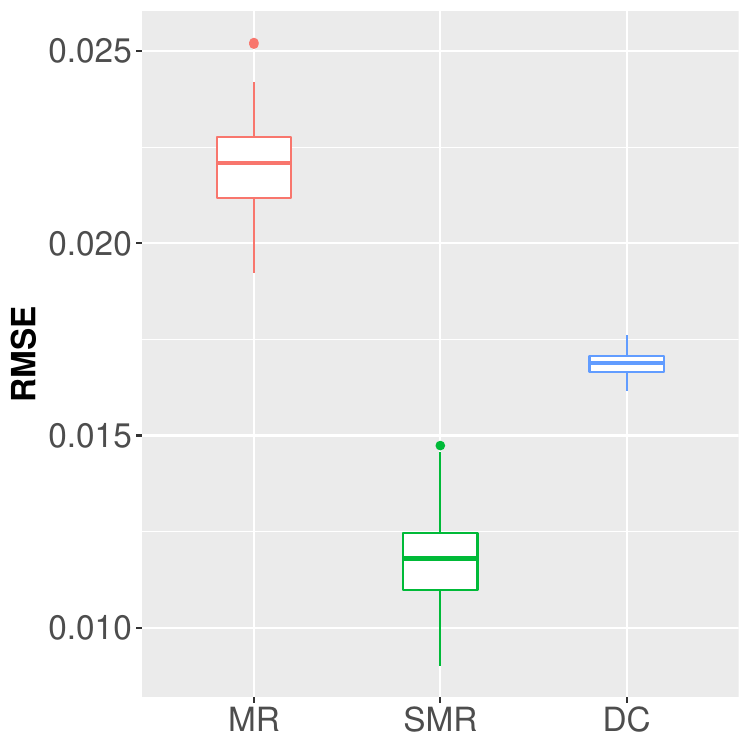}
		\caption{{\bf EXP}}
	\end{subfigure}
	~ 
	\begin{subfigure}[b]{0.25\textwidth}
		\includegraphics[width=\textwidth]{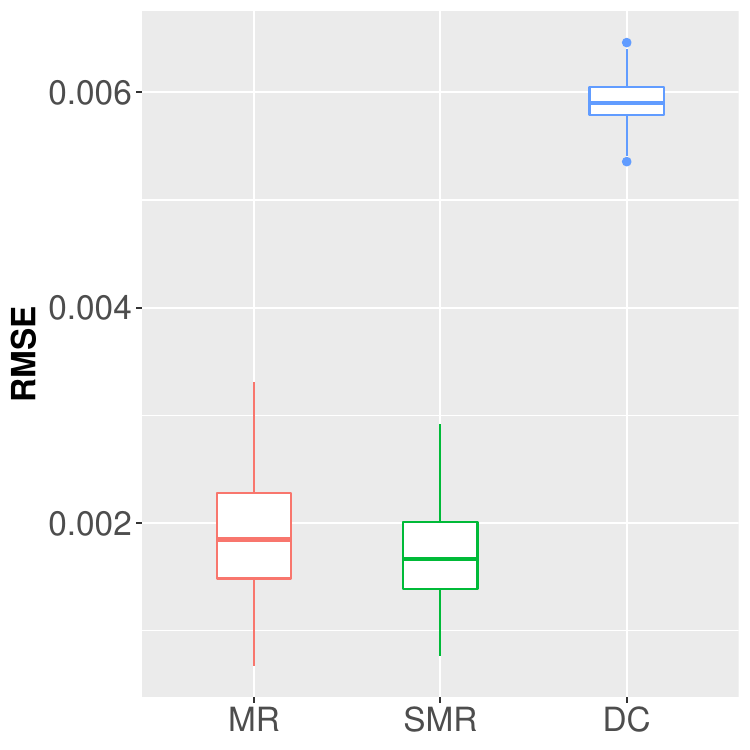}
		\caption{{\bf BETA}}
	\end{subfigure}   
	\caption{Boxplots of RMSEs from full data estimates based on 100 simulations under logistic model with $N=10^6$: full data estimate, MR and SMR using $k$-means partition with $K=1000$, and DC with 1000 points per block}\label{fig:logitkm2}
\end{figure}

We simulate 100 datasets of size $N= 10^6$ for each of the seven distributions of covariates in the logistic regression model~\eqref{eq:logistic}. We apply both SMR and the divide-and-conquer (DC) algorithm proposed by \cite{lin2011} to estimate the parameter values. Table~\ref{tab:logitdd} shows that SMR based on a $k$-means partition with $K=1000$ outperforms the divide-and-conquer method with $1000$ blocks in most of simulation settings. Actually, the SMR estimates are comparable with the full data estimates in terms of RMSEs from the true parameter value. In Figure~\ref{fig:logitkm} and  Figure~\ref{fig:logitkm2}, we plot the corresponding boxplots. For comparison purpose, we also list the corresponding MR estimates, which are overall not as good as DC's.       

SMR is ideal for massive data stored in multiple nodes, since it exchanges only the representative data points and estimated parameter values between nodes. It can perform well even with limited network bandwidth. 

On the contrary, divide-and-conquer methods typically operate on random partitions. Each random data block for divide-and-conquer might consist of data points from many different nodes, which requires heavy communications between nodes. It may not be feasible when raw data transfer is prohibited.

\subsection{SMR vs Support Points for Logistic Models}\label{sec:support}

In order to compare the performance of the proposed methods and the support points techniques \citep{mak2018b}, we run a simulation study under the logistic regression model with 7 covariates. We use R function {\tt sp} in package {\tt support} for searching support points for a given dataset. Since it takes {\tt sp} more than one hour to generate 20,000 support points from 1,000,000 data points, we reduce the simulation setup to choosing 1,000 support points from 10,000 original data points, which is the simulation setup used by \cite{wang2017}.

\begin{table}[!h]
	\centering
	\caption{Average (std) of  RMSEs ($10^{-3}$) of 100 simulations for logistic models with $N=10^4$: simple random sample (SRS), support points (Support), A-optimal sampler (A-opt), MR, SMR (with 1000 subsample/support/representative points),   }\label{tab:logitddkmrep}
	\begin{tabular}{c|rrrrr}\hline
		Simulation setup &  SRS  & Support &  A-opt &  MR & SMR\\ \hline
		{\bf mzNormal} & 111 (3.0)   &  118 (3.4)   &   93 (2.6)   &  22 (0.4)   &  16 (0.4)   \\
		{\bf nzNormal} & 235 (7.5)   &  237 (7.9)   &  115 (3.6)   &  41 (1.1)   &  31 (1.0)   \\
		{\bf ueNormal} & 65 (3.2)   &  65 (3.3)   &  45 (1.8)   &  186 (1.0)   &  11 (0.5)   \\
		{\bf mixNormal} & 160 (4.7)   &  162 (4.6)   &  101 (3.7)   &  28 (0.6)   &  22 (0.6)   \\
		{\bf $T_3$} & 493 (15.9)   &  526 (14.3)   &  472 (15.5)   &  92 (2.6)   &  87 (2.6)   \\
		{\bf EXP} & 193 (5.2)   &  187 (4.8)   &  137 (4.0)   &  22 (0.6)   &  17 (0.5)   \\
		{\bf BETA} & 241 (6.5)   &  233 (6.6)   &  180 (5.2)   &  23 (0.7)   &  19 (0.6)   \\ \hline
	\end{tabular}    
\end{table}

In Table~\ref{tab:logitddkmrep}, we list the average RMSE ($\sum_{i=1}^7 (\tilde{\beta}_i - \hat{\beta}_i)^2/7)^{1/2}$ over 100 simulations, where $\hat{\beta}_i$ is estimated by the full dataset consisting of 10,000 data points, $\tilde{\beta}_i$ is estimated by 1000 simple random samples (SRS), support points (Support), A-optimal sample points (Aopt), mean representatives (MR), or the proposed score-matching representatives (SMR).

In terms of RMSE, 
SMR still performs the best. The difference between MR and SMR is not as large as in Table~\ref{tab:logitdd} since the sample size $N=10^4$ is smaller in this simulation study.

\subsection{MR and SMR with Finer Partition}\label{sec:part}

According to Theorems~\ref{thm:mrglm} and \ref{thm:smrglm}, the estimate $\tilde{\boldsymbol\beta}$ obtained by MR or SMR converges to the full data estimate $\hat{\boldsymbol\beta}$ as $\tilde{\Delta} \rightarrow 0$. That is, with a finer partition, $\tilde{\boldsymbol\beta}$ gets closer to $\hat{\boldsymbol\beta}$ (but not necessarily closer to the true parameter $\boldsymbol\beta$ once the dataset is given).
Table~\ref{tab:dd} and Table~\ref{tab:km} show that with finer and finer partitions, both MR and SMR estimates get closer to $\hat{\boldsymbol\beta}$, while SMR is more robust to the block size than MR.

\begin{table}[!h]
	\centering
	\caption{Average (std) of RMSEs ($10^{-3}$) of 100 simulations for logistic model given equal-depth partitions with different $m$}\label{tab:dd}
	\begin{threeparttable}
		\begin{tabular}{crrr|rr}
			\hline
			\multicolumn{1}{c}{ $m$ }&\multicolumn{3}{c}{\bf RMSE from true $\boldsymbol\beta$} & \multicolumn{2}{c}{\bf RMSE from full $\hat{\boldsymbol\beta}$} \\
			\cline{2-6}
			& \multicolumn{1}{c}{\bf Full data} &  \multicolumn{1}{c}{\bf MR} & \multicolumn{1}{c}{\bf SMR} &  \multicolumn{1}{c}{\bf MR} & \multicolumn{1}{c}{\bf SMR} \\
			\hline
			2 & 3.72 (1.08) & 69.82 (0.88) & 4.75 (1.40) & 69.67 (0.48) & 3.00 (1.00) \\ 
			3 & 3.72 (1.08) & 33.35 (1.05) & 4.20 (1.16) & 33.09 (0.36) & 1.90 (0.57) \\ 
			4 & 3.72 (1.08) & 20.44 (1.00) & 3.94 (1.15) & 20.06 (0.25) & 1.44 (0.39) \\ 
			5 & 3.72 (1.08) & 13.89 (1.06) & 3.85 (1.07) & 13.34 (0.20) & 1.01 (0.31) \\ 
			\hline
		\end{tabular}
		\begin{tablenotes}
			\item {\footnotesize Sample size $N=10^6$; Covariate distribution: {\bf mzNormal}}
		\end{tablenotes}
	\end{threeparttable}
\end{table}

\begin{table}[!h]
	\centering
	\caption{Average (std) of RMSEs ($10^{-3}$) of 100 simulations for logistic model given $k$-means partitions with different $K$}\label{tab:km}
	\begin{threeparttable}
		\begin{tabular}{crrr|rr}
			\hline
			\multicolumn{1}{c}{ $K$ }&\multicolumn{3}{c}{\bf RMSE from true $\boldsymbol\beta$} & \multicolumn{2}{c}{\bf RMSE from full $\hat{\boldsymbol\beta}$} \\
			\cline{2-6}
			& \multicolumn{1}{c}{\bf Full data} &  \multicolumn{1}{c}{\bf MR} & \multicolumn{1}{c}{\bf SMR}   & \multicolumn{1}{c}{\bf MR} & \multicolumn{1}{c}{\bf SMR}   \\
			\hline
			500 & 3.72 (1.08) & 21.14 (1.05) & 4.26 (1.26) & 20.76 (0.33) & 2.14 (0.69) \\ 
			1000 & 3.72 (1.08) & 17.95 (1.00) & 4.14 (1.22) & 17.51 (0.30) & 1.92 (0.49) \\ 
			2000 & 3.72 (1.08) & 15.17 (0.96) & 4.09 (1.12) & 14.66 (0.31) & 1.67 (0.51) \\ 
			3000 & 3.72 (1.08) & 13.75 (1.00) & 4.08 (1.19) & 13.17 (0.29) & 1.60 (0.47) \\ 
			\hline
		\end{tabular}
		\begin{tablenotes}[noitemsep]
			\item {\footnotesize Sample size $N=10^6$; covariate distribution: {\bf mzNormal}}
		\end{tablenotes}
	\end{threeparttable}
\end{table}

\subsection{CPU Time of MR and SMR}\label{sec:cpu}

\begin{figure}[!h]
	\centering
	\begin{subfigure}[b]{0.45\textwidth}
		\includegraphics[width=\textwidth]{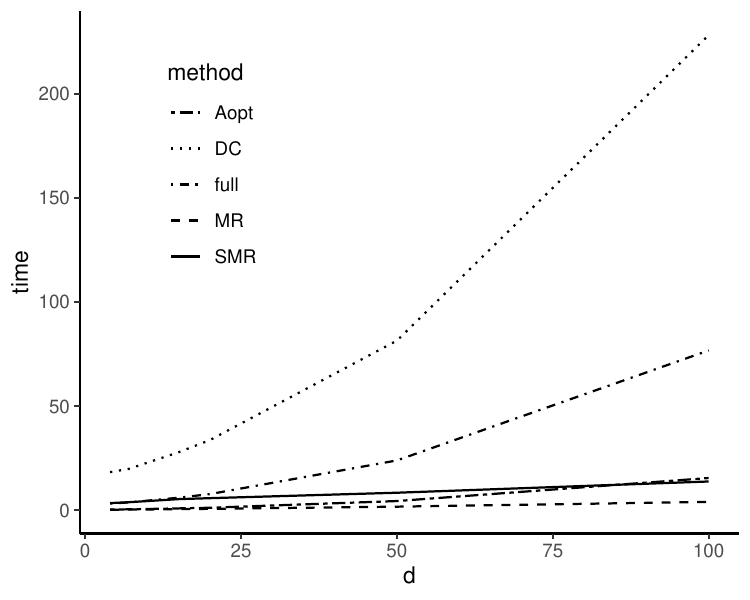}
		\caption{ CPU time against $d$ covariates with $N=10^6$}
	\end{subfigure}
	~ 
	\begin{subfigure}[b]{0.45\textwidth}
		\includegraphics[width=\textwidth]{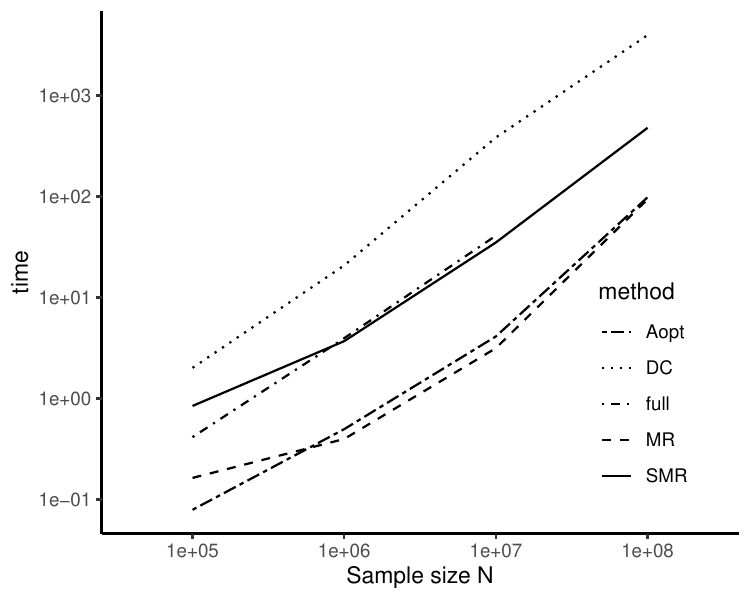}
		\caption{CPU time against $N$  with $d=7$}
	\end{subfigure}    		 	
	\caption{Average CPU time (secs) over 100 simulations  for logistic  model with sample size $N$ and $d$ covariates ({\bf mzNormal})}\label{fig:glmtime}
\end{figure}

In this section, we provide Table~\ref{tab:glmtime1} and Figure~\ref{fig:glmtime} mentioned in Section~\ref{sec:cpu}. All computations are carried out on a single thread of a MAC Pro running macOS 10.15.6 with 3.5 GHz 6-Core Intel Xeon E5 and 32GB 1866 MHz DDR3 memory.

\begin{table}[!h]
	\centering
	\caption{Average CPU time (secs) of MR, SMR, A-optimal, Divide-and-conquer over 100 simulations for logistic model with $d=7$ covariates }\label{tab:glmtime1}
	\begin{threeparttable}
		\begin{tabular}{crrrrr}
			\hline
			\multicolumn{1}{c}{\bf Simulation setup}&\multicolumn{1}{c}{\bf  Full data} & \multicolumn{1}{c}{\bf  MR} & \multicolumn{1}{c}{\bf SMR} & \multicolumn{1}{c}{\bf A-opt} & \multicolumn{1}{c}{\bf DC}\\ 
			\hline
			{\bf mzNormal} & 5.22 & 0.72 & 3.25 & 0.55 & 20.49 \\ 
			{\bf nzNormal} & 6.45 & 0.67 & 3.19 & 0.57 & 21.58 \\ 
			{\bf ueNormal} & 6.50 & 0.67 & 3.64 & 0.57 & 21.48 \\ 
			{\bf mixNormal} & 5.85 & 0.68 & 3.17 & 0.54 & 21.05 \\ 
			$\boldsymbol{T}_3$ & 3.42 & 0.68 & 3.36 & 0.55 & 19.55 \\ 
			{\bf EXP} & 4.59 & 0.71 & 3.07 & 0.55 & 20.04 \\ 
			{\bf BETA} & 3.96 & 0.67 & 3.01 & 0.57 & 19.69 \\ 
			\hline
		\end{tabular}
		\begin{tablenotes}
			\item {\footnotesize Distribution: {\bf mzNormal}; $N=10^6$; MR and SMR (1-iteration): given a $k$-means partition ($K=1000$); A-opt: subsample size $20,000$; DC: given a random partition with $1000$ blocks}
		\end{tablenotes}
	\end{threeparttable}
\end{table}

\subsection{Subset Clustering Strategy}\label{sec:subset}

When there is no natural partition available or when sub-partitions are needed,  a partitioning procedure is required by representative approaches. Based on our simulation study and justification (see Section~\ref{sec:simulatelogistic} and Section~\ref{sec:linear}), a clustering procedure, such as $k$-means, is more efficient than grid partitions. However, the computational cost of $k$-means clustering is quite heavy especially for large  sample size $N$. We recommend the {\it subset clustering strategy} (see the last paragraph of Section~\ref{sec:simulatelogistic}), which can significantly reduce the clustering cost for representative approaches. More specifically, the $K$ cluster centers are determined by a clustering algorithm on a much smaller subset of size $M$, which will not be changed as the full sample size $N$ increases. Once the $K$ centers are determined, each data point with predictor vector ${\mathbf X} \in \mathbb{R}^p$ is assigned to the cluster whose center is closest to ${\mathbf X}$. Since both $K$ and $M$ will not change with $N$ and they are much less than $N$, the time complexity of the overall clustering procedure is $O(Np)$. 

\begin{figure}[!h]
	\centering
	\includegraphics[width=.7\textwidth]{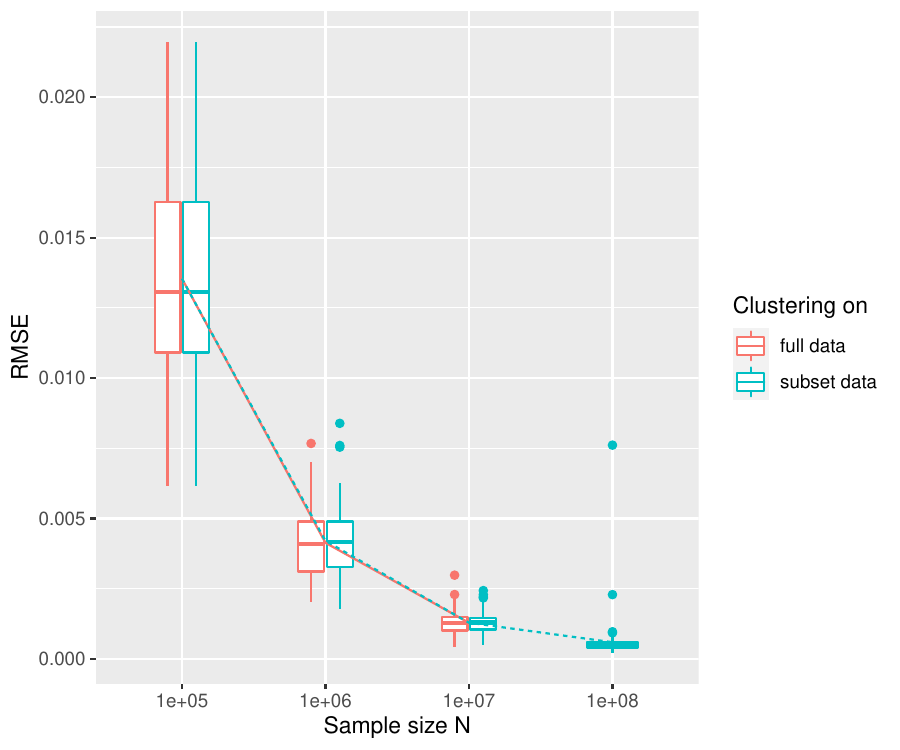}
	\caption{Full data clustering (size $N$) vs. subset clustering (size $M=10^5$): Boxplots of RMSEs of SMR based on $k$-means clustering ($K=1000$) over 100 simulations for logistic  model with $d=7$ covariates and distribution {\bf mzNormal} (full data clusterings are not available at $N=10^8$)}\label{fig:subsetpart}
\end{figure}

To illustrate the performance of the subset clustering strategy compared with a full data clustering for MR and SMR, we carry two simulation studies as follows. In the first simulation study, we generate datasets with sample size $N=10^5, 10^6, 10^7, 10^8$ and $d=7$ covariates following {\bf mzNormal} for the logistic model described in Section~\ref{sec:simulatelogistic}. For each $N$, we generate $100$ independent replicates. The size of a randomly selected subset for locating $K=1000$ clustering centers is fixed at $M=10^5$. Figure~\ref{fig:subsetpart} shows in terms of the average SMR RMSE over 100 simulations the difference between the subset clustering and the full data clustering is negligible (the result of full data clustering is not available at $N=10^8$ due to too much computational time.  

Table~\ref{tab:splittime} shows the computational time on clustering. For full data clustering, the time cost is proportional to the sample size $N$. For subset clustering, we break the overall time cost into two parts. The first part is for locating the $K$ cluster centers based on the subset of size $M$, whose time cost is fairly short and roughly constant. The second part is also full data clustering but with given $K$ cluster centers, whose time cost is also proportional to $N$ but much less than the original full data clustering.
Overall, the subset clustering strategy reduces the clustering cost significantly.

\begin{figure}[H]
	\centering
	\includegraphics[width=\textwidth]{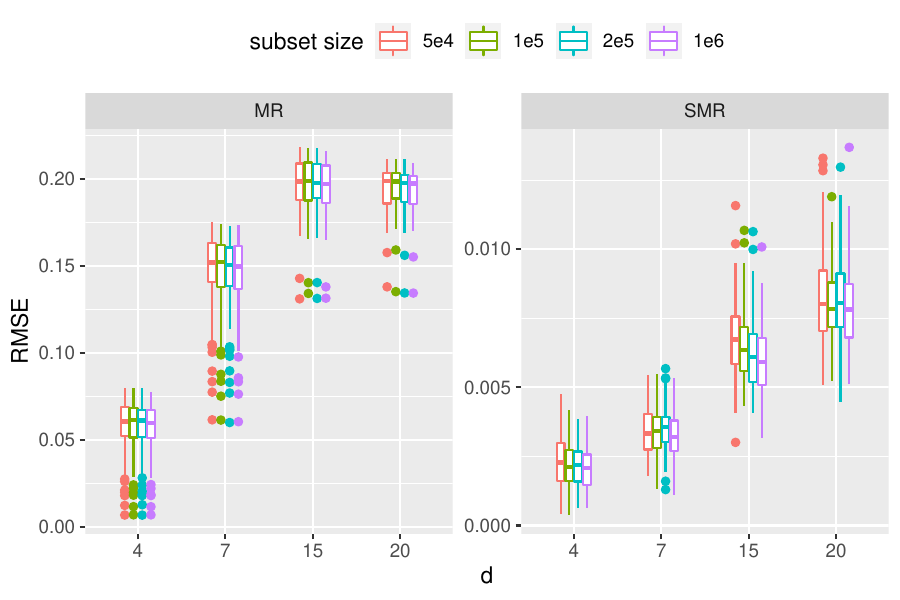}
	\caption{Boxplots of MR and SMR RMSEs over 100 simulations based on subset clustering with different subset size $M$ and dimension $d$ of covariates for logistic  model with size $N=10^6$ and distribution {\bf mzNormal}, using $k$-means partitions ($K=1000$)}\label{fig:subset}
\end{figure}

To further explore the interaction between the dimension of predictors and the subset clustering strategy, we simulate datasets with $d=4,7,15,20$ covariates and sample size $N=10^6$ for logistic main-effects models. Since the dimension of covaraites changes, for each simulation we randomly generate the regression coefficients $\boldsymbol\beta$ such that $\|\boldsymbol\beta\|=3$. We also adopt different subset size $M=5\times10^4, 10^5, 2\times10^5, 10^6$. Figure~\ref{fig:subset} shows that in terms of MR and SMR RMSEs, the subset clustering strategy is not sensitive to the initial subset size $M$ across different dimensions of predictors.

\begin{table}[H]
	\centering
	\caption{Average CPU time (secs) of full data clustering (size $N$) and subset clustering (size $M=10^5$) over 100 simulations  for logistic  model, $d=7$ with {\bf mzNormal}, $k$-means partition ($K=1000$)}\label{tab:splittime}
	\begin{threeparttable}
		\begin{tabular}{rrrr}
			\hline
			N & full data centers & subset cluster centers & full data clustering given centers\\ 
			\hline
			$10^5$ & 4.96 & 4.97 & 1.02 \\ 
			$10^6$ & 49.81 & 5.01 & 10.08 \\ 
			$10^7$ & 478.85 & 4.81 & 96.48 \\ 
			$10^{8*}$ & 4812.69 & 6.46 & 973.98 \\ 
			\hline
		\end{tabular}
		\begin{tablenotes}
			\item {\footnotesize Full data centers: Finding cluster centers based on the full data of size $N$. }
			\item {\footnotesize Subset cluster centers: Finding cluster centers based on the subset of size $M$.}
			\item {\footnotesize *: CPU time with $N=10^8$ is calculated over 30 replicates. }
		\end{tablenotes}
	\end{threeparttable}
\end{table}

\subsection{More Proofs}\label{sec:proofs}

\begin{lemma}[\citeauthor{kann1983}, 1983, Theorem 2.2]\label{lemma:ks}
	Suppose $X$ is a finite dimensional space and $f_n:X\rightarrow \mathbb{R}$, $n=0, 1, \ldots$, are strictly convex. Suppose $f_n \rightarrow f_0$ uniformly and $x^*_n={\rm \arg\min}_x f_n(x)$ exists uniquely. Then $x^*_n \rightarrow x^*_0$ as $n$ goes to $\infty$.
\end{lemma}

\begin{proof}
	of {\bf Theorem~\ref{thm:mrglm}}:
	
	According to \citeauthor{pmcc1989} (1989, Section~2.5), the log-likelihood of a GLM with data $\mathscr{D} = \{({\mathbf X}_i, y_i), i=1, \ldots, N\}$, which is fixed throughout the proof, is given by 
	\[
	l(\boldsymbol\beta; {\mathbf y}, \boldsymbol{X})  = 
	\sum_{i=1}^N \left[\frac{y_i \theta({\mathbf X}_i^T \boldsymbol\beta)-b(\theta({\mathbf X}_i^T \boldsymbol\beta))}{a(\phi)}+c(y_i,\phi)\right]
	\]
	where $\theta(\cdot) = (b')^{-1}(g^{-1}(\cdot))$, $a(\cdot)$, $b(\cdot)$ and $c(\cdot, \cdot)$ are known functions, and $\phi$ is the dispersion parameter.
	Given a data partition $\{I_1, \ldots, I_K\}$ of $I=\{1, \ldots, N\}$, which will become finer and finer later, the log-likelihood contributed by the  $k$th block $\mathscr{D}_k = \{({\mathbf X}_i, y_i), i \in I_k\}$ with block size $n_k=\abs{I_k}$ is essentially  
	\[l_k(\boldsymbol\beta) = \sum_{i \in I_k} a(\phi)^{-1}\left[y_i\theta({\mathbf X}_i^T \boldsymbol\beta)- b(\theta({\mathbf X}_i^T \boldsymbol\beta))\right] \]
	while the log-likelihood contributed by the weighted $k$th representative $(n_k, \tilde{\mathbf X}_k, \tilde{y}_k)$ is
	\begin{align}
		\tilde{l}_k(\boldsymbol\beta) =& n_k a(\phi)^{-1}\left[\tilde{y}_k\theta(\tilde{\mathbf X}_k^T \boldsymbol\beta)- b(\theta(\tilde{\mathbf X}_k^T \boldsymbol\beta))\right] \nonumber \\
		=& a(\phi)^{-1}\left(n_k \tilde{y}_k- \sum_{i \in I_k} y_i\right) \theta(\tilde{\eta}_k) + \sum_{i \in I_k} a(\phi)^{-1}\left[y_i\theta(\tilde{\mathbf X}_k^T \boldsymbol\beta)- b(\theta(\tilde{\mathbf X}_k^T \boldsymbol\beta))\right] 
		\nonumber\\
		=& \sum_{i \in I_k} a(\phi)^{-1}\left[y_i\theta(\tilde{\mathbf X}_k^T \boldsymbol\beta)- b(\theta(\tilde{\mathbf X}_k^T \boldsymbol\beta))\right] \nonumber
	\end{align}
	since $\tilde{y}_k = n_k^{-1} \sum_{i\in I_k} y_i$. Then the log-likelihood based on the full data is $l(\boldsymbol\beta) = \sum_{k} l_k(\boldsymbol\beta)$, and the log-likelihood based on the weighted representative data is $\tilde{l}(\boldsymbol\beta) = \sum_k \tilde{l}_k (\boldsymbol\beta)$. 
	
	Recall that the derivative $\partial l/\partial \boldsymbol\beta$ is simply the score function~\eqref{eq:score}.
	By plugging in the first order Taylor expansion of $\tilde{l}_k$ about $\tilde{\mathbf X}_k$ at ${\mathbf X}_i$ and the Cauchy-Schwarz inequality, we have
	\begin{eqnarray*}
		& &\abs{\tilde{l}_k(\boldsymbol\beta)-l_k(\boldsymbol\beta)}\\ 
		&=& \sum_{i \in I_k} \left\{\left[(y_i- G({\mathbf X}_i^T \boldsymbol\beta))\nu({\mathbf X}_i^T \boldsymbol\beta)\right](\tilde{\mathbf X}_k-{\mathbf X}_i)^T \boldsymbol\beta  +o(\|\tilde{\mathbf X}_k-{\mathbf X}_i\|)\right\}\\
		&\leq& \left(\sum_{i \in I_k} (y_i- G({\mathbf X}_i^T \boldsymbol\beta))^2\nu({\mathbf X}_i^T \boldsymbol\beta)^2 \cdot \sum_{i \in I_k}\|\tilde{\mathbf X}_k-{\mathbf X}_i\|^2\cdot \|\boldsymbol\beta\|^2  \right)^{1/2} \\
		& &\ + \sum_{i \in I_k}  o(\|\tilde{\mathbf X}_k-{\mathbf X}_i\|)\\
		&\leq&  n_k \tilde\Delta \|\boldsymbol\beta\| \left(n_k^{-1} \sum_{i \in I_k}(y_i- G({\mathbf X}_i^T \boldsymbol\beta))^2\nu({\mathbf X}_i^T \boldsymbol\beta)^2  \right)^{1/2} + \sum_{i \in I_k}  o(\tilde\Delta)
	\end{eqnarray*}
	Denote $F_k=(n_k^{-1} \sum_{i \in I_k}(y_i- G({\mathbf X}_i^T \boldsymbol\beta))^2\nu({\mathbf X}_i^T \boldsymbol\beta)^2)^{1/2}$. Then for sufficiently small $\tilde\Delta$ and all $\boldsymbol\beta \in B$, we have 
	\begin{align}
		\abs{\tilde{l}(\boldsymbol\beta)-l(\boldsymbol\beta)} &\leq \sum_{k=1}^K n_k \tilde\Delta \|\boldsymbol\beta\| F_k +  \sum_{i =1}^N  o(\tilde\Delta)\nonumber \\
		& \leq N \tilde\Delta  \|\boldsymbol\beta\|\cdot \underset{k}{\max}F_k +N o(\tilde\Delta)\nonumber \\
		& \leq M \tilde\Delta \label{eq:4110}
	\end{align}
	for some $M>0$, which depends on the data, which is given and fixed, but not the representatives or $\tilde\Delta$. That is, $\tilde{l}(\boldsymbol\beta)$ converges to $l(\boldsymbol\beta)$ uniformly for all $\boldsymbol\beta \in B$ as $\tilde{\Delta}$ goes to $0$.
	
	The strict concavity of $l(\boldsymbol\beta)$ implies the existence and uniqueness of $\hat{\boldsymbol\beta} \in B$, such that $\hat{\boldsymbol\beta}={\arg\max}_{\boldsymbol\beta} l(\boldsymbol\beta)$.  Let $\tilde{\boldsymbol\beta}$ maximizes $\tilde{l}(\boldsymbol\beta)$. For sufficiently small $\tilde\Delta$, $\tilde{l}(\boldsymbol\beta)$ is also strictly concave, which guarantees the existence and uniqueness of $\tilde{\boldsymbol\beta}$.
	By Lemma~\ref{lemma:ks}, $\tilde{\boldsymbol\beta}$ converges to $\hat{\boldsymbol\beta}$ as $\tilde\Delta \to 0$.
	
	Since $l(\boldsymbol\beta)$ is twice differentiable and $\pdv{l}{\boldsymbol\beta} (\hat{\boldsymbol\beta})=0$, the second-order Taylor expansion of $l(\boldsymbol\beta)$ at $\hat{\boldsymbol\beta}$ is
	\[
	l(\boldsymbol\beta)= l(\hat{\boldsymbol\beta}) + \frac{1}{2}(\boldsymbol\beta-\hat{\boldsymbol\beta})^T H(\hat{\boldsymbol\beta}) (\boldsymbol\beta-\hat{\boldsymbol\beta}) +o(\|\boldsymbol\beta-\hat{\boldsymbol\beta}\|^2)
	\]
	where $H(\boldsymbol\beta) = \pdv{l}{\boldsymbol\beta}{\boldsymbol\beta^T}$ is the Hessian matrix. Let $\lambda_1$ be the smallest eigenvalue of $H(\hat{\boldsymbol\beta})$. Since $l(\boldsymbol\beta)$ is strictly concave, then $\lambda_1>0$.
	For small enough $\|\boldsymbol\beta-\hat{\boldsymbol\beta}\|$,
	\begin{align}\label{eq:4111}
		|l(\boldsymbol\beta)-l(\hat{\boldsymbol\beta})| > \frac{\lambda_1}{4}  \|\boldsymbol\beta-\hat{\boldsymbol\beta}\|^2
	\end{align}
	We claim that
	\begin{equation}\label{eq:betadifference}
		\|\tilde{\boldsymbol\beta}-\hat{\boldsymbol\beta}\|\leq \left(8M\lambda_1^{-1}\right)^{1/2} \tilde{\Delta}^{1/2}
	\end{equation}
	Actually, if $\|\tilde{\boldsymbol\beta}-\hat{\boldsymbol\beta}\|^2 > 8M\lambda_1^{-1} \tilde\Delta$, then we have
	\begin{equation}\label{eq:4112} l(\hat{\boldsymbol\beta})- l(\tilde{\boldsymbol\beta})> 2M \tilde\Delta
	\end{equation}
	due to \eqref{eq:4111} and $l(\hat{\boldsymbol\beta}) \geq l(\tilde{\boldsymbol\beta})$.
	From \eqref{eq:4110} and \eqref{eq:4112}, we have
	\begin{equation}\label{eq:tildel}
		\tilde{l}(\tilde{\boldsymbol\beta}) \leq l(\tilde{\boldsymbol\beta}) + M\tilde{\Delta} < l(\hat{\boldsymbol\beta})- 2M\tilde{\Delta} + M\tilde{\Delta} =l(\hat{\boldsymbol\beta}) - M\tilde{\Delta}
	\end{equation} 
	On the other hand, since $\tilde{\boldsymbol\beta}$ maximizes $\tilde{l}(\boldsymbol\beta)$, we have
	\[
	\tilde{l}(\tilde{\boldsymbol\beta}) \geq \tilde{l}(\hat{\boldsymbol\beta})
	\geq l(\hat{\boldsymbol\beta}) - M\tilde{\Delta}
	\]
	due to \eqref{eq:4110}. That leads to a contradiction with \eqref{eq:tildel}. Thus \eqref{eq:betadifference} is justified and 
	\[
	\|\tilde{\boldsymbol\beta}-\hat{\boldsymbol\beta}\|=O( \tilde\Delta^{1/2}) 
	\]
\end{proof}

\medskip
Recall that $\mathbf{X}_i = (h_1({\mathbf x}_i), \ldots, h_p({\mathbf x}_i))^T$ in general. Denote the $k$th representative $\tilde{\mathbf X}_k = (\tilde{X}_{k1}, \ldots, \tilde{X}_{kp})^T$. 

\begin{corollary}\label{cor:center}
	All three center representative estimates $\tilde{\boldsymbol\beta} \rightarrow \hat{\boldsymbol\beta}$ as $\Delta \to 0$.
\end{corollary}

\begin{proof} of
	{\bf Corollary~\ref{cor:center}}:
	For mean representatives,  $\tilde{\mathbf X}_k = n_k^{-1} \sum_{i\in I_k} \mathbf{X}_i$. Then 
	\[
	\max_{i\in I_k} \|\mathbf{X}_i - \tilde{\mathbf X}_k\| = \max_{i\in I_K} \|n_k^{-1} \sum_{j\in I_k} (\mathbf{X}_i - \mathbf{X}_j)\| \leq \max_{i,j\in I_k} \|\mathbf{X}_i - \mathbf{X}_j\|
	\] 
	Therefore, $\tilde{\Delta} = \max_k \max_{i\in I_k} \|\mathbf{X}_i - \tilde{\mathbf X}_k\| \leq \max_k \max_{i,j\in I_k} \|\mathbf{X}_i - \mathbf{X}_j\| = \Delta$. Thus $\Delta \rightarrow 0$ also implies $\tilde{\boldsymbol\beta} \rightarrow \hat{\boldsymbol\beta}$. 
	
	For median representatives, $\tilde{X}_{kl} = {\rm median}(\{h_l({\mathbf x}_i) \mid i\in I_k\})$, $l=1, \ldots, p$. Then
	\begin{eqnarray*}
		\max_{i\in I_k} \|{\mathbf X}_i - \tilde{\mathbf X}_k\| &=& \max_{i\in I_k} \left\{\sum_{l=1}^p [h_l(\mathbf{x}_i) - \tilde{X}_{kl}]^2\right\}^{1/2}\\
		&\leq & \max_{i\in I_k} \left\{\sum_{l=1}^p \max_{j\in I_k} [h_l(\mathbf{x}_i) - h_l(\mathbf{x}_j)]^2\right\}^{1/2}\\
		&\leq & \max_{i\in I_k} \left[\sum_{l=1}^p \max_{j\in I_k} \norm{\mathbf{X}_i - \mathbf{X}_j}^2\right]^{1/2}\\
		&=& p^{1/2} \cdot \max_{i,j\in I_k} \norm{\mathbf{X}_i - \mathbf{X}_j}
	\end{eqnarray*}
	Therefore, $\tilde{\Delta} \leq p^{1/2} \Delta$.
	If $\Delta \rightarrow 0$, then $\tilde{\boldsymbol\beta} \rightarrow \hat{\boldsymbol\beta}$ for median representatives. 
	
	For mid-point representatives, the $k$th block $I_k$ is typically defined by the grid points $-\infty < a_{kl} \leq b_{kl} < \infty$ such that, $i\in I_k$ if and only if $a_{kl} \leq h_l(\mathbf{x}_i) \leq b_{kl}$, $l=1, \ldots, p$. In this case, we redefine $\Delta = \left(\sum_{l=1}^p (b_{kl} - a_{kl})^2\right)^{1/2}\geq \tilde{\Delta}$. Then $\tilde{\boldsymbol\beta} \rightarrow \hat{\boldsymbol\beta}$ as $\Delta \to 0$. 
\end{proof}

When all the covariates are categorical or have finite discrete values, one may partition the data according to distinct ${\mathbf X}_i$'s. In this case, $\Delta = 0$.

\begin{corollary}\label{cor:cate}
	Let $\Delta$ be $\max_k \max_{i,j\in I_k} \|\mathbf{X}_i - \mathbf{X}_j\|$ for mean and median representatives, and $\left(\sum_{l=1}^p (b_{kl} - a_{kl})^2\right)^{1/2}$ for mid-point representative. Under the conditions of Theorem~\ref{thm:mrglm},  $\tilde{\boldsymbol\beta} \rightarrow \hat{\boldsymbol\beta}$ as $\Delta \to 0$ and  $\|\tilde{\boldsymbol\beta}-\hat{\boldsymbol\beta}\|=O(\Delta^{1/2})$.
	If $\Delta = 0$, then the mid-point, median, and mean representative approaches are the same and all satisfy $\tilde{\boldsymbol\beta} = \hat{\boldsymbol\beta}$.
\end{corollary}

\begin{proof} of
	{\bf Corollary~\ref{cor:cate}}:
	Since $\tilde{\Delta} \leq \Delta$ for mean and mid-point representatives, and $\tilde{\Delta} \leq p^{1/2} \Delta$ for median representatives, then under the conditions of Theorem~\ref{thm:mrglm}, $\tilde{\boldsymbol\beta} \rightarrow \hat{\boldsymbol\beta}$ as $\Delta \to 0$ and  $\|\tilde{\boldsymbol\beta}-\hat{\boldsymbol\beta}\|=O(\Delta^{1/2})$.
	
	If $\Delta=0$, then in each block all the predictor variables are the same, that is, ${\mathbf X}_i\equiv \bar{\mathbf X}_k$ for $i \in I_k$, $k=1,\dots, K$.  Therefore, $\tilde{l}_k(\boldsymbol\beta) = l_k(\boldsymbol\beta)$, $k=1, \ldots, K$ and $\tilde{\boldsymbol\beta} = \hat{\boldsymbol\beta}$.
\end{proof}

\begin{proof} of
	{\bf Theorem~\ref{thm:smrglm}}:
	Since $\tilde{y}_k = n_k^{-1} \sum_{i\in I_k} y_i + O(\tilde{\Delta})$, following a similar argument for Theorem~\ref{thm:mrglm}, we can obtain $\tilde{\boldsymbol\beta}^{(t)}$ converges to  $\hat{\boldsymbol\beta}$ as $\tilde{\Delta}$ goes to $0$ and $\|\tilde{\boldsymbol\beta}^{(t)} - \hat{\boldsymbol\beta}\| = O(\tilde{\Delta}^{1/2})$.
\end{proof}

\begin{proof} of
	{\bf Theorem~\ref{thm:smr1}}:
	Let $s(\boldsymbol\beta) = s(\boldsymbol\beta; {\mathbf y}, {\bm X})$ as in \eqref{eq:score} be the score function based on the full data satisfying $s(\hat{\boldsymbol\beta})=0$. Let 
	\[
	\tilde{s}(\boldsymbol\beta) = \sum_{k=1}^K n_k (\tilde{y}_k^{(t+1)} - G(\boldsymbol\beta^T \tilde{\mathbf X}_k^{(t+1)})) \nu(\boldsymbol\beta^T \tilde{\mathbf X}_k^{(t+1)})  \tilde{\mathbf X}_k^{(t+1)}
	\] 
	be the score function based on the representative data points for the $(t+1)$th iteration satisfying $\tilde{s}(\tilde{\boldsymbol\beta}^{(t)})=s(\tilde{\boldsymbol\beta}^{(t)})$ and  $\tilde{s}(\tilde{\boldsymbol\beta}^{(t+1)}) = 0$. 
	Consider their first-order Taylor expansions at $\tilde{\boldsymbol\beta}^{(t)}$: 
	\begin{align}
		s(\hat{\boldsymbol\beta})&= s(\tilde{\boldsymbol\beta}^{(t)}) + H(\tilde{\boldsymbol\beta}^{(t)}) (\hat{\boldsymbol\beta}-\tilde{\boldsymbol\beta}^{(t)}) + o(\| \hat{\boldsymbol\beta}-\tilde{\boldsymbol\beta}^{(t)}\|)\label{eq:htaylor1}\\
		\tilde{s}(\tilde{\boldsymbol\beta}^{(t+1)})&= \tilde{s}(\tilde{\boldsymbol\beta}^{(t)}) + \tilde{H}(\tilde{\boldsymbol\beta}^{(t)}) (\tilde{\boldsymbol\beta}^{(t+1)}-\tilde{\boldsymbol\beta}^{(t)}) + o(\| \tilde{\boldsymbol\beta}^{(t+1)}-\tilde{\boldsymbol\beta}^{(t)}\|)\label{eq:htaylor2}
	\end{align}
	where
	\begin{align*}
		H(\tilde{\boldsymbol\beta}^{(t)}) &=\pdv{s}{\boldsymbol\beta}\Big|_{\boldsymbol\beta=\tilde{\boldsymbol\beta}^{(t)}} = \sum_{k=1}^K \sum_{i \in I_k} \left[(y_i-G(\eta_i))\nu'(\eta_i)-G'(\eta_i)\nu(\eta_i)\right] \boldsymbol{\mathrm{X}}_i \boldsymbol{\mathrm{X}}_i^T\\
		\tilde{H}(\tilde{\boldsymbol\beta}^{(t)}) &=\pdv{\tilde{s}}{\boldsymbol\beta}\Big|_{\boldsymbol\beta=\tilde{\boldsymbol\beta}^{(t)}} \\
		&= \sum_{k=1}^K n_k \left[(\tilde{y}_k^{(t+1)}-G(\tilde{\eta}_k))\nu'(\tilde{\eta}_k)-G'(\tilde{\eta}_k)\nu(\tilde{\eta}_k)\right] \tilde{\mathbf X}_k ^{(t+1)} \left(\tilde{\mathbf X}_k^{(t+1)}\right)^T
	\end{align*}
	with $\eta_i=\boldsymbol{\mathrm{X}}_i^T \tilde{\boldsymbol\beta}^{(t)}$ and $\tilde{\eta}_k = \left(\tilde{\mathbf X}_k^{(t+1)}\right)^T \tilde{\boldsymbol\beta}^{(t)}$. 
	Subtracting \eqref{eq:htaylor1} from \eqref{eq:htaylor2}, we obtain
	\[
	\tilde{\boldsymbol\beta}^{(t+1)} - \tilde{\boldsymbol\beta}^{(t)} = \tilde{H}(\tilde{\boldsymbol\beta}^{(t)})^{-1} H(\tilde{\boldsymbol\beta}^{(t)}) (\hat{\boldsymbol\beta} - \tilde{\boldsymbol\beta}^{(t)}) + o(\| \tilde{\boldsymbol\beta}^{(t+1)}-\tilde{\boldsymbol\beta}^{(t)}\|)+o(\| \hat{\boldsymbol\beta}-\tilde{\boldsymbol\beta}^{(t)}\|)
	\]
	Therefore,
	\begin{eqnarray*}
		& & \tilde{\boldsymbol\beta}^{(t+1)} - \hat{\boldsymbol\beta}\\
		&=& (\tilde{\boldsymbol\beta}^{(t+1)}-\tilde{\boldsymbol\beta}^{(t)}) - (\hat{\boldsymbol\beta}-\tilde{\boldsymbol\beta}^{(t)})\\
		&=& \left(I - \tilde{H}(\tilde{\boldsymbol\beta}^{(t)})^{-1} H(\tilde{\boldsymbol\beta}^{(t)})\right) (\tilde{\boldsymbol\beta}^{(t)}- \hat{\boldsymbol\beta}) + o(\| \tilde{\boldsymbol\beta}^{(t+1)}-\tilde{\boldsymbol\beta}^{(t)}\|)+o(\| \hat{\boldsymbol\beta}-\tilde{\boldsymbol\beta}^{(t)}\|)
	\end{eqnarray*}
	According to Theorem~\ref{thm:smrglm}, as $\tilde{\Delta} \rightarrow 0$, $\|\tilde{\boldsymbol\beta}^{(t)} - \hat{\boldsymbol\beta}\| = O(\tilde{\Delta}^{1/2})$ and $\|\tilde{\boldsymbol\beta}^{(t+1)} - \hat{\boldsymbol\beta}\| = O(\tilde{\Delta}^{1/2})$. Then 
	$\|\tilde{\boldsymbol\beta}^{(t+1)}-\tilde{\boldsymbol\beta}^{(t)}\| \leq \|\tilde{\boldsymbol\beta}^{(t+1)} - \hat{\boldsymbol\beta}\|$ $+$ $\|\tilde{\boldsymbol\beta}^{(t)} - \hat{\boldsymbol\beta}\| = O(\tilde{\Delta}^{1/2})$. Thus  
	\[
	\tilde{\boldsymbol\beta}^{(t+1)} - \hat{\boldsymbol\beta} = \left(I - \tilde{H}(\tilde{\boldsymbol\beta}^{(t)})^{-1} H(\tilde{\boldsymbol\beta}^{(t)})\right) (\tilde{\boldsymbol\beta}^{(t)}- \hat{\boldsymbol\beta}) + o(\tilde{\Delta}^{1/2}) 
	\]
	According to condition~\ref{itm:cond2}, $\tilde{y}_k^{(t+1)} = \bar{y}_k + O(\tilde{\Delta})$. Since $\mathbf{X}_i = \tilde{\mathbf X}_k^{(t+1)}  + O(\tilde{\Delta})$ for $i \in I_k$, it can be verified that
	$H(\tilde{\boldsymbol\beta}^{(t)}) = \tilde{H}(\tilde{\boldsymbol\beta}^{(t)})  +O(\tilde{\Delta})$. Therefore, $I - \tilde{H}(\tilde{\boldsymbol\beta}^{(t)})^{-1} H(\tilde{\boldsymbol\beta}^{(t)}) = I - \tilde{H}(\tilde{\boldsymbol\beta}^{(t)})^{-1} \left(\tilde{H}(\tilde{\boldsymbol\beta}^{(t)})  +O(\tilde{\Delta})\right) =  O(\tilde{\Delta})$.
	Let $\rho(\tilde{\Delta})$ be the largest eigenvalue of $I-\tilde{H}(\tilde{\boldsymbol\beta}^{(t)})^{-1} H(\tilde{\boldsymbol\beta}^{(t)})$. Then $\rho(\tilde{\Delta}) = O(\tilde{\Delta})$, which is strictly less than $1$ for sufficiently small $\tilde{\Delta}$. Therefore,  
	\begin{align}
		\|\tilde{\boldsymbol\beta}^{(t+1)}-\hat{\boldsymbol\beta}\| \leq \rho(\tilde{\Delta}) \|\tilde{\boldsymbol\beta}^{(t)}-\hat{\boldsymbol\beta}\| +o(\tilde{\Delta}^{1/2})\label{eq:converge}
	\end{align}
	It guarantees that $\tilde{\boldsymbol\beta}^{(t)} \rightarrow \hat{\boldsymbol\beta}$ as $t\rightarrow \infty$ and $\tilde{\Delta} \rightarrow 0$.
\end{proof}

\begin{proof} of
	{\bf Corollary~\ref{cor:cate2}}:
	If $\tilde{\Delta}=0$, by definition we have ${\mathbf X}_i=\tilde{\mathbf X}_k$ and $\eta_i=\tilde{\eta}_k$ for all $i\in I_k$. Therefore, $\tilde{y}_k=\bar{y}_k$ and $\tilde{\mathbf X}_k = \bar{\mathbf X}_k$ for both MR and SMR.
	
	If $\Delta = 0$, then ${\mathbf X}_i = {\mathbf X}_j$ and $\eta_i = \eta_j$ for all $i,j \in I_k$ and thus $\tilde{y}_k = \bar{y}_k$ according to \eqref{eq:ytilde}. Since $\eta_i = \bar{\eta}_k$ for all $i\in I_k$, then $\bar{\eta}_k$ is a solution for solving \eqref{eq:ssb}. Since we always choose a solution closest to $\bar{\eta}_k$ when the solutions are not unique, the proposed SMR have $\tilde{\eta}_k = \bar{\eta}_k$. In this case, we have $\tilde{\mathbf X}_k = \bar{\mathbf X}_k$ by \eqref{eq:xtilde}.
	
	In both cases, SMR and MR estimates are the same. By Corollary~\ref{cor:cate}, we know both of them equal to the full data estimate $\hat{\boldsymbol\beta}$.
\end{proof}

\subsection{More on  Airline on-time Performance Data}\label{sec:cases}

In the simulation study (see Table~\ref{tab:case}) based on the oracle regression coefficients $\boldsymbol\beta$, MR and SMR perform about the same, which is mainly due to the tiny oracle coefficient $7.955\times 10^{-5}$ of the only continuous predictor DISTANCE. 

In order to verify when SMR is better than MR, we inflate the oracle coefficient of DISTANCE by $10$ times to get a new oracle $\boldsymbol\beta'$ (see Table~\ref{tab:oracle}). Then the maximum contribution of the predictor DISTANCE is about $4$, which is expected to play a more important role in predicting {\tt Late Arrival}. We redo the simulation study in Section~\ref{sec:case} with the new oracle $\boldsymbol\beta'$ and list the corresponding results in Table~\ref{tab:case10}. Clearly the SMR estimates show consistent advantage over MR's estimate throughout all data sizes. Both MR and SMR estimates based on $387$-month data are better than the last available full-data estimate based on the $120$-month data. 

\begin{figure}[bt]
	\centering
	\includegraphics[width=\textwidth]{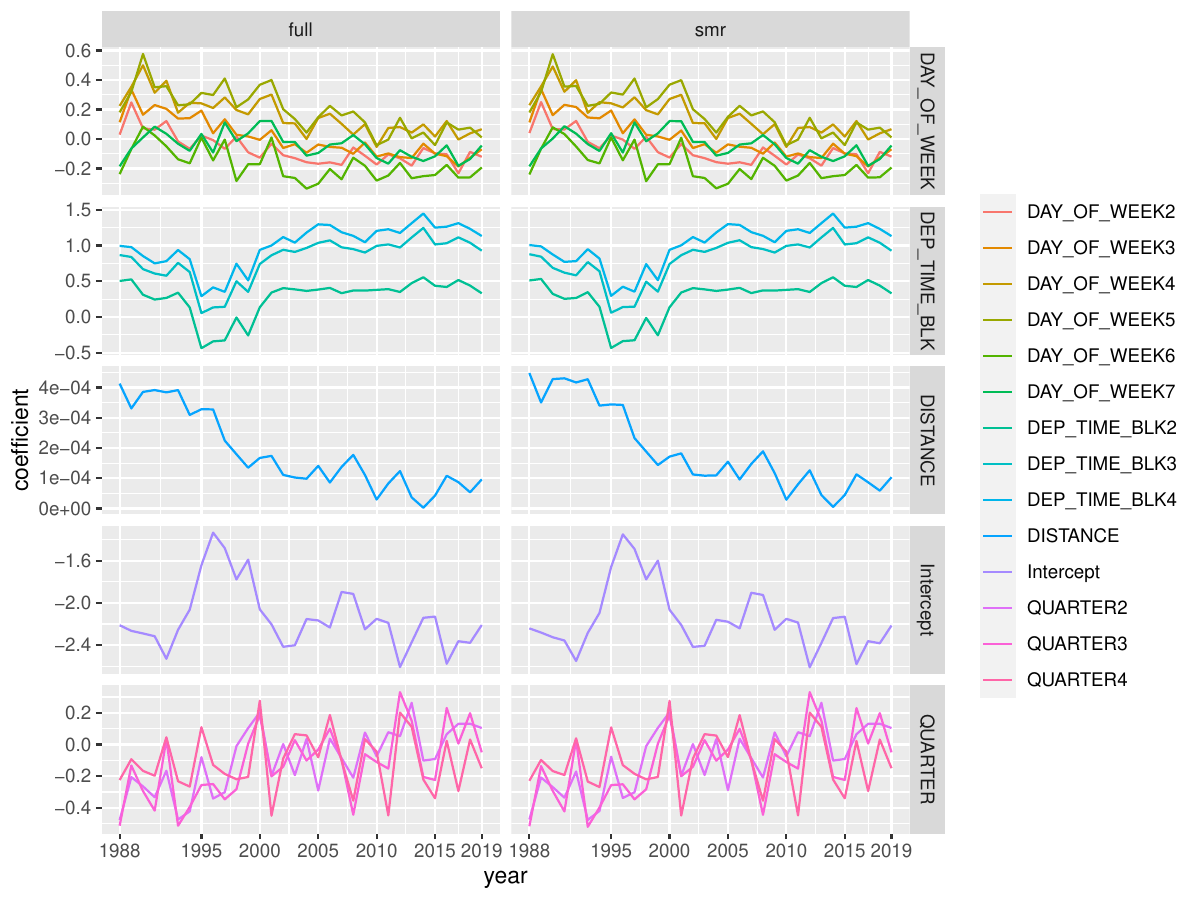}
	\caption{Regression coefficients fitted yearly by original GLM (full) or SMR algorithms}\label{fig:flight}
\end{figure}

The rest of this section provides detailed description of the Airline raw data (Table~\ref{tab:flightorg}) and working data (Table~\ref{tab:casevar}), the MR and SMR estimates based on all available $387$ months (Table~\ref{tab:flight}), and regression coefficients fitted yearly based on the original GLM (full) or SMR algorithm (Figure~\ref{fig:flight}).

\begin{table}[!h]
	\centering
	\caption{Oracle coefficients of predictors}\label{tab:oracle}
	\begin{threeparttable}
		\begin{tabular}{crr}
			\hline
			\bf{Predictor} & Oracle $\boldsymbol\beta$ & Inflated oracle $\boldsymbol\beta'$\\ 
			\hline
			{\tt Intercept} & -2.322 & -2.322 \\ 
			{\tt QUARTER2} & 7.083e-02 & 7.083e-02 \\ 
			{\tt QUARTER3} & 3.215e-02 & 3.215e-02 \\ 
			{\tt QUARTER4} & -1.396e-01 & -1.396e-01 \\ 
			{\tt DAY\_OF\_WEEK2} & -1.251e-01 & -1.251e-01 \\ 
			{\tt DAY\_OF\_WEEK3} & -1.178e-01 & -1.178e-01 \\ 
			{\tt DAY\_OF\_WEEK4} & 4.755e-02 & 4.755e-02 \\ 
			{\tt DAY\_OF\_WEEK5} & 4.443e-02 & 4.443e-02 \\ 
			{\tt DAY\_OF\_WEEK6} & -2.268e-01 & -2.268e-01 \\ 
			{\tt DAY\_OF\_WEEK7} & -1.022e-01 & -1.022e-01 \\ 
			{\tt DEP\_TIME\_BLK2} & 4.229e-01 & 4.229e-01 \\ 
			{\tt DEP\_TIME\_BLK3} & 1.019e+00 & 1.019e+00 \\ 
			{\tt DEP\_TIME\_BLK4} & 1.230e+00 & 1.230e+00 \\ 
			{\tt DISTANCE} & 7.955e-05 & {\bf 7.955e-04} \\ 
			\hline
		\end{tabular}
	\end{threeparttable}
\end{table}

\begin{table}[!h]
	\centering
	\caption{Average (std) of RMSEs ($10^{-3}$) from oracle $\boldsymbol\beta'$ for Airline on-time performance data}\label{tab:case10}
	\begin{threeparttable}
		\begin{tabular}{crrr}
			\hline
			{\bf Number of months} &  {\bf Full} & {\bf MR} &  {\bf SMR} \\ 
			\hline
			60 months& 10.289 (3.814) & 11.235 (5.937) & 10.793 (5.298) \\ 
			120 months& 8.309 (4.306) & 9.557 (5.221) & 8.919 (5.028) \\ 
			240 months& - & 9.019 (5.016) & 8.482 (4.754) \\ 
			387 months& - & 7.916 (4.282) & 7.512 (4.162) \\  
			\hline
		\end{tabular}
		\begin{tablenotes}[noitemsep]
			\item {\footnotesize ``-": Full data estimates for 240 months and 387 months are not available due to memory limitation}
		\end{tablenotes}
	\end{threeparttable}
\end{table}

\begin{table}[!h]
	\centering
	\caption{Description of fields in the airline raw data}\label{tab:flightorg}
	\begin{threeparttable}
		\resizebox{\columnwidth}{!}{
			\begin{tabular}{  p{4.5cm}  p{11cm} } 
				\hline
				\bf{Field Name}   & \bf{Description}\\
				\hline
				{\tt YEAR} &  Year, from 1987 to 2017 \\
				{\tt QUARTER} & Quarter (1-4)  \\
				{\tt MONTH} &  Month \\
				{\tt DAY\textunderscore OF\textunderscore MONTH} & Day of Month  \\
				{\tt DAY\textunderscore OF\textunderscore WEEK} &  Day of Week \\
				{\tt FL\textunderscore DATE} & Flight Date (yyyymmdd)  \\
				{\tt ORIGIN\textunderscore AIRPORT\textunderscore ID} &  Origin Airport, Airport ID. An identification number assigned by US DOT to identify a unique airport. Use this field for airport analysis across a range of years because an airport can change its airport code and airport codes can be reused. \\
				{\tt DEST\textunderscore AIRPORT\textunderscore ID} &  Destination Airport, Airport ID. An identification number assigned by US DOT to identify a unique airport. Use this field for airport analysis across a range of years because an airport can change its airport code and airport codes can be reused. \\
				{\tt CRS\textunderscore DEP\textunderscore TIME} & CRS Departure Time (local time: hhmm)  \\
				{\tt DEP\textunderscore DELAY} & Difference in minutes between scheduled and actual departure time. Early departures show negative numbers.  \\
				{\tt DEP\textunderscore DELAY\textunderscore GROUP} & Departure Delay intervals, every (15 minutes from $<-15$ to $>180$)  \\
				{\tt DEP\textunderscore TIME\textunderscore BLK} &  CRS Departure Time Block, Hourly Intervals \\
				{\tt CRS\textunderscore ARR\textunderscore TIME} & CRS Arrival Time (local time: hhmm)  \\
				{\tt ARR\textunderscore DELAY} &  Difference in minutes between scheduled and actual arrival time. Early arrivals show negative numbers. \\
				{\tt ARR\textunderscore DELAY\textunderscore GROUP} &  Arrival Delay intervals, every (15-minutes from $<-15$ to $>180$) \\
				{\tt ARR\textunderscore TIME\textunderscore BLK} &  CRS Arrival Time Block, Hourly Intervals \\
				{\tt CANCELLED} &  Cancelled Flight Indicator (1=Yes) \\
				{\tt CANCELLATION\textunderscore CODE} & Specifies The Reason For Cancellation  \\
				{\tt DIVERTED} & Diverted Flight Indicator (1=Yes)  \\
				{\tt CRS\textunderscore ELAPSED\textunderscore TIME} & CRS Elapsed Time of Flight, in Minutes  \\
				{\tt DISTANCE} & Distance between airports (miles)  \\
				{\tt DISTANCE\textunderscore GROUP} &  Distance Intervals, every 250 Miles, for Flight Segment  \\
				\hline
			\end{tabular}
		}
	\end{threeparttable}
\end{table}

\begin{table}[!h]
	\caption{Description of fields in the airline working data}\label{tab:casevar}
	\centering
	\begin{threeparttable}
		\begin{tabular}{  p{3cm}  p{7cm} } 
			\hline
			\bf{Field Name} & \bf{Description}\\
			\hline
			{\tt ArrDel15} & binary response variable: arrival delay indicator, 15 minutes or more (1=Yes)  \\		
			{\tt QUARTER} & season,``1": January 1-March 31, ``2": April 1-June 30, ``3": July 1-September 30,\\
			& ``4": October 1-December 31  \\
			{\tt DayOfWeek} &  day of week, ``1": Monday, ``2": Tuesday, ``3": Wednesday, ``4": Thursday, ``5": Friday,\\
			& ``6": Saturday, ``7": Sunday \\
			{\tt DepTimeBlk}& CRS departure time block, ``1": 12:00 AM - 05:59 AM, ``2": 06:00 AM - 11:59 AM,\\ & ``3": 12:00 PM - 05:59 PM, ``4": 06:00 PM - 11:59 PM\\
			{\tt DISTANCE} & distance between airports, in miles  \\ 
			\hline  
		\end{tabular}
	\end{threeparttable}
\end{table}

\begin{table}[!h]
	\centering
	\caption{MR and SMR estimates based on all $387$ months}\label{tab:flight}
	\begin{threeparttable}
		\begin{tabular}{crr}
			\hline
			\bf{Predictor}& MR& SMR \\ 
			\hline
			{\tt Intercept} & -1.984 & -1.981 \\ 
			{\tt QUARTER2} & -8.628e-02 & -8.584e-02 \\ 
			{\tt QUARTER3} & -1.163e-01 & -1.163e-01 \\ 
			{\tt QUARTER4} & -8.917e-02 & -8.906e-02 \\ 
			{\tt DAY\_OF\_WEEK2} & -7.486e-02 & -7.587e-02 \\ 
			{\tt DAY\_OF\_WEEK3} & -3.527e-03 & -4.684e-03 \\ 
			{\tt DAY\_OF\_WEEK4} & 1.468e-01 & 1.461e-01 \\ 
			{\tt DAY\_OF\_WEEK5} & 1.775e-01 & 1.773e-01 \\ 
			{\tt DAY\_OF\_WEEK6} & -1.844e-01 & -1.854e-01 \\ 
			{\tt DAY\_OF\_WEEK7} & -4.481e-02 & -4.562e-02 \\ 
			{\tt DEP\_TIME\_BLK2} & 1.288e-01 & 1.274e-01 \\ 
			{\tt DEP\_TIME\_BLK3} & 6.739e-01 & 6.720e-01 \\ 
			{\tt DEP\_TIME\_BLK4} & 8.679e-01 & 8.665e-01 \\ 
			{\tt DISTANCE} & 1.732e-04 & 1.726e-04 \\ 
			\hline
		\end{tabular}
	\end{threeparttable}
\end{table}

\end{document}